\documentclass[amstex,epsf,amssymb,usenatbib]{mn2e}

\usepackage{epsf,graphics}
\usepackage{amsmath,amssymb,natbib}
\usepackage{rotating,times,pictex,graphicx,latexsym,color,longtable}
\newcommand{\mic}{\,$\mu$m}

\title{Molecular jets in the DR21/W75N high-mass star-forming region}
    
\author[M.D. Smith, et al.]
{Michael Smith $^{1}$\thanks{E-mail: m.d.smith@kent.ac.uk, }
{Chris J. Davis $^{2}$\thanks{E-mail: c.j.davis@ljmu.ac.uk}}
{Jonathan H. Rowles $^{1}$}
\& {Michael Knight $^{1}$}
\\
$^1$Centre for Astrophysics \& Planetary Science, The University of Kent, Canterbury, Kent CT2 7NH, U.K. \\
$^2$Astrophysics Research Institute, Liverpool John Moores University, Liverpool, U.K.}                                                                                                                                                             
\date{Accepted .....
      Received ..... ;
      in original form .....}
                                                                                                                                                             
\pagerange{\pageref{firstpage}--\pageref{lastpage}}
\pubyear{2013}
                                                                                                                                                             
\begin{document}
                                                                                                                                                             
\maketitle
                                                                                                                                                             
\label{firstpage}
                                                                                                                                                             
\begin{abstract}

Molecular jets have been discovered in large numbers, spread throughout star formation regions. They can usually be traced back to embedded driving protostars.
We here investigate a squadron of such molecular hydrogen jets in the DR21/W75N region through echelle spectroscopy  of the near infrared
 v\,=\,1-0\,S(1) emission line centred at 2.122~$\mu$m. We detect 79 components,   a number of which possess radial velocities in excess of 80~km~s$^{-1}$.   
 The majority of the components exhibit blue shifts.The regions closer to DR21 exhibit more blue-shifted components
  suggesting that extinction is important across individual flows and is higher near DR21. We provide a classification scheme for the resulting collection of 
  position-velocity diagrams, including other published data.   One  prominent class is  associated with pairs of shocks well separated in radial velocity. 
  We use hydrodynamic simulations with molecular cooling and chemistry to show that these are associated with Mach discs and bow shocks. We also 
  employ a steady-state bow shock model to interpret other revealing position-velocity diagrams. We consider mechanisms which can generate vibrationally-excited 
  hydrogen molecules  moving at speeds well beyond the breakdown speed permitted for shock excitation. We conclude that the molecules have formed 
  within the jets well before being excited by  internal shocks triggered by impacts with the ambient clouds.  We also note the relatively high number of high 
  blue-shifted radial velocity components and argue that these must be  associated with high-density molecular jets from Class 0 protostars which are obscured 
  unless we are selectively viewing  within a conical cavity containing the jet. 
  \end{abstract}
                                                                                                                                                             
\section{Introduction}              
\label{intro}
   
Collections of collimated molecular  outflows are detected in molecular clouds, emerging from a large fraction of  embedded protostars. They can take the form of jets 
and aligned series of  knots  \citep{2010A&A...511A..24D,2011MNRAS.413..480F,2013ApJ...767..147G}  or  extended lobes \citep{2010MNRAS.408.1516C,2012MNRAS.425.2641N}. 
They may be a consequence of the   magnetohydrodynamic mechanism  which is believed to regulate the mass of the star and the planet-forming disc \citep{1994ApJ...429..781S, 2012ajb..book.....S}.
 In addition, by transporting mass, momentum and energy away from the core, as well as stirring it, jets can both accelerate and regulate the infall process \citep{2010ApJ...709...27W}.  
 
Surveys for jets over wide fields are now providing large unbiased sets of data with which we can analyse their general properties \citep{2012MNRAS.421.3257I}. In particular, 
molecular outflows can be searched for  in the   near-infrared through wide-field cameras with narrow-band filters tuned to vibrational emission lines, especially the K-band 2.12\,$\mu$m line.
 Collections of jets have thus been made and their properties analysed,.as first achieved on a large scale by \citet{1998A&A...332..307S} for the Orion Molecular Cloud.
    
 Such a survey was also performed for the massive star-forming complex DR21/W75N in Cygnus~X  using the Wide Field Camera (WFCAM) on the United Kingdom Infrared 
 Telescope (UKIRT) \citep{2007MNRAS.374...29D}  It revealed H$_2$ structure in the form of knots and bow shocks associated with more than 50 distinct flows. 
 Most appear to be driven by embedded, low-mass protostars. The orientations of the outflows, particularly from the few higher mass sources in the region (DR21, DR21(OH), W75N and ERO 1), display a preference to be roughly orthogonal to the chain of dusty cores that runs north-south through DR21. The outflows are, however,  widely scattered with no enhanced outflow activity associated with clusters of protostars.
  
Prominent individual outflows in the DR21/W75N region have long been studied. One of the most spectacular flows known is associated with DR21~\citep{1991ApJ...366..474G, 1992ApJ...392..602G} 
which is bipolar and extremely bright in H$_2$ line emission~\citep{1991ApJ...366..474G,1996A&A...310..961D, 1998MNRAS.297..687S}. The DR21 outflow is split into north-eastern and south 
western lobes which are separated by part of the dense molecular ridge. A distinct outflow aligned perpendicular to this was identified through bipolar CO lobes \citep{1991ApJ...374..540G}. 
One proposal for generating the main outflow is that of aligned  interacting outflows from several massive protostars \citep{2014arXiv1404.5463P}.
An alternative is that of a single explosive disintegration event that occurred 10,000 years ago \citep{2013ApJ...765L..29Z}, supported by the properties of a number of  submillimetre 
CO filaments stretching radially away from the central region. An outflow fainter in H$_2$ emanates from W75N \citep{1985ApJ...293..508F}. This could be an evolved  wind-driven flow 
with sweeping bow shocks that entrain ambient gas, forming a massive CO molecular outflow~\citep{1998AJ....115.1118D, 1998MNRAS.299..825D}. although the mass flow rate 
again suggests more than driving source is responsible \citep{2001ApJ...546..345S,2003ApJ...584..882S}.

  To evaluate models for feeding and feedback requires a quantitative study of  physical and dynamical  properties. However, many of the derived results then  require knowledge of the jet speed. These include all the flow rates. the outflow age and time lapse between outbursts \citep{2012MNRAS.425.1380I}. Moreover, with spatial information, we can distinguish explosive events from continuously-driven jets: if the structures are produced in a single explosive event, we could expect a linear relationship between the radial speed and the projected distance from the protostar.

  Jet speeds of different kinds can be inferred from proper motion and radial velocity measurements. For the shock-excited H$_2$, proper motions measure the displacement of shock waves or, more accurately, the changing location of the brightest section of a shock front. On the other hand, the radial velocity measures the instant line-of-sight speed of the hot material. For distant regions such as DR21, only radial velocities can be detected within a reasonable time frame. When spatially resolved, position-velocity diagrams can also help locate the protostar   through the  identification of  jet and counter-jet pairs.
     
Here, we determine the radial velocities for a collection of jets identified by \citet{2007MNRAS.374...29D}.  Most are well collimated, and at least five qualify as parsec-scale flows. Linear features are well suited to analysis by long-slit spectroscopy. Resulting position-velocity diagrams  can then be interpreted in terms of the underlying configuration (bow shocks, Mach discs, internal oblique shocks, internal  advance-reverse pairs, magnetic reconnection events, turbulent boundary layers or combinations) and the underlying physics (J-type, C-type or hybrid). We can thus infer whether it is the varying driving source or the impacted environment that is mostly responsible for the radiating portions of the material jet. 

A major issue associated with the main DR21 outflow  is the speed with which much of the molecular hydrogen moves. Although close to the plane of the sky, line profiles are as wide as 140\,km~s$^{-1}$ \citep{1991ApJ...366..474G} and peak radial velocities display blueshifts exceeding  80\,km~s$^{-1}$ \citep{1996A&A...310..961D}. These components exceed the  theoretical H$_2$ dissociation limit for acceleration within J-shocks \citep{1995A&A...296..789S} and C-shocks unless associated with very low Alfv\'en Mach numbers  \citep{1991MNRAS.248..730S}.  In this study, we will demonstrate that such enigmatic molecular speeds 
are also associated with the  outflows driven from the low-mass protostars in the vicinity.                                                                                                      

\begin{figure*}
\beginpicture
\setcoordinatesystem units <0.2992cm,0.29776cm> point at 0 0
\setplotarea x from 0 to 53.333 , y from 0 to 53.7333
\put {\includegraphics[width=16cm]{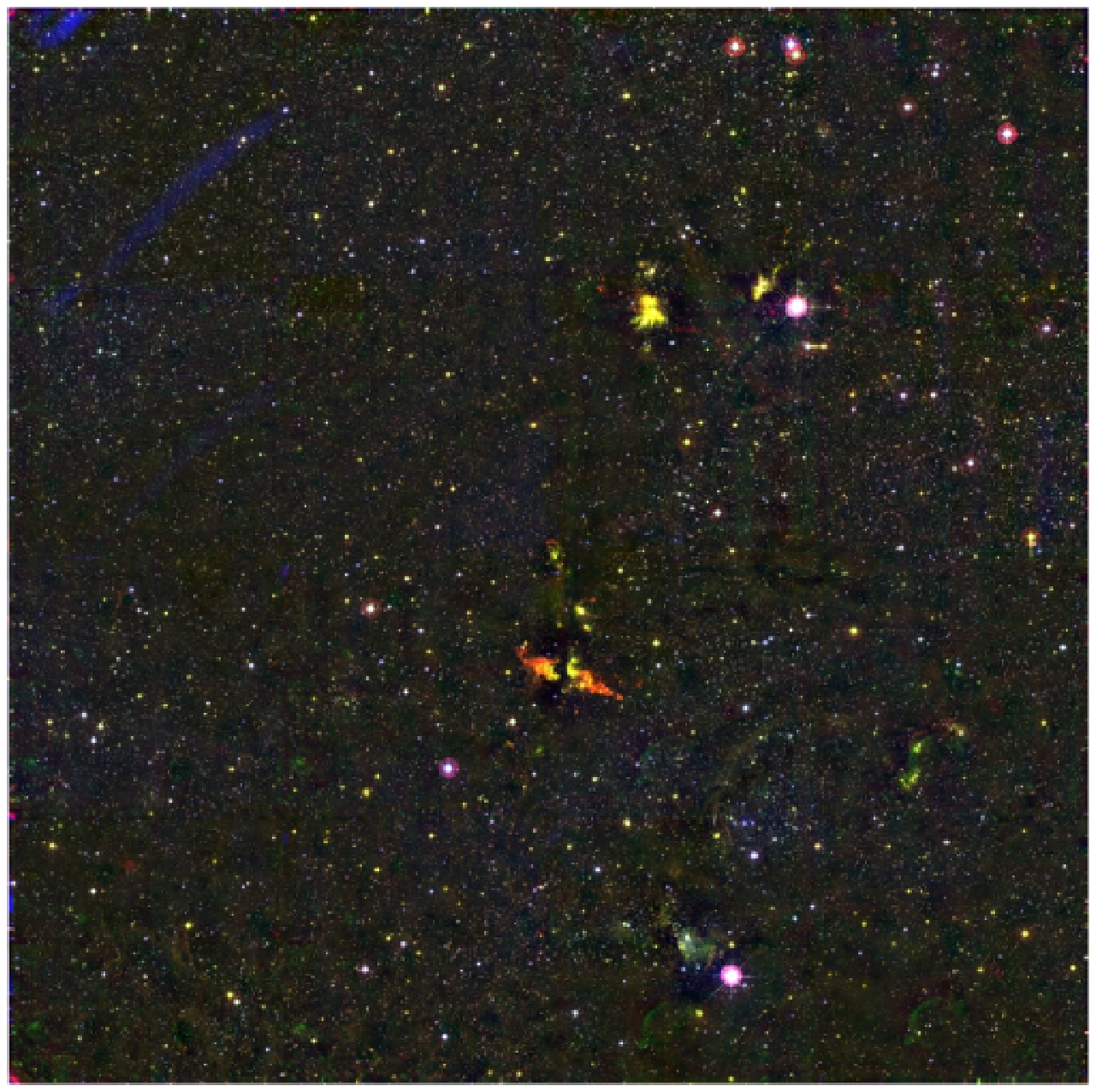}} at 26.666 26.8666
\axis left label {$Dec$\,[']}
ticks out long numbered from 0 to 53 by 10
       short unlabeled from 0 to 53 by 2 /
\axis right label {}
ticks out long unlabeled from 0 to 53 by 10
        short unlabeled from 0 to 53 by 2 /
\axis bottom label {$RA$\,[']}
ticks out long numbered from 0 to 53 by 10
       short unlabeled from 0 to 53 by 2 /
\axis top label {}
ticks out long unlabeled from 0 to 53 by 10
        short unlabeled from 0 to 53 by 2 /

{\color{white}

\plot 23.5 23 31.5 23 31.5 15 23.5 15 23.5 23 /
\plot 25 20 33 20 33 28 25 28 25 20 /
\plot 28 36 36 36 36 44 28 44 28 36 /
\plot 35 36.75 43 36.75 43 44.75 35 44.75 35 36.75 /
\plot 44 10 52 10 52 18 44 18 44 10 /

\put {\Large A} at 22 20
\put {\Large B} at 23 26
\put {\Large C} at 26 39
\put {\Large D} at 45 40.75
\put {\Large E} at 42 14
}
\endpicture
\caption{\label{wfcam} The DR21/W75N region showing the five sub-regions studied in this paper. The image is a composite of J-band (blue), K-band (green) and H$_2$ (red) observations presented by \citet{2007MNRAS.374...29D}.}
\end{figure*}

The entire WFCAM field shown in Fig.~\ref{wfcam}  contains interacting molecular cores and filaments within and across a north-south molecular ridge \citep{1978ApJ...223..840D, 1985ApJ...293..508F,2007MNRAS.374...54K,2010A&A...520A..49S}. The ridge extends over at least half a degree. More recently, these cores are evident in 850-$\mu $m dust-continuum maps \citep{2006ApJ...636..332V} and, in turn, have been resolved into smaller clumps \citep{1990A&A...239..305W, 1992ApJ...388..467M, 1993MNRAS.260..337C, 2001ApJ...546..345S}. Streamers of dense material that seem to emanate away from the brighter infrared sources located along the molecular ridge have been revealed in \textit{Spitzer} images at 5.8 and 8.0 $\mu $m~\citep{2004ApJS..154..333M}. A number of `Extremely Red Objects' (EROs) have also been identified along the ridge \citep{2004ApJS..154..333M}. A convincing interpretation was advanced by \citet{2010A&A...520A..49S} 
in which the DR21 ridge and extending filaments were formed in the convergence of large-scale flows, generating a cloud now in a state of global gravitational collapse

\begin{figure*}
  \begin{center}
  \epsfxsize=7.5cm       \epsfbox{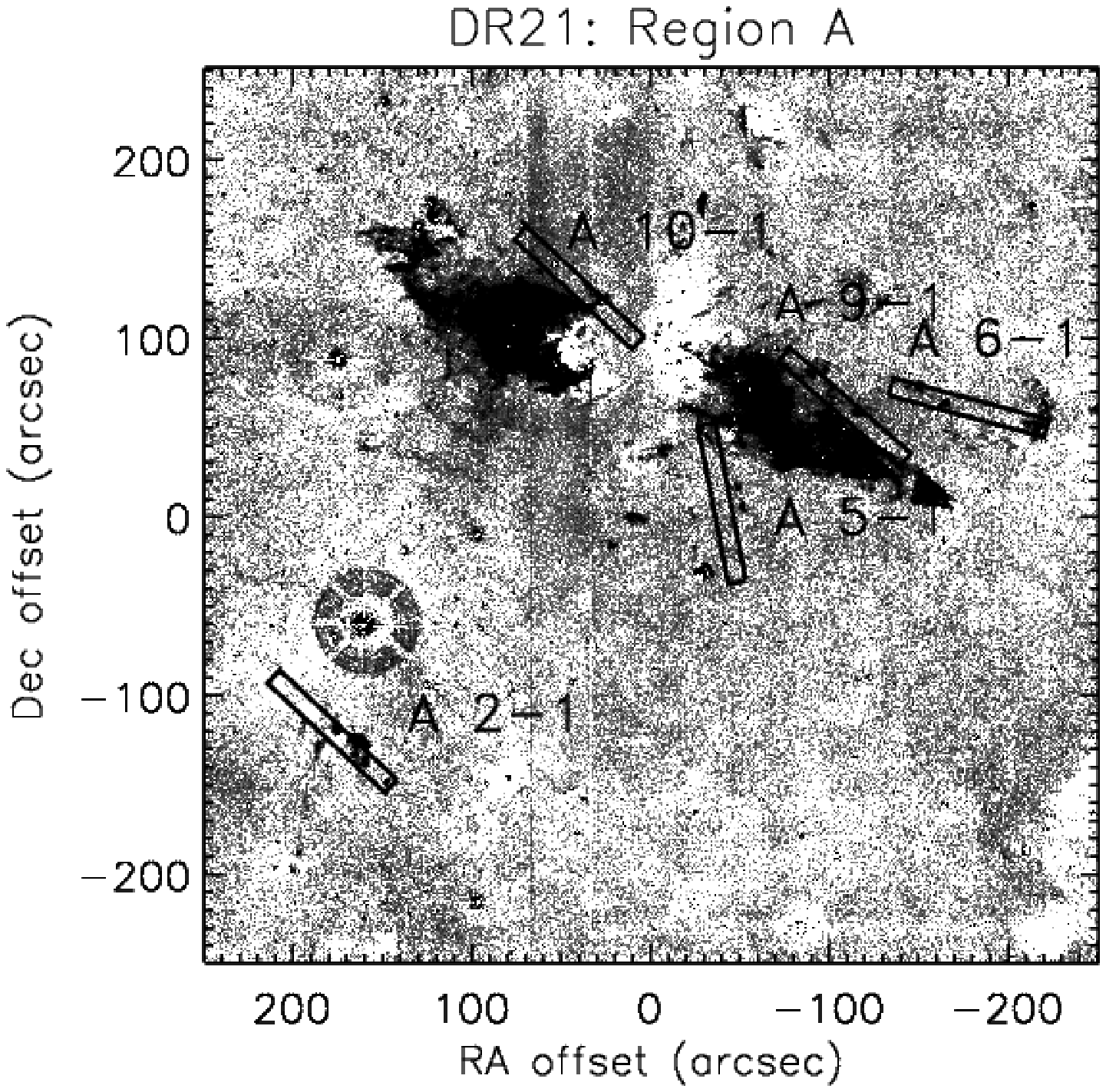}
  \epsfxsize=7.5cm       \epsfbox{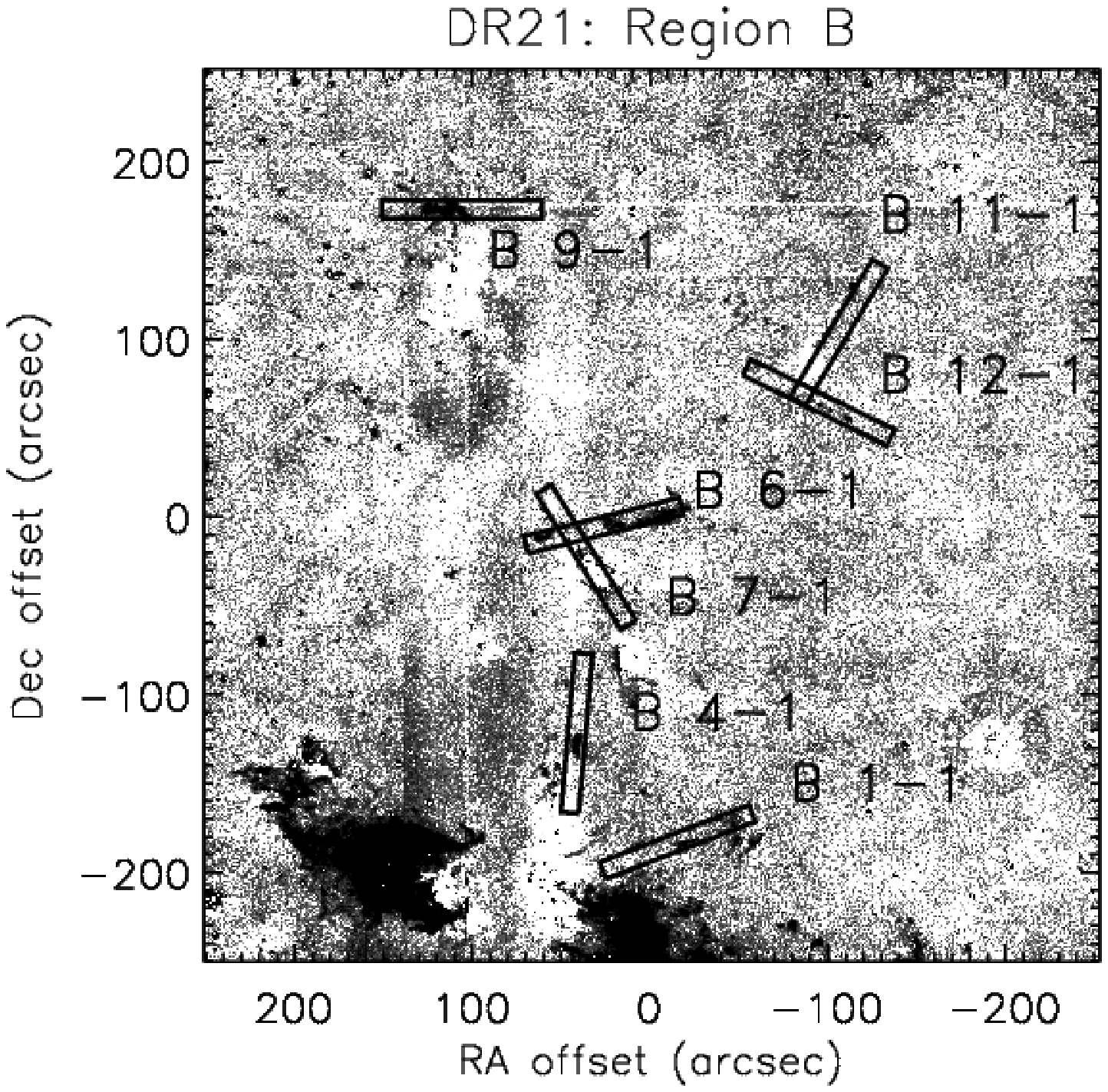}
  \epsfxsize=7.5cm    \epsfbox{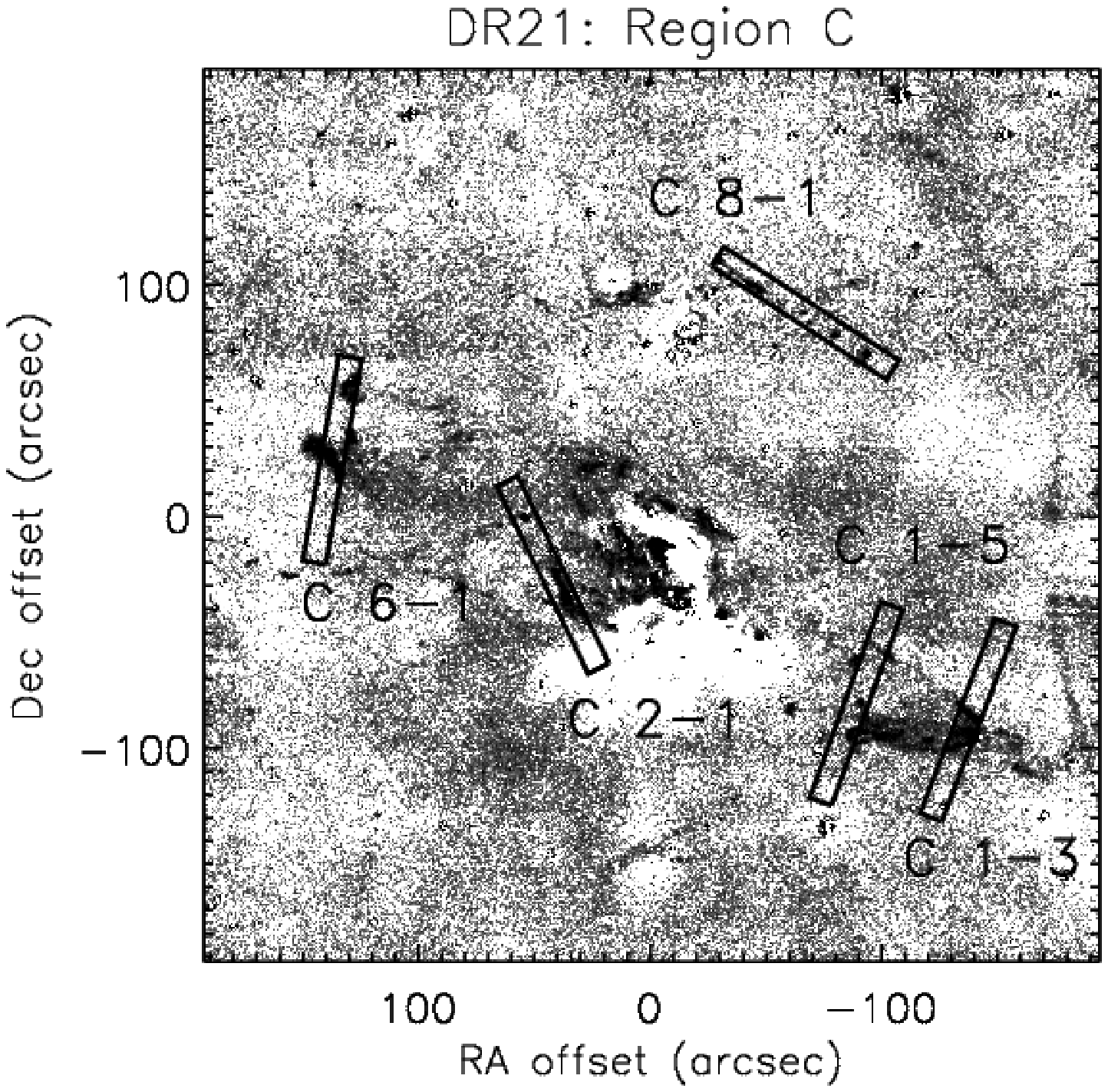}
   \epsfxsize=7.5cm      \epsfbox{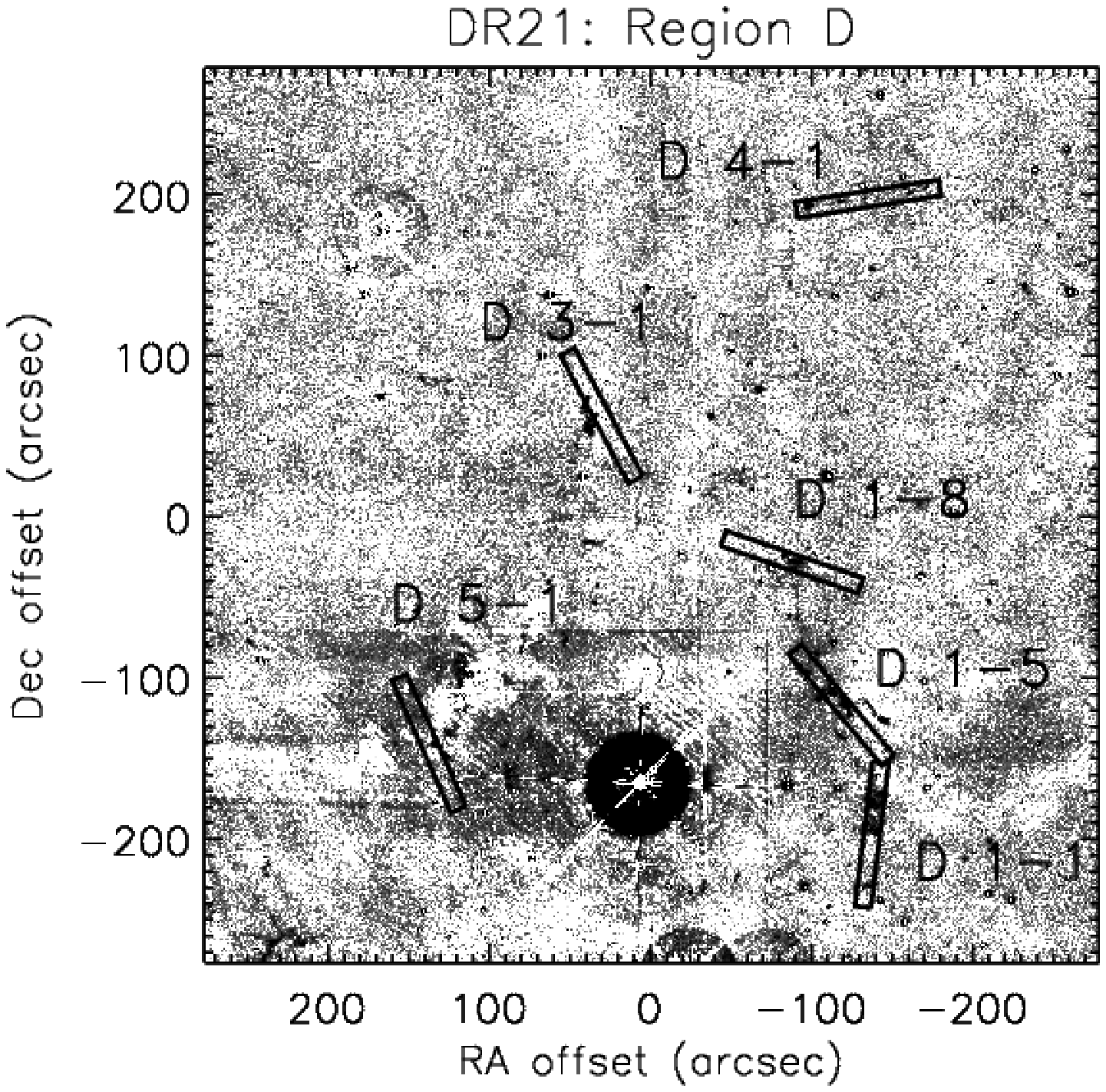}
   \epsfxsize=7.5cm    \epsfbox{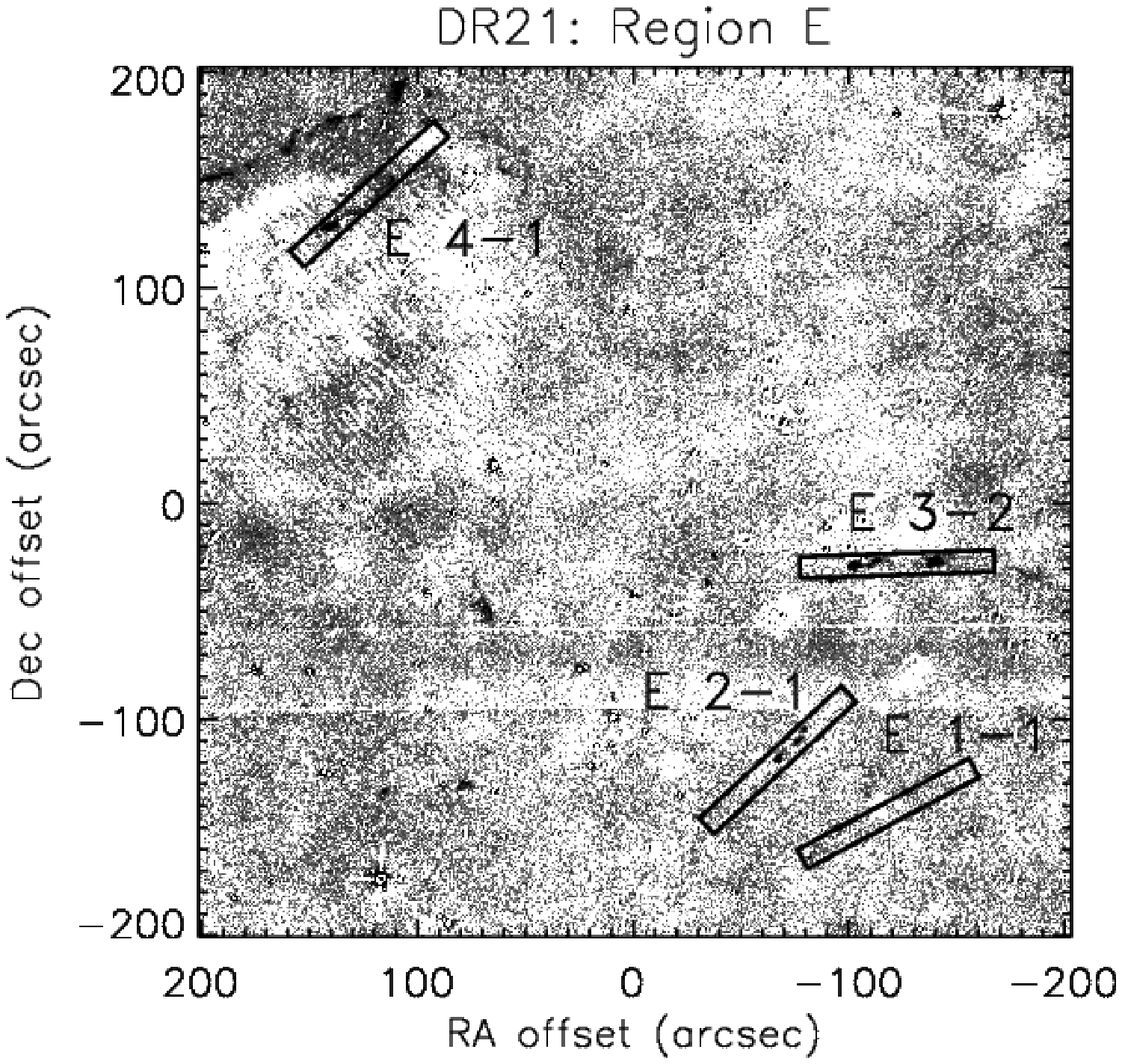}
\caption[]
 {Slit positions overlaid on continuum-subtracted H$_2$ images of the five regions A-E  taken from a larger  H$_2$ and K-band composite~\citep{2007MNRAS.374...29D}. The width of
 the slit has been increased for illustrative purposes.}
   \label{fiveregions}
  \end{center}
\end{figure*}

In this paper we discuss high-resolution, near-IR spectroscopic observations from the DR21/W75N region.  In all, 27 position-velocity (PV) diagrams were obtained from 
within the five regions delineated in Fig.~\ref{wfcam}. The Regions A through E are shown in Fig.~\ref{fiveregions}.  
We adopt the distance of 1.4~kpc to DR21/W75N based on parallax  measurements of VLBA masers \citep{2012A&A...539A..79R}.

\begin{figure*}
  \begin{center}
  \epsfxsize=7.5cm       \epsfbox{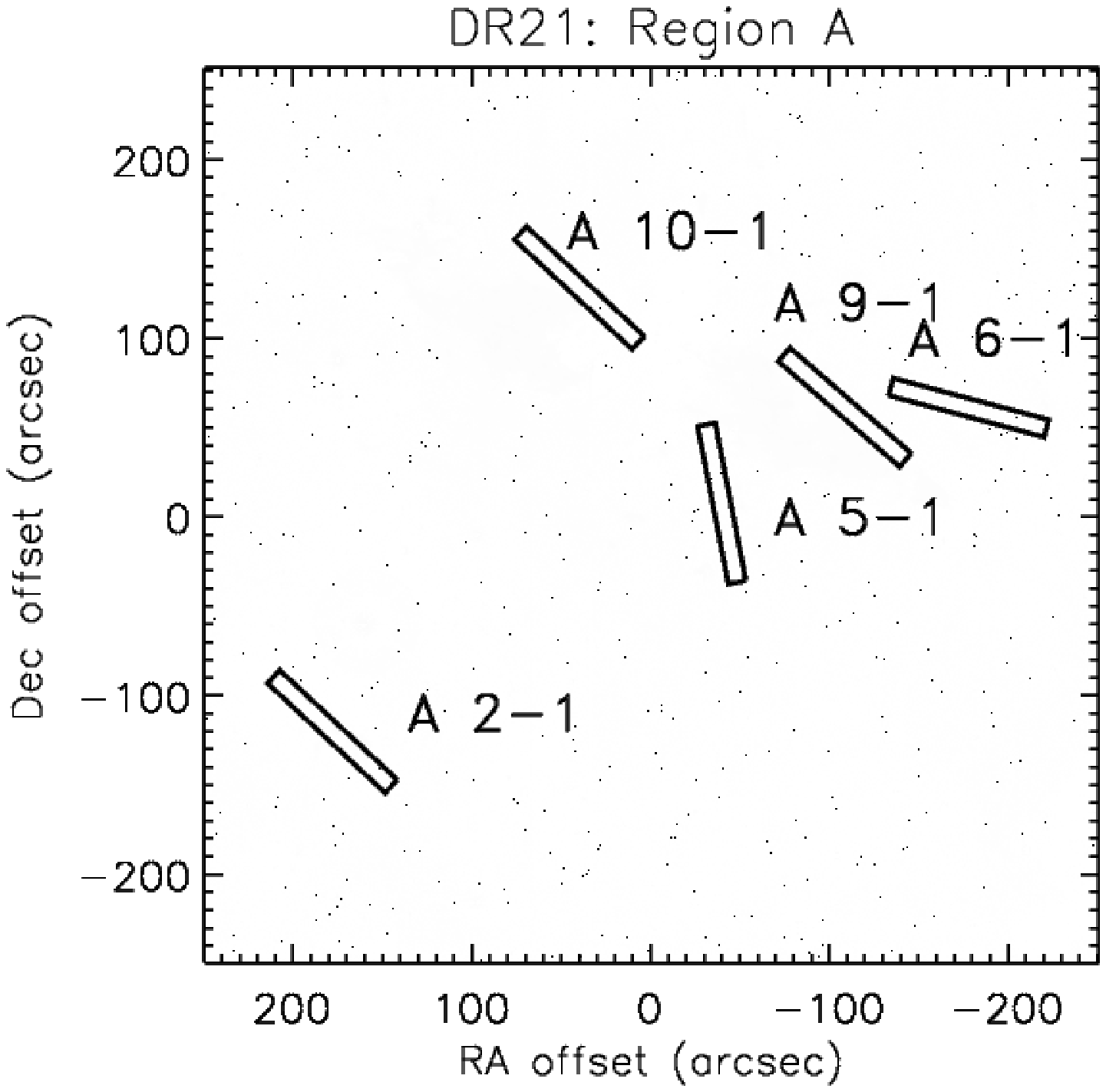}
  \epsfxsize=7.5cm       \epsfbox{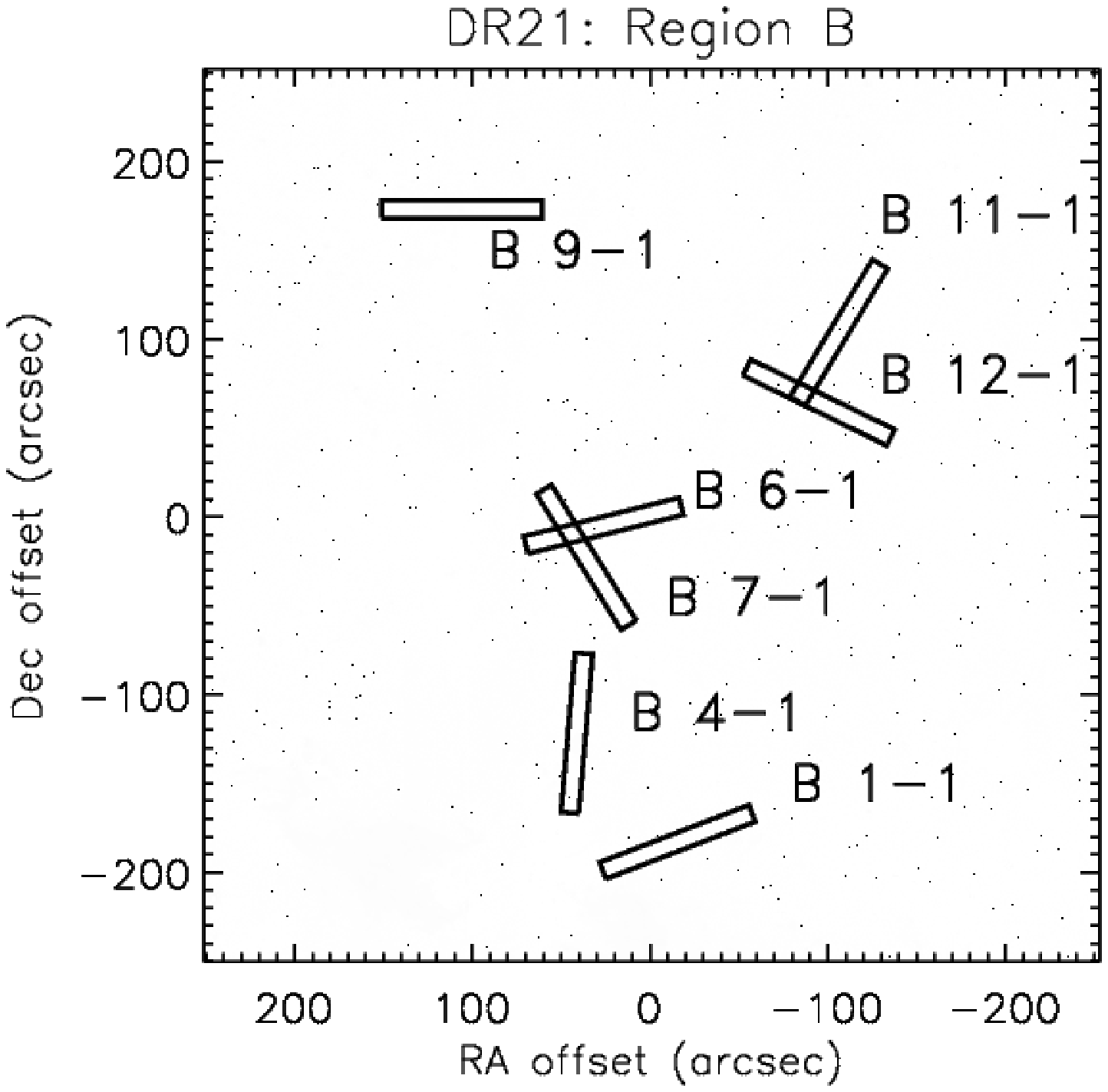}
  \epsfxsize=7.5cm    \epsfbox{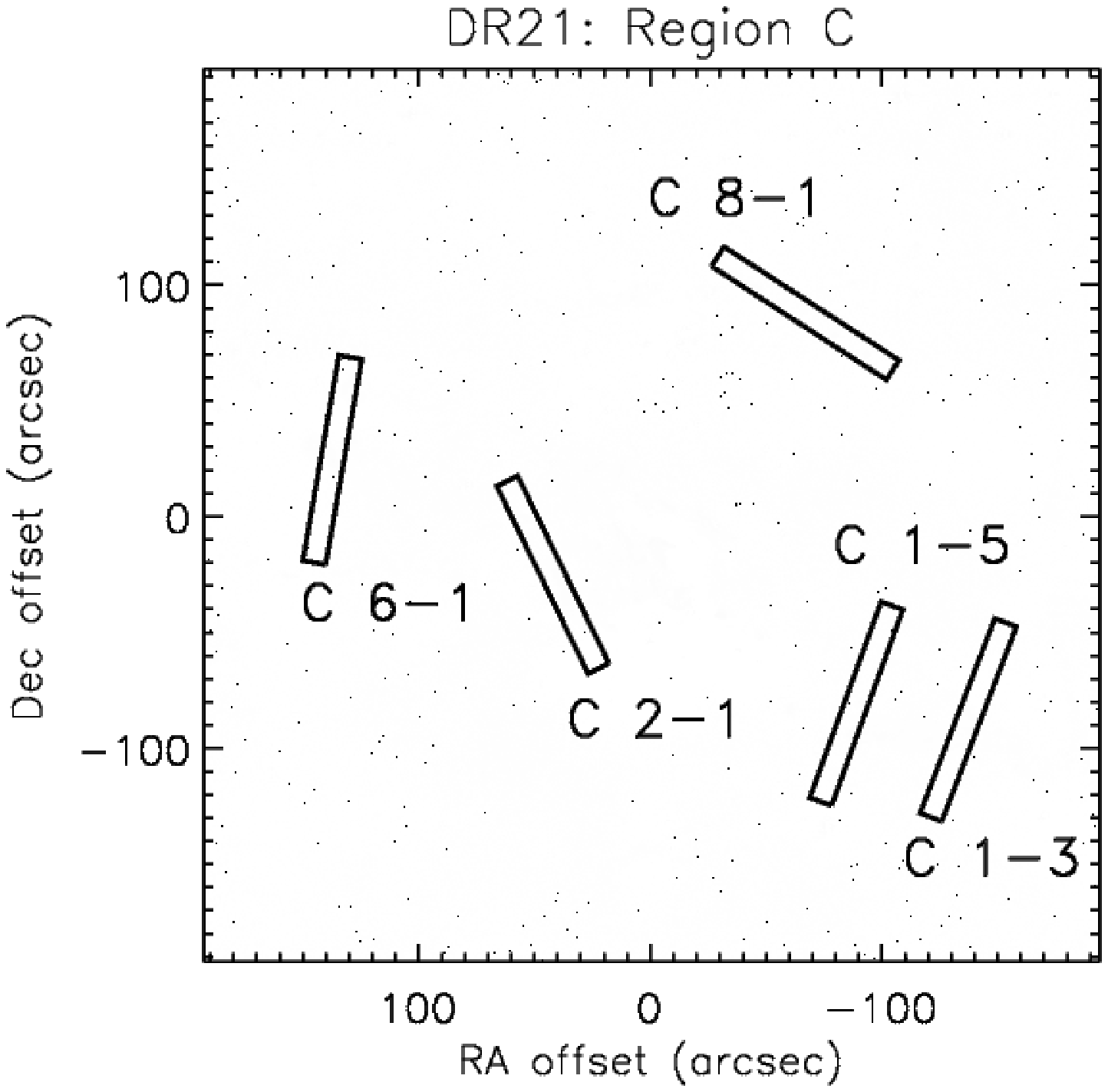}
   \epsfxsize=7.5cm      \epsfbox{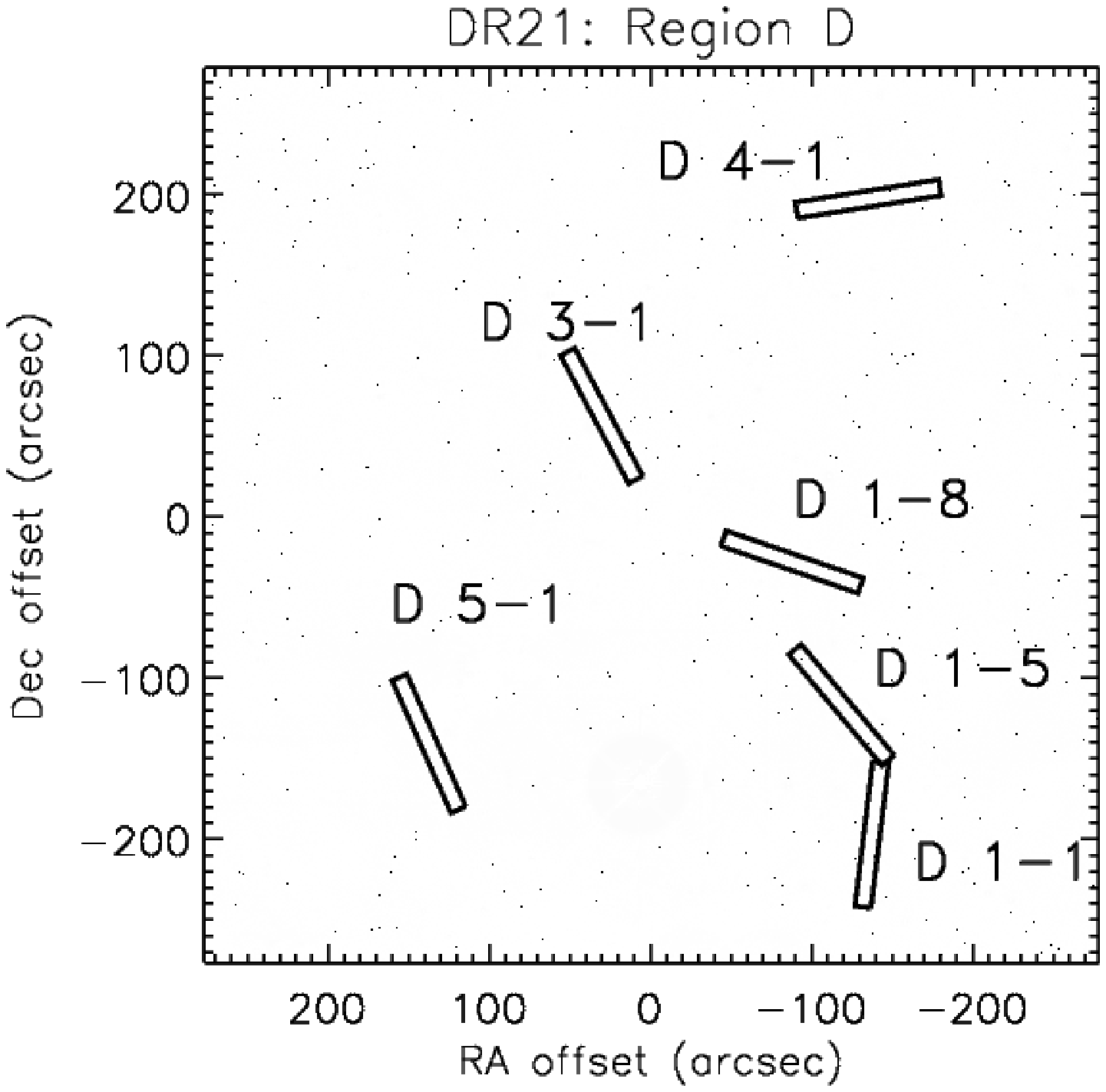}
   \epsfxsize=7.5cm    \epsfbox{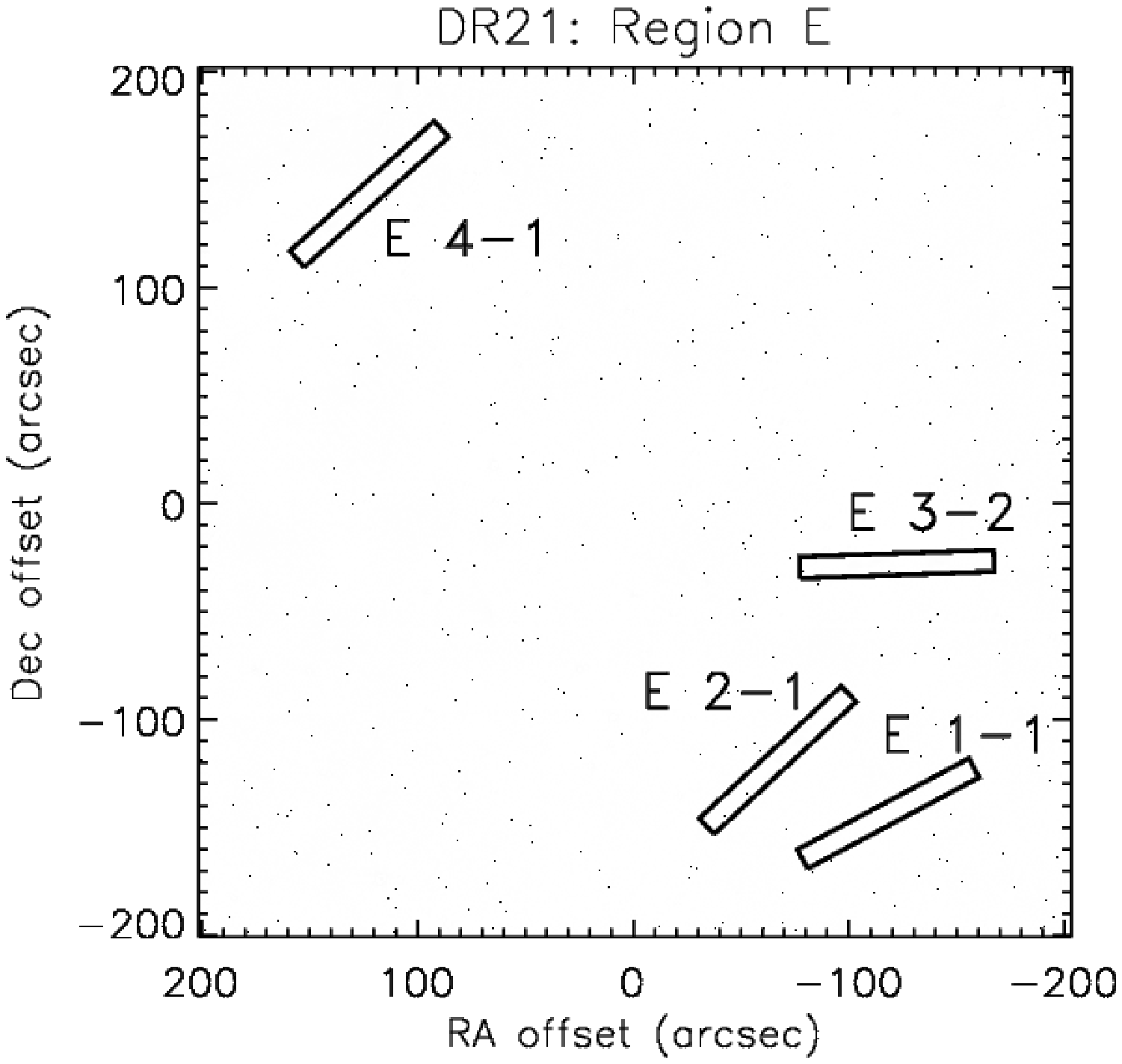}
\caption[]
 {Finding charts for the slit positions in the five regions A-E as superimposed in Fig.~\ref{fiveregions}. The width of
 the slit has been increased for illustrative purposes.}
   \label{fivefinding}
  \end{center}
\end{figure*}

\section{Observations}
\label{observations}

 A large number of collimated molecular outflows in the DR21/W75N ridge was revealed by \citet{2007MNRAS.374...29D} through UKIRT/WFCAM narrow-band imaging.
 Follow-up echelle spectra at  2.122\mic\ were obtained at the 3.8\,m United Kingdom Infrared Telescope (UKIRT), on Mauna Kea, 
 Hawaii, during August, 2006.    The cooled grating spectrometer CGS\,4 \citep{1990SPIE.1235...25M,1994ExA.....3...17W} was employed.  This
instrument is equipped with a 256$\times$256 pixel InSb array and a 31 lines/mm echelle grating.  The pixel scale is 0.41\arcsec
$\times$0.88\arcsec\ (0.41\arcsec\ in the dispersion direction).  A 2-pixel-wide slit was used, resulting in a velocity resolution of
$\sim 16$\,km s$^{-1}$ (this being the Full Width at Half Maximum (FWHM) of unresolved sky or arc lamp lines).  The slit lengths are 
about 80-90\arcsec\  long. It is possible to orient a slit at any angle on the sky.  At the above wavelength, the 
 H$_2$  v\,=\,1-0\,S(1) ($\lambda_{\rm vac} = 2.1218334 \mu$m; \citet{1982ApJ...263..999B}) is well centred on the array.  The wavelength coverage 
was approximately  2.115-2.128$\mu$ (these ranges shifted slightly from night-to-night, depending on the
setting of the echelle grating angle).

For each target  an exposure at a ``blank sky'' position was followed by five on-source ``object'' exposures, the
former being subtracted from all of the object frames.  This sequence of six frames was repeated two or three times to build up
signal-to-noise.  To avoid saturation on the bright continuum, relatively short exposures of 
60\,sec ($\times$2 co-adds) were used. The total integration time on each source was 30~mins in H$_2$. The surface 
brightness of the target H$_2$ features is of the order of 1-100~$\times $10$^{-19}$~W~m$^{-2}$~arcsec$^{-2}$.

With the above exposure times, at the high spectral resolution of the instrument ($R \sim 18,750$), there is essentially no detectable
thermal emission (or associated shot noise) from the sky or background between the well-resolved OH sky lines.  Thus, the 
process of ``sky subtraction'' mainly serves to remove the OH lines and any ``warm''
pixels that remain after bad-pixel masking and flat-fielding (all data frames are bad-pixel-masked using a mask taken at the start of each
night, and flat-fielded, using an observation of an internal blackbody lamp obtained before each set of target observations as 
described in detail by  \citet{2005MNRAS.360..104D}.  At 2.122\mic\
the H$_2$  line is well separated from any bright OH sky lines.  Thus, any residual sky lines
in the reduced spectral images resulting from imperfect sky subtraction do not influence the data.  

Co-added H$_2$ spectral images were wavelength calibrated using OH sky lines, there being four lines well spaced across the
array in both cases \citep{2000A&A...354.1134R}.  The first raw object frame observed  was used as the
reference frame.  The employed {\em STARLINK:FIGARO} software packages also correct for distortion along the columns in each image
(i.e. along arc or sky lines) via a polynomial fit to the OH lines in each row.  Examination of the OH lines in distortion-corrected and
wavelength-calibrated raw frames showed that the velocity calibration along the length of the slit, measured from Gaussian fits to the sky
lines, is extremely good for the H$_2$ data.  The relative velocity calibration along the slit and between adjacent slits and
targets is estimated to be accurate to $\sim 3$ km s$^{-1}$ . Velocities were calibrated with respect to the {\em kinematic} Local Standard of Rest [LSR].  
They were not corrected for the systemic velocity in each region.

Finally, spectra of the standard star  BS7672, a G1V bright dwarf star which has a visual magnitude of 5.80, 
were obtained and reduced in a similar fashion.  In each case an extracted, wavelength-calibrated spectrum was 
``grown'' into a 2-D image so that division by this image would correct each reduced spectral image for telluric absorption.  


\section{Results}
\label{results}

The entire field is displayed in Fig.~\ref{wfcam}. Here, H$_2$ line emission from shock waves  appears red, and from fluorescence appears yellow while embedded or background 
stars also appear yellow. Note the proximity of Regions A and B to DR21 and Regions C and D to W75N. Region E is located at the eastern edge of the Lynds dark cloud L906.

The 27 slit positions in the  Regions A through E are  displayed in Fig.~\ref{fiveregions} and finding charts displayed in Fig.~\ref{fivefinding}. Suggested H$_2$ structures and the possible driving sources for all the associated flows
are listed in Table~\ref{tab_sources} in addition to the  date of observation, Right Ascension and Declination, and the position angle of the slit. These identifications and
descriptions were derived through a combination of near-infrared, infrared and sub millimetre observations \citep{2007MNRAS.374...29D}.
Close-up 90\arcsec~$\times$~90\arcsec\  images, in which the exact locations of the slits relative to the H$_2$  structures can be discerned, are provided in Appendix~A.

\begin{figure*}
  \begin{center}
  \epsfysize=5.3cm   \epsfbox{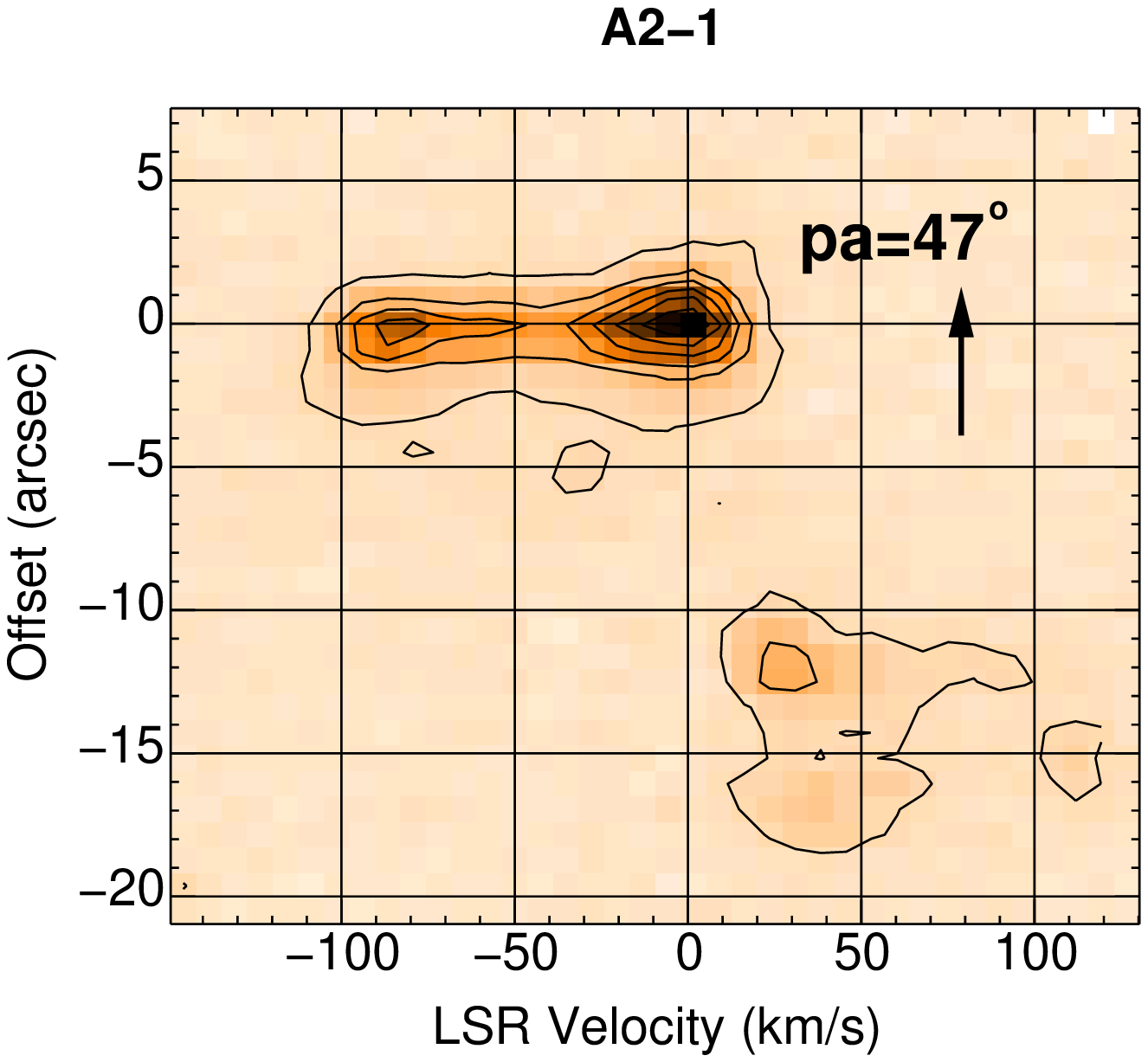}
  \epsfysize=5.3cm   \epsfbox{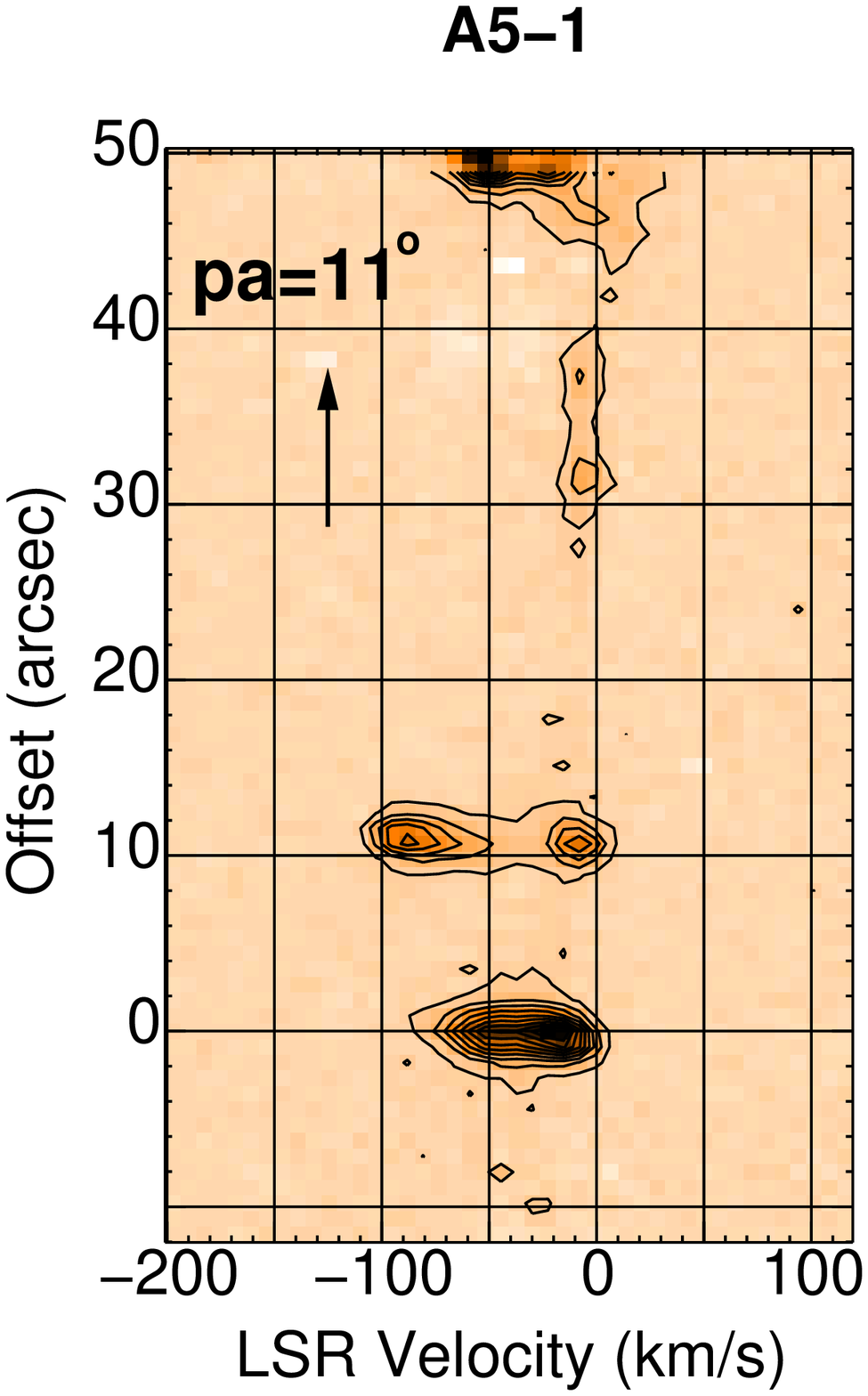}
  \epsfysize=5.3cm   \epsfbox{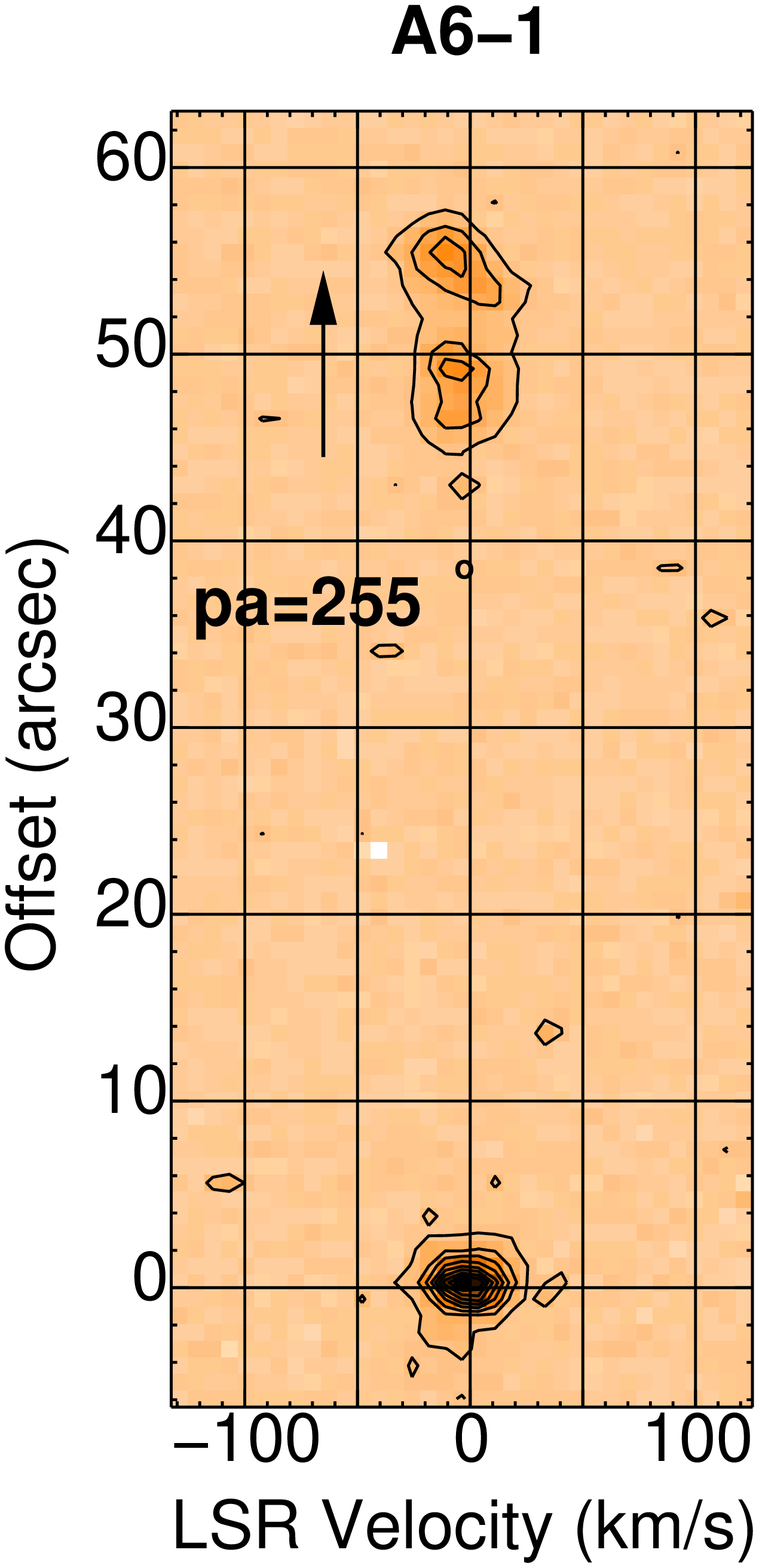}
\newline
  \epsfysize=5.3cm   \epsfbox{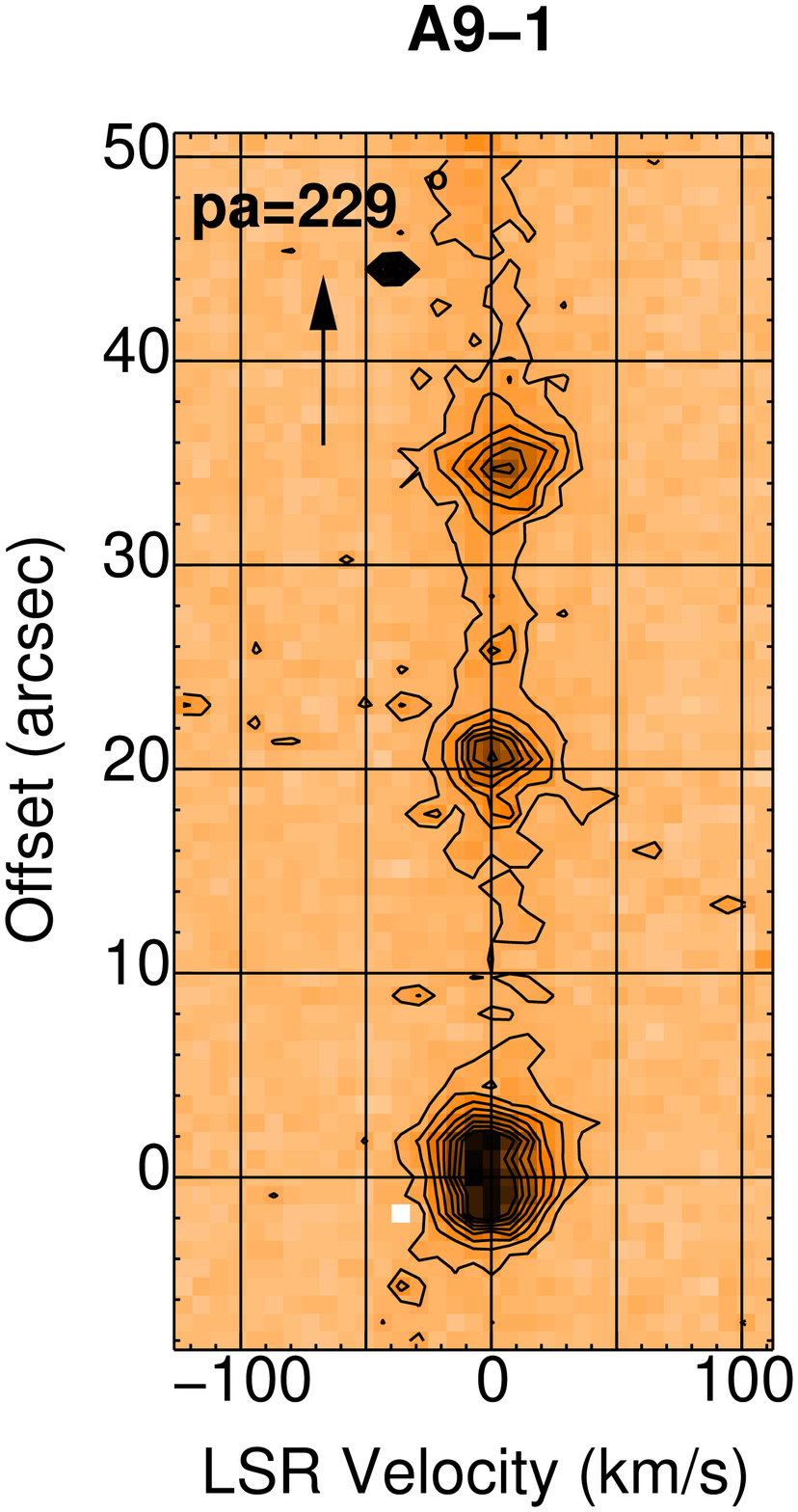}
  \epsfysize=5.3cm   \epsfbox{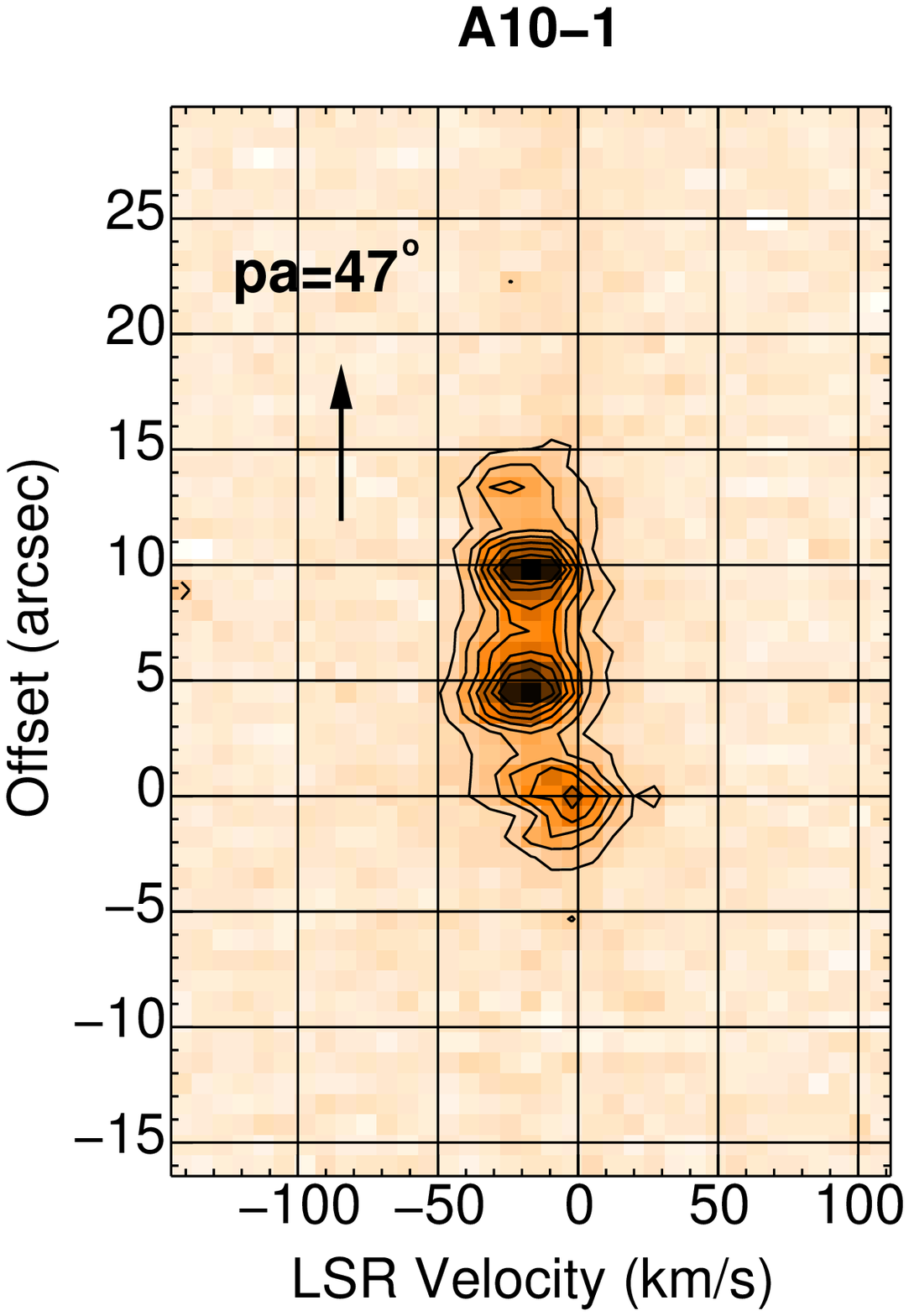}     
\caption[]
 {Position-Velocity diagrams for slits in Region A. Contour levels are linearly spaced from an arbitrary lower signal level which is two to three times the noise level.}
   \label{regiona}
  \end{center}
\end{figure*}

\begin{figure*}
  \begin{center}
  \epsfysize=5.3cm      \epsfbox{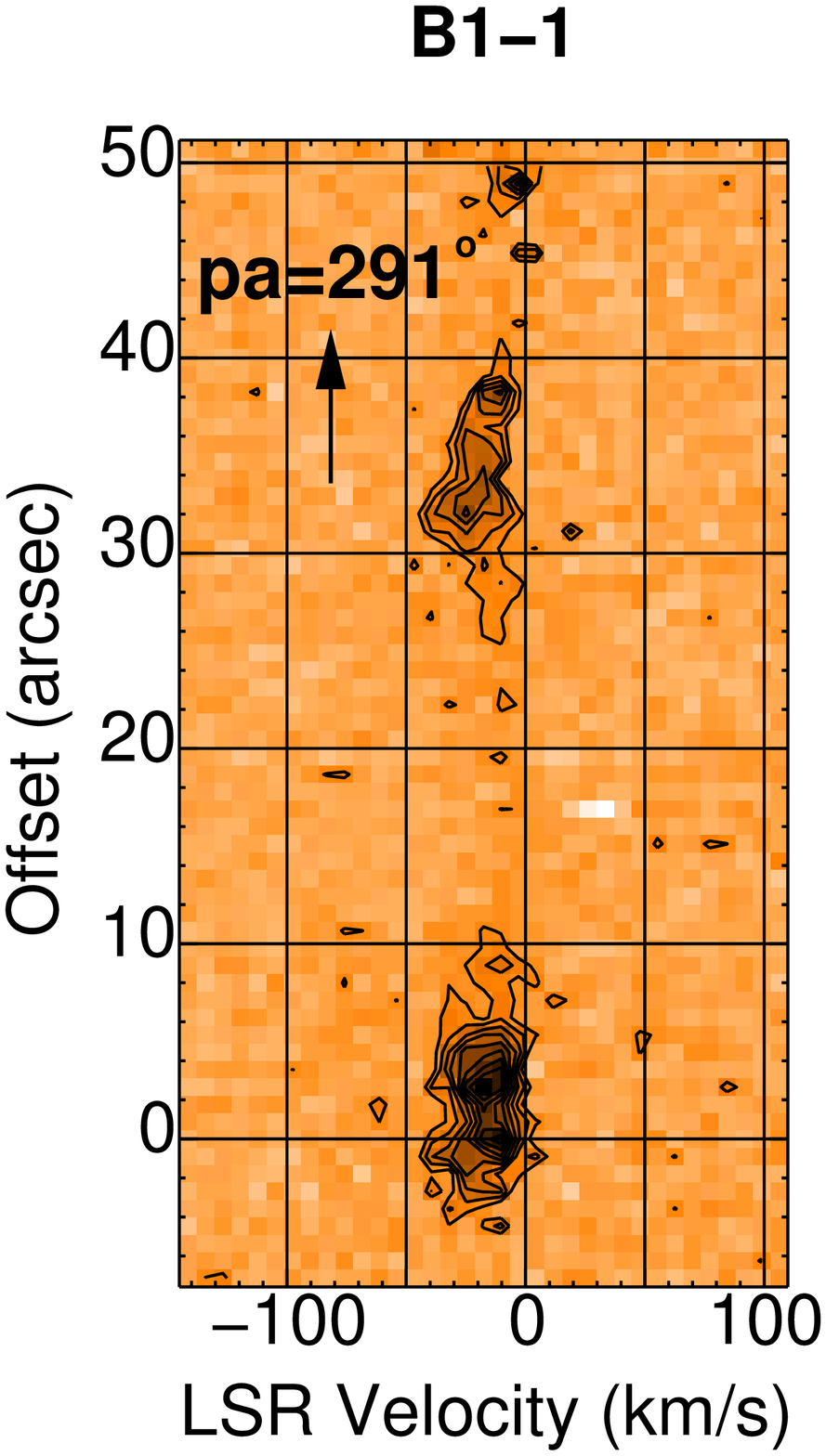}
   \epsfysize=5.3cm     \epsfbox{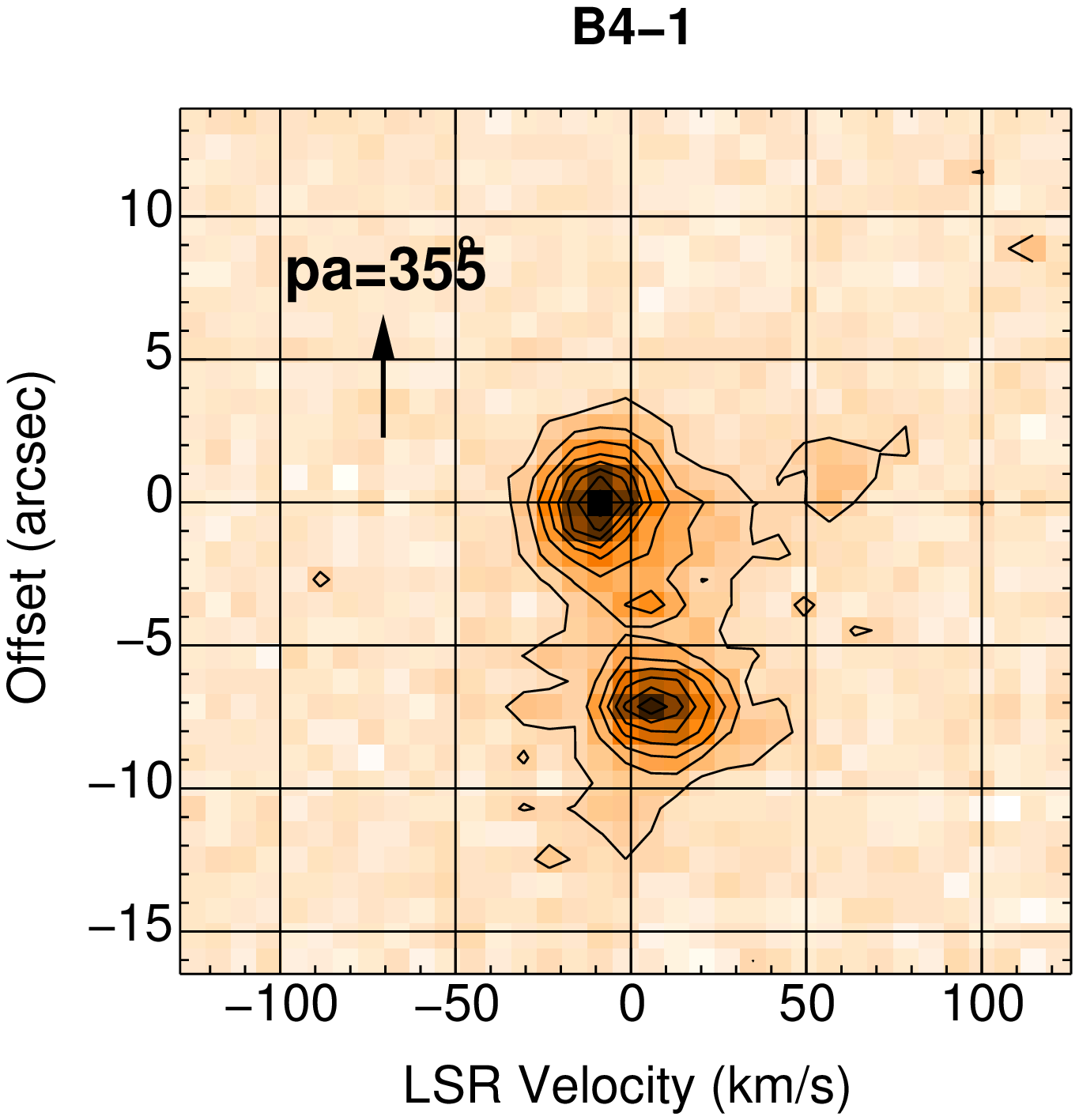}
    \epsfysize=5.3cm    \epsfbox{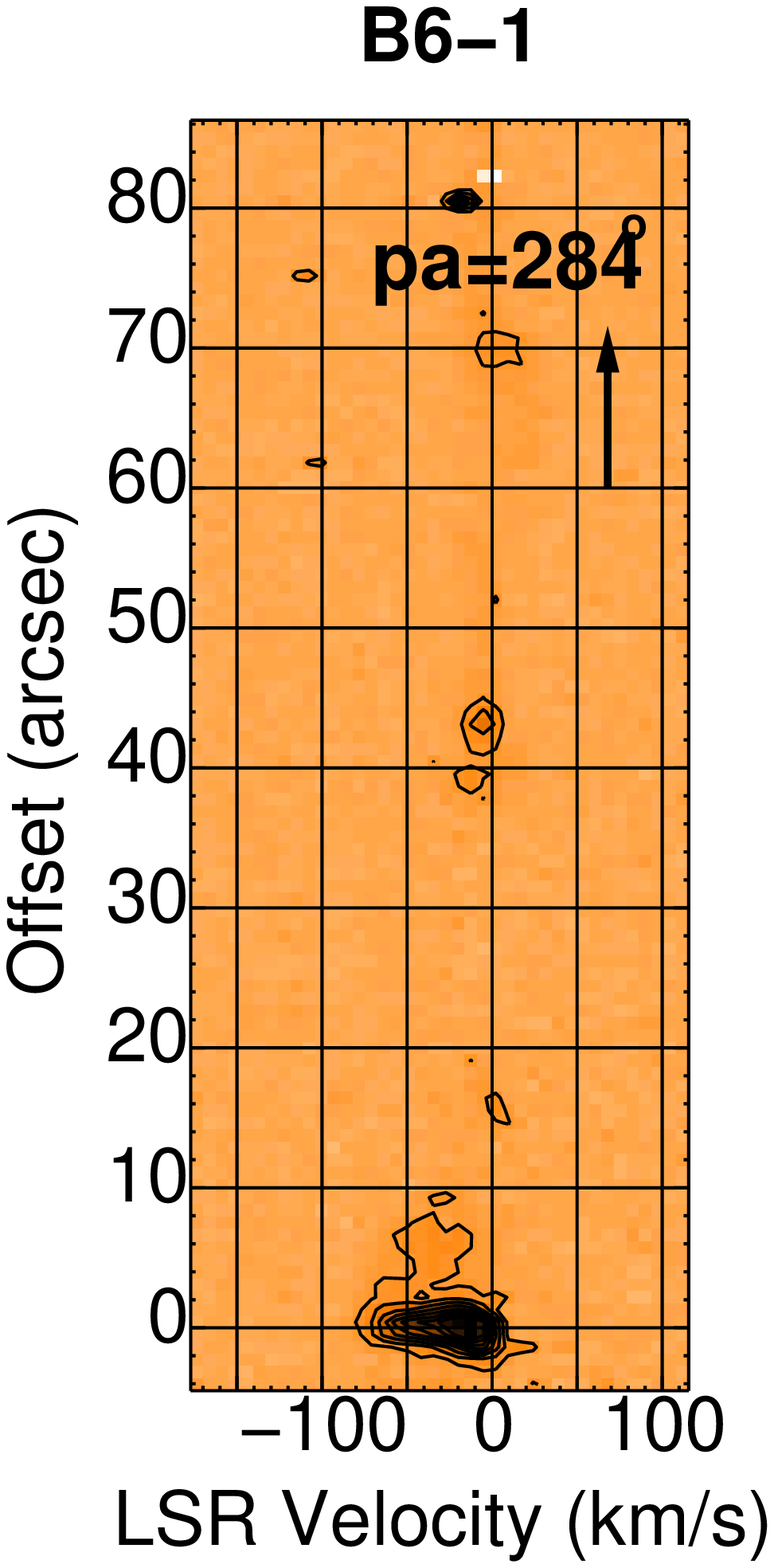}
   \epsfysize=5.3cm     \epsfbox{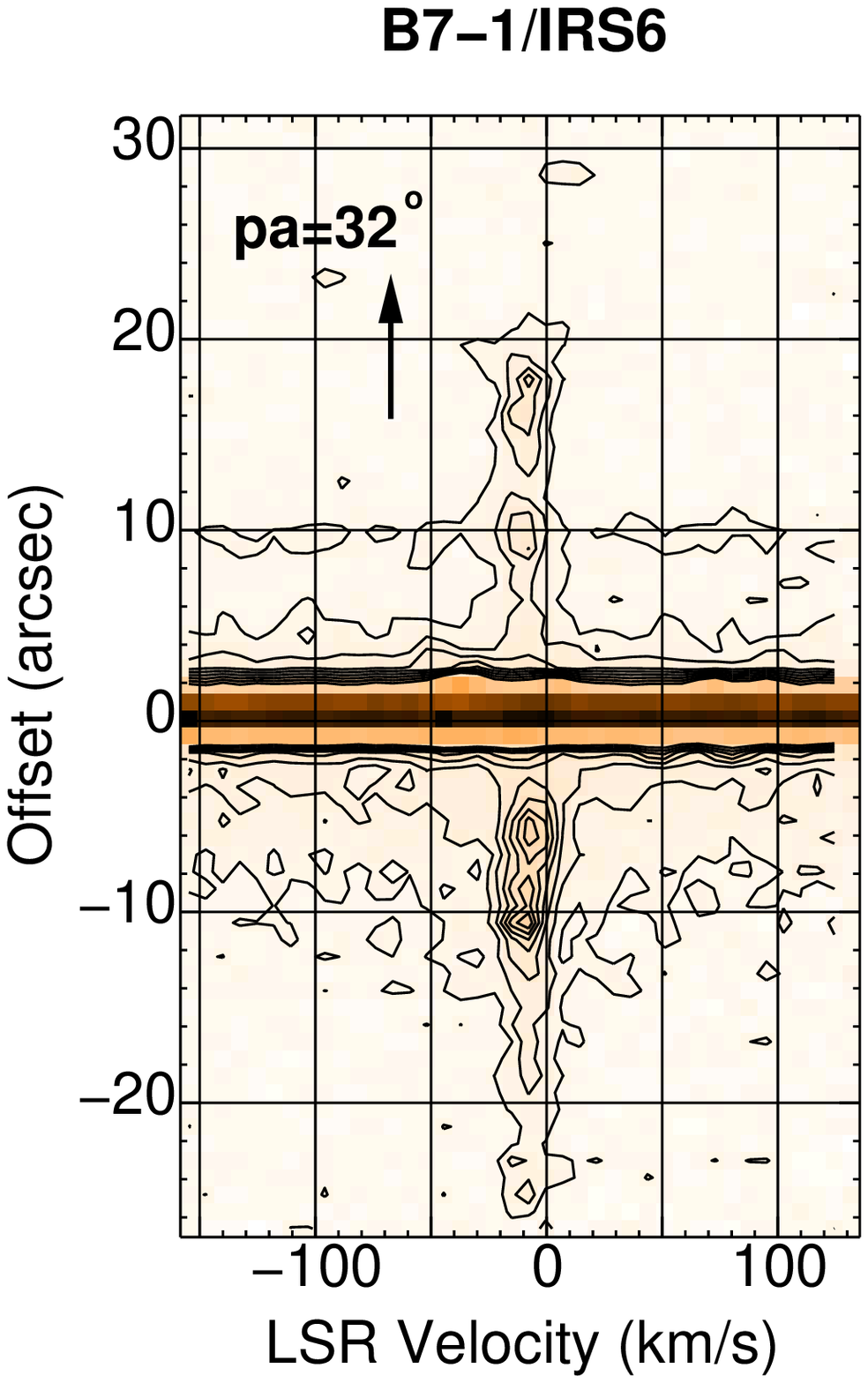}
\newline
    \epsfysize=5.3cm    \epsfbox{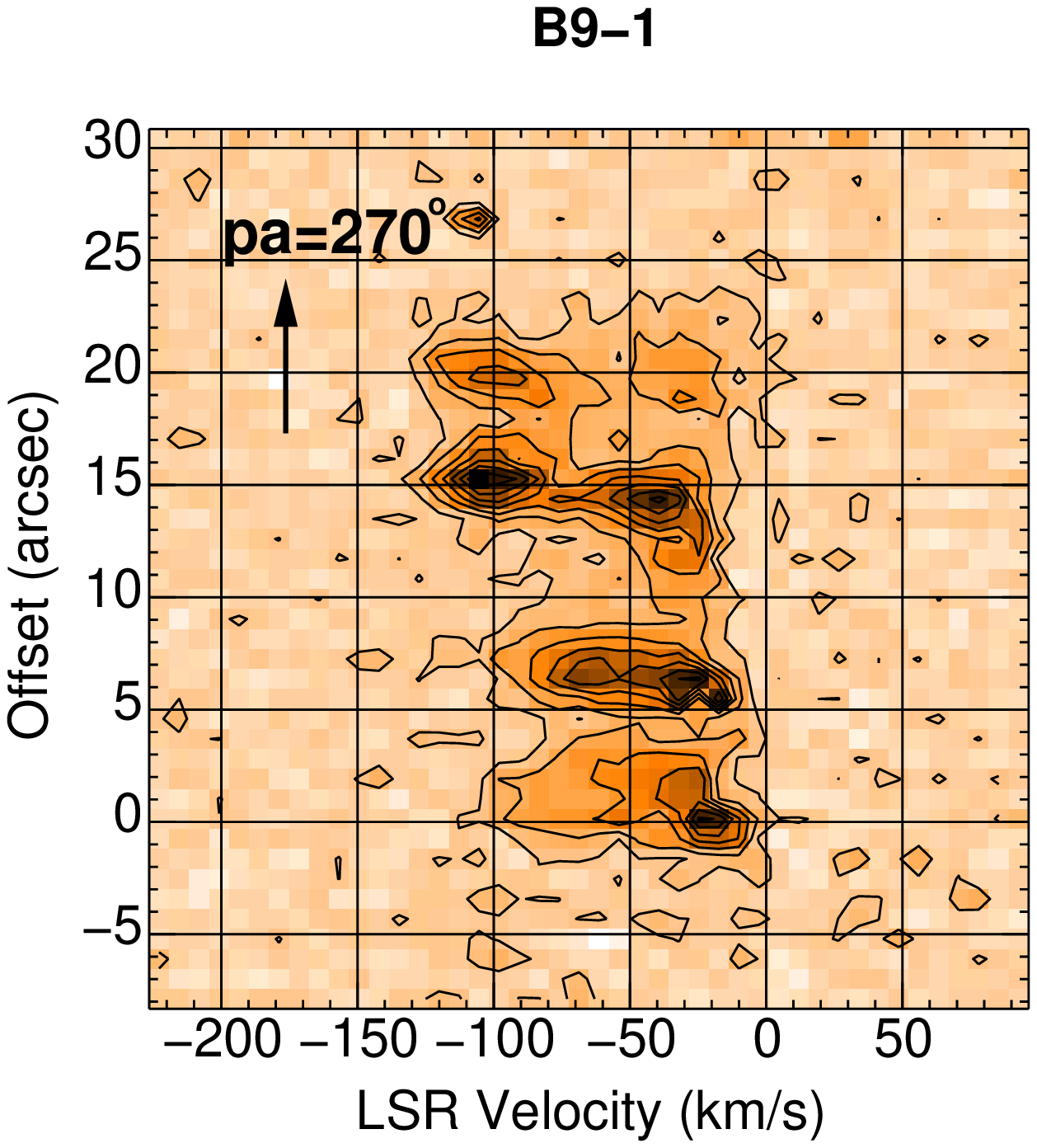}
   \epsfysize=5.3cm      \epsfbox{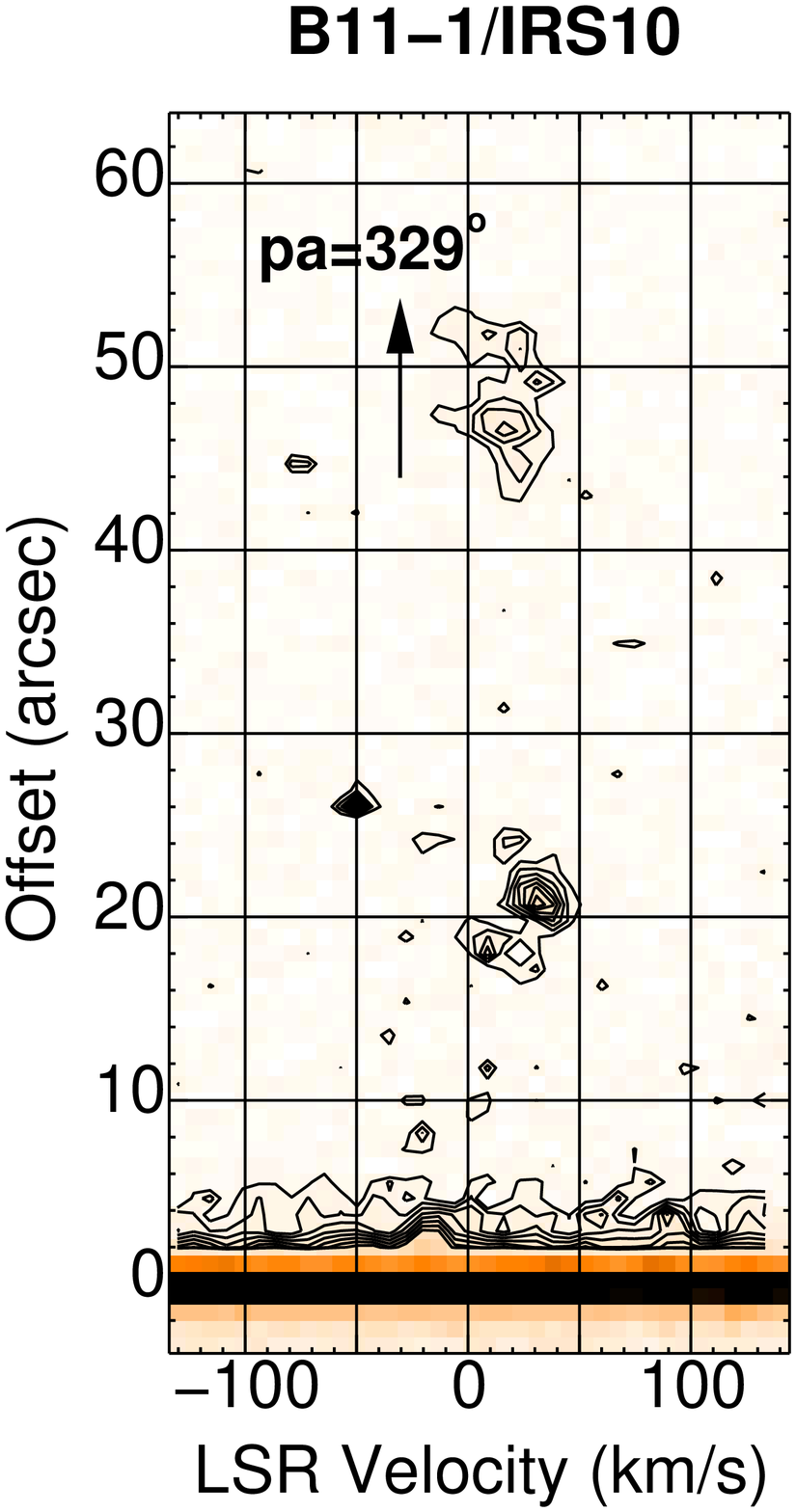}        
   \epsfysize=5.3cm     \epsfbox{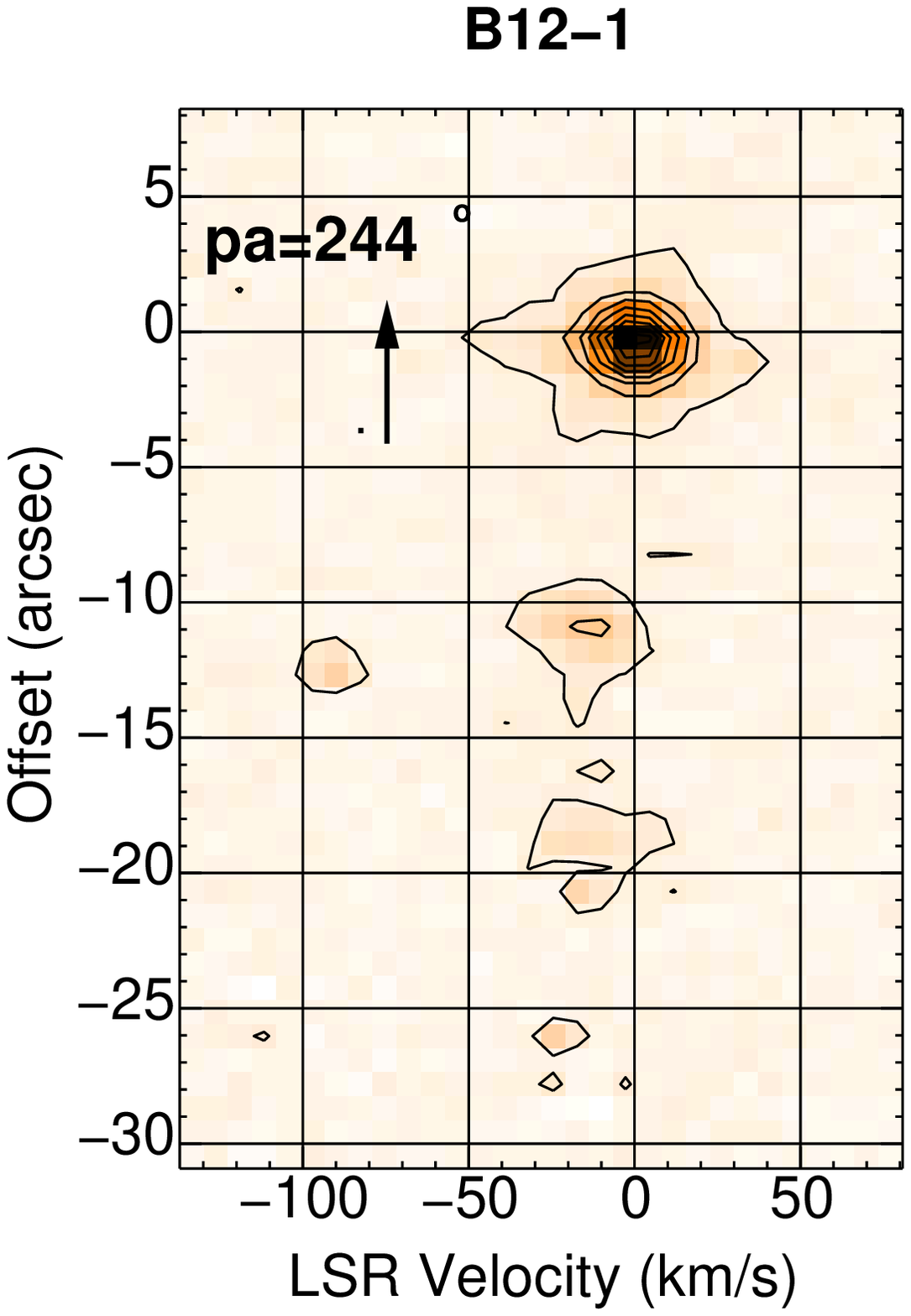}
\caption[]
 {Position-Velocity diagrams for slits in Region B. Contour levels are linearly spaced from an arbitrary lower signal level which is two to three times the noise level.}  \label{regionb}
  \end{center}
\end{figure*}

\begin{figure*}
  \begin{center}
  \epsfysize=5.3cm           \epsfbox{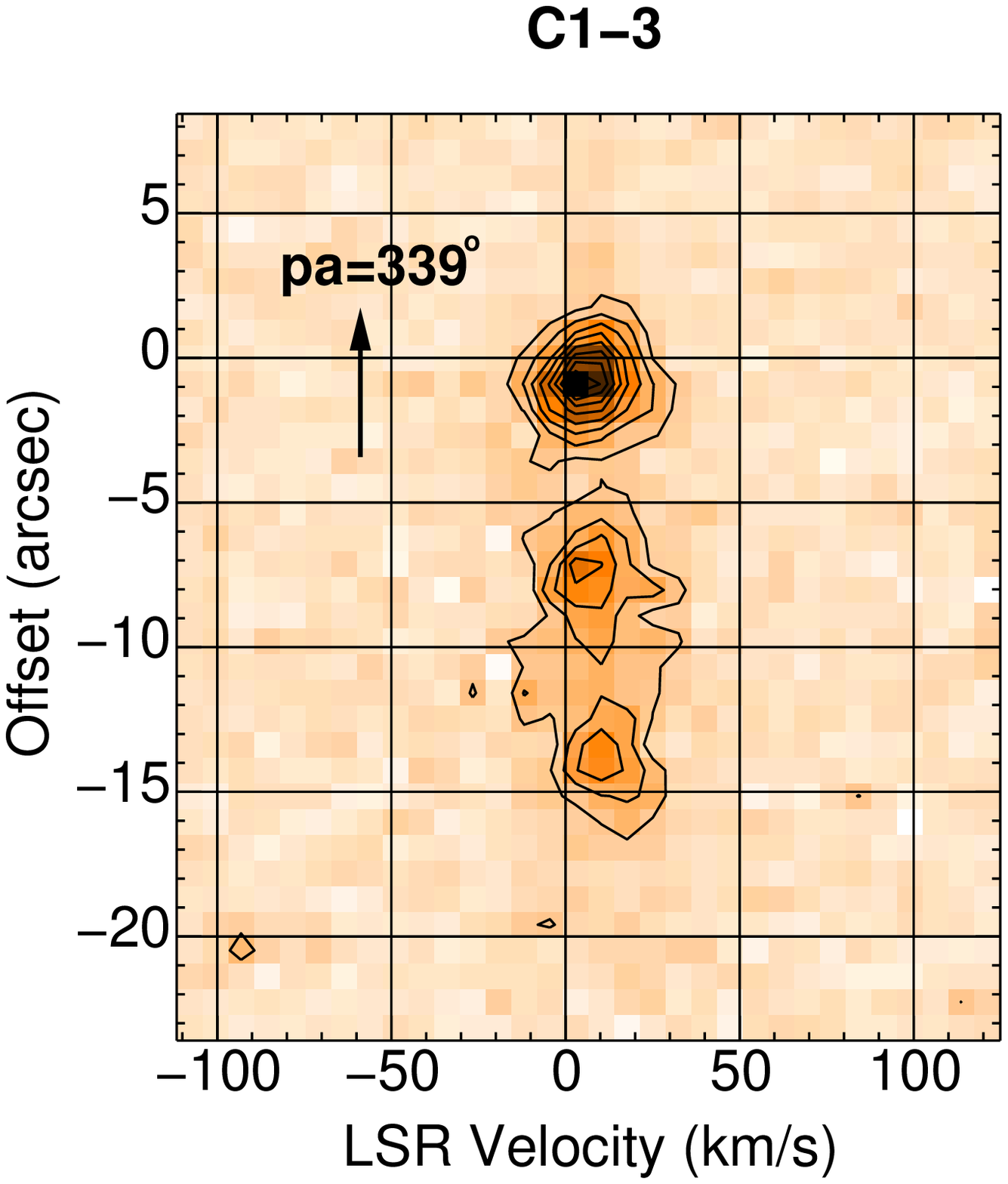}
  \epsfysize=5.3cm           \epsfbox{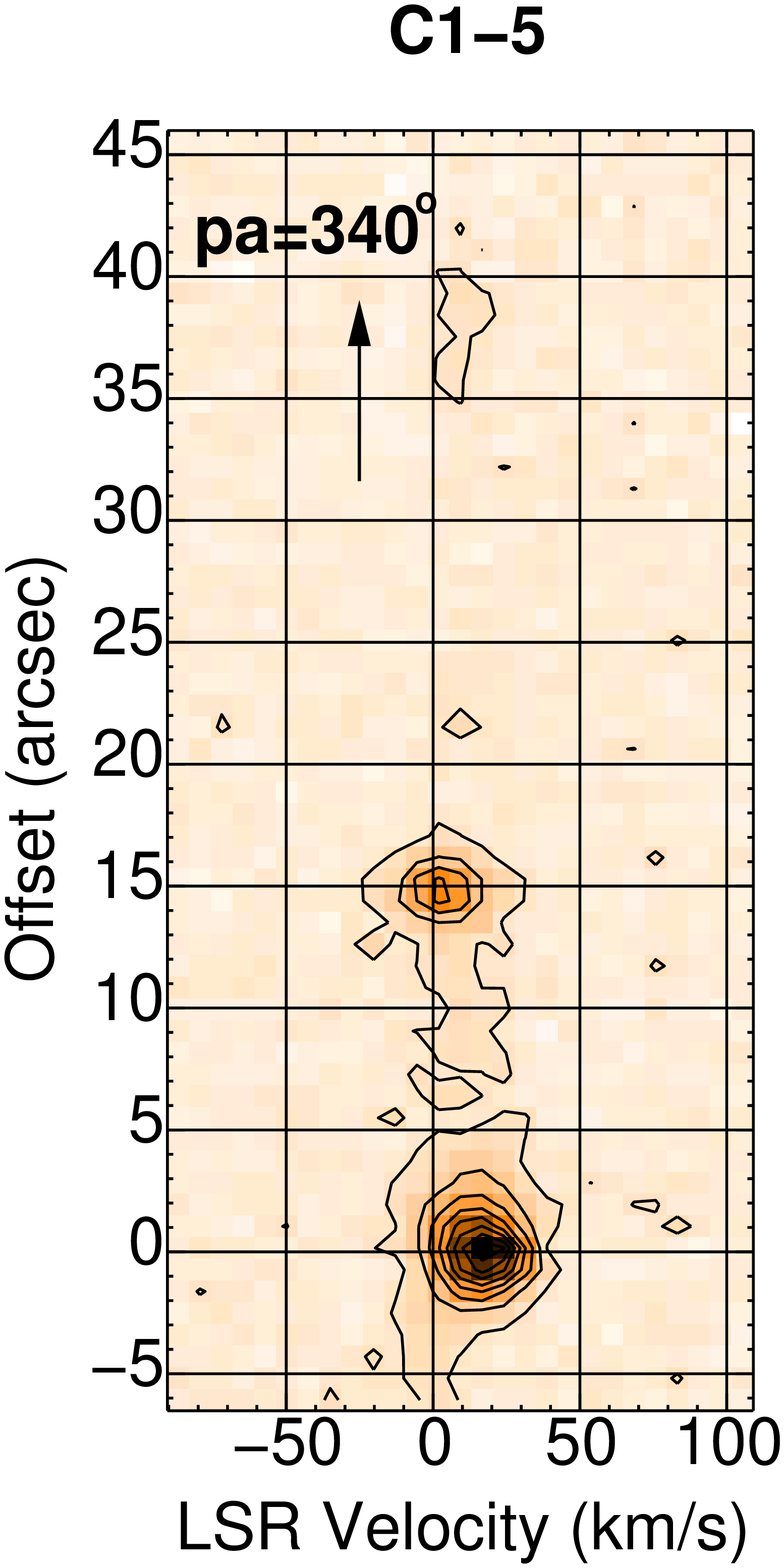}
   \epsfysize=5.3cm          \epsfbox{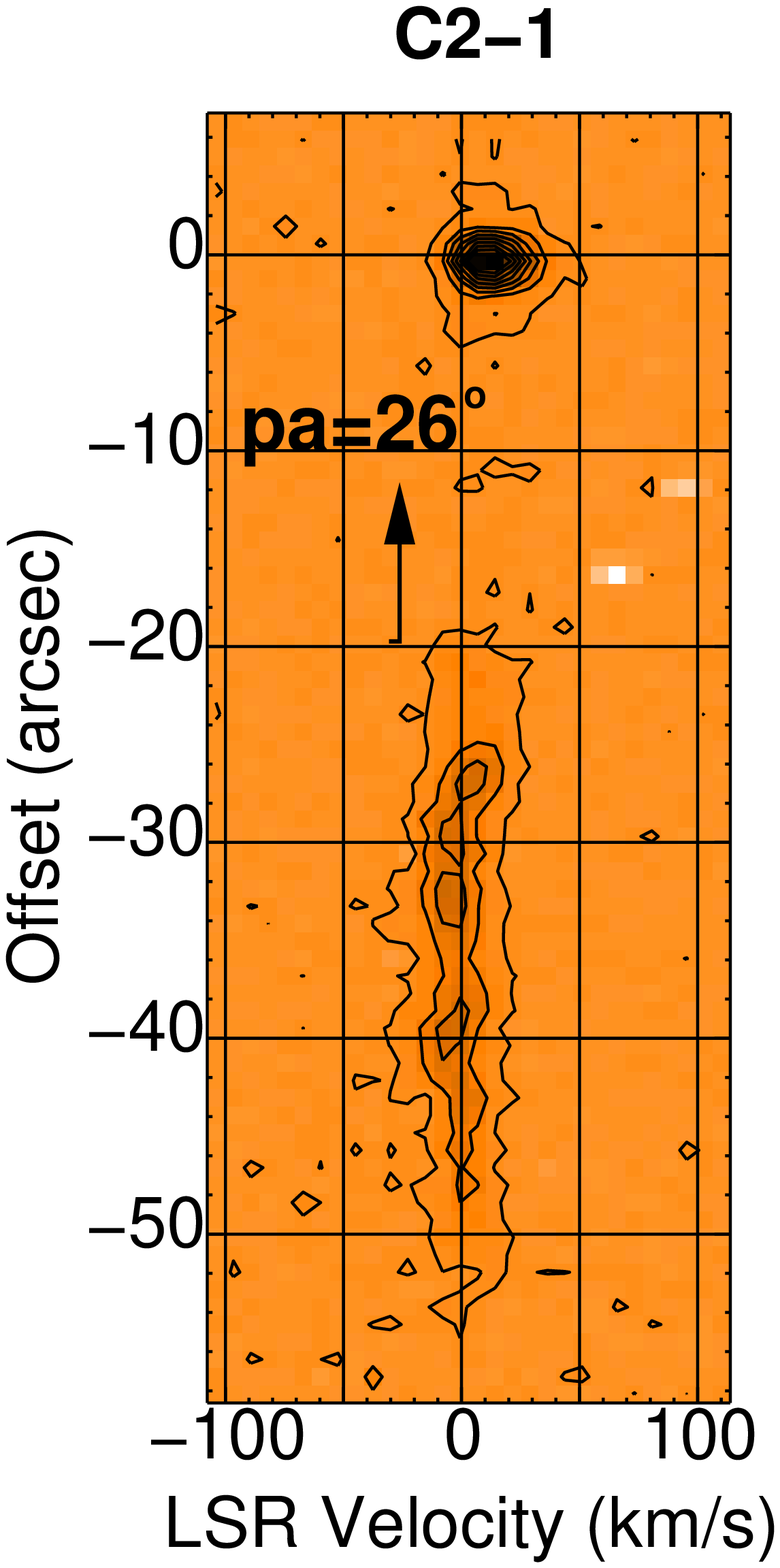}
\newline
   \epsfysize=5.3cm          \epsfbox{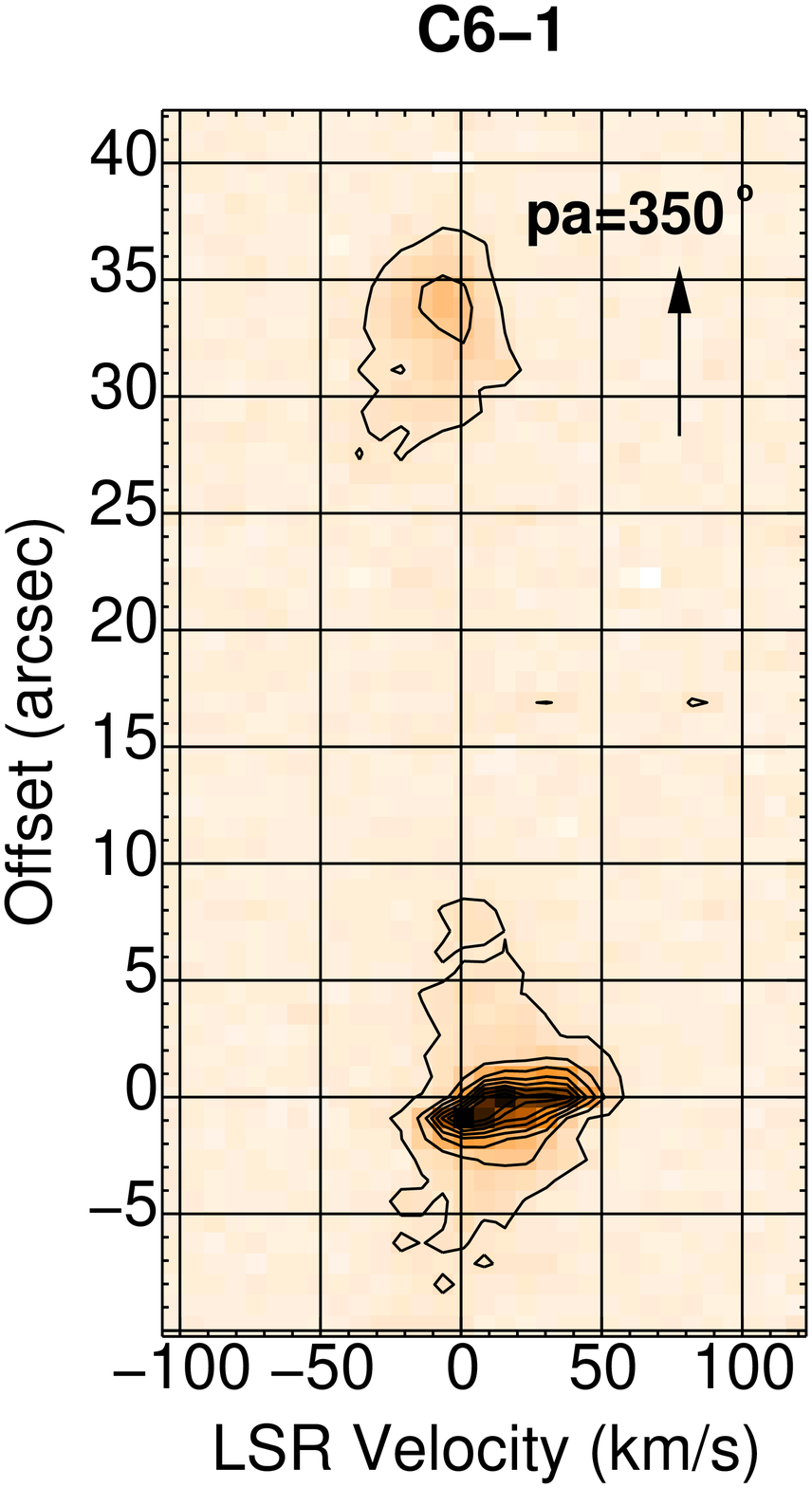}
    \epsfysize=5.3cm         \epsfbox{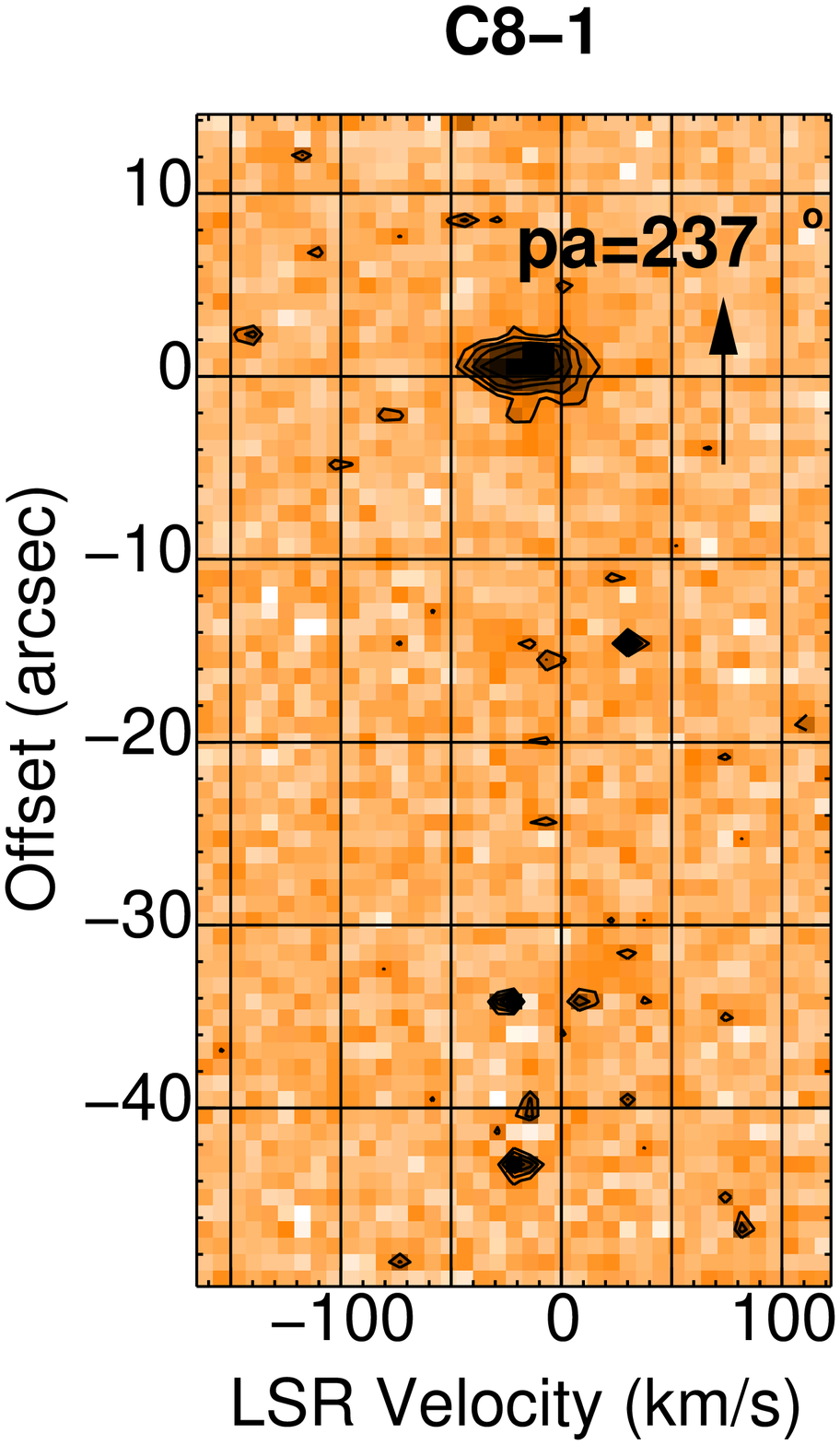}
\caption[]
 {Position-Velocity diagrams for slits in Region C. Contour levels are linearly spaced from an arbitrary lower signal level which is two to three times the noise level.}   \label{regionc}
  \end{center}
\end{figure*}

\begin{figure*}
  \begin{center}
  \epsfysize=5.3cm           \epsfbox{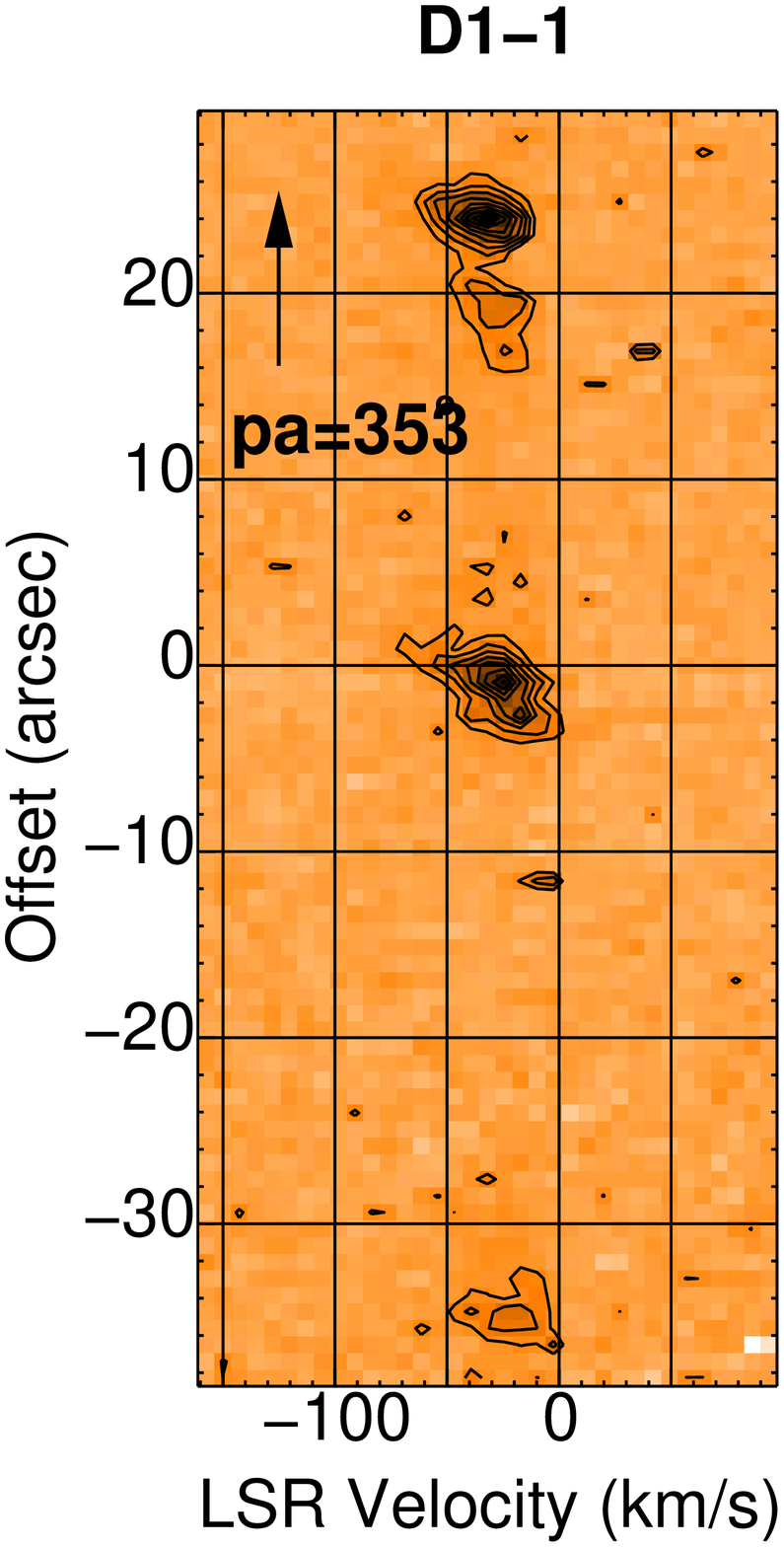} 
   \epsfysize=5.3cm          \epsfbox{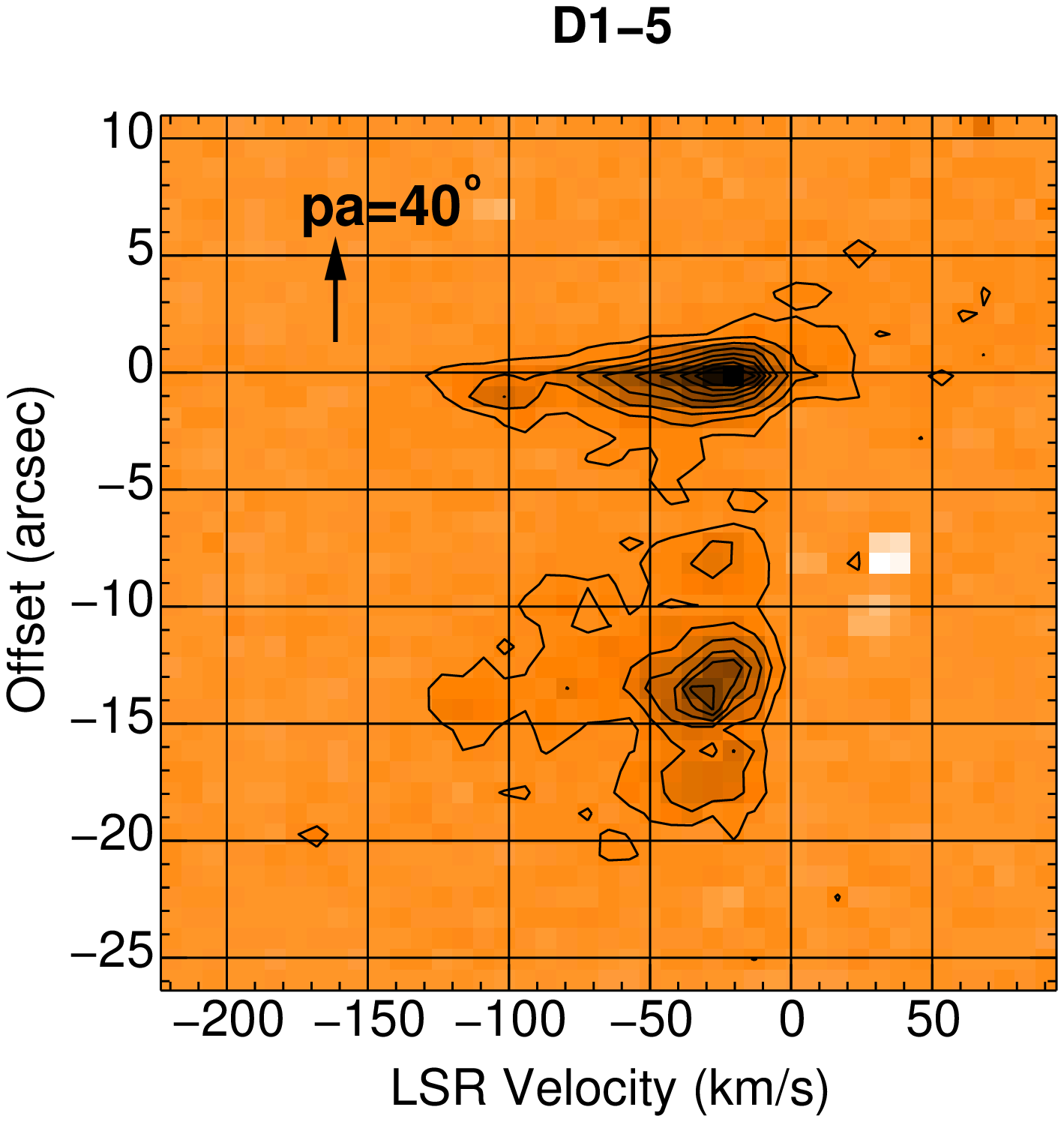}
    \epsfysize=5.3cm         \epsfbox{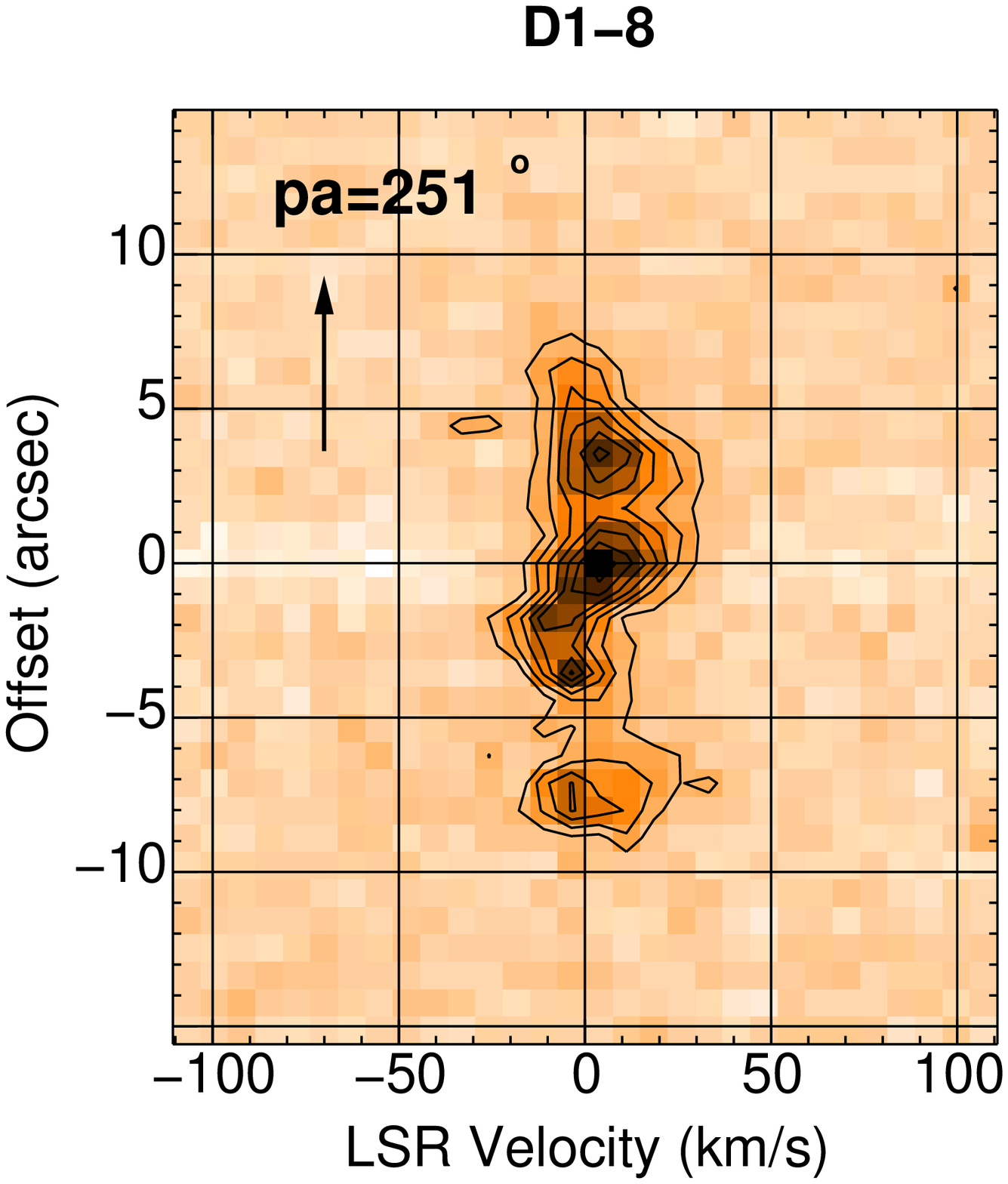}
\newline
    \epsfysize=5.3cm         \epsfbox{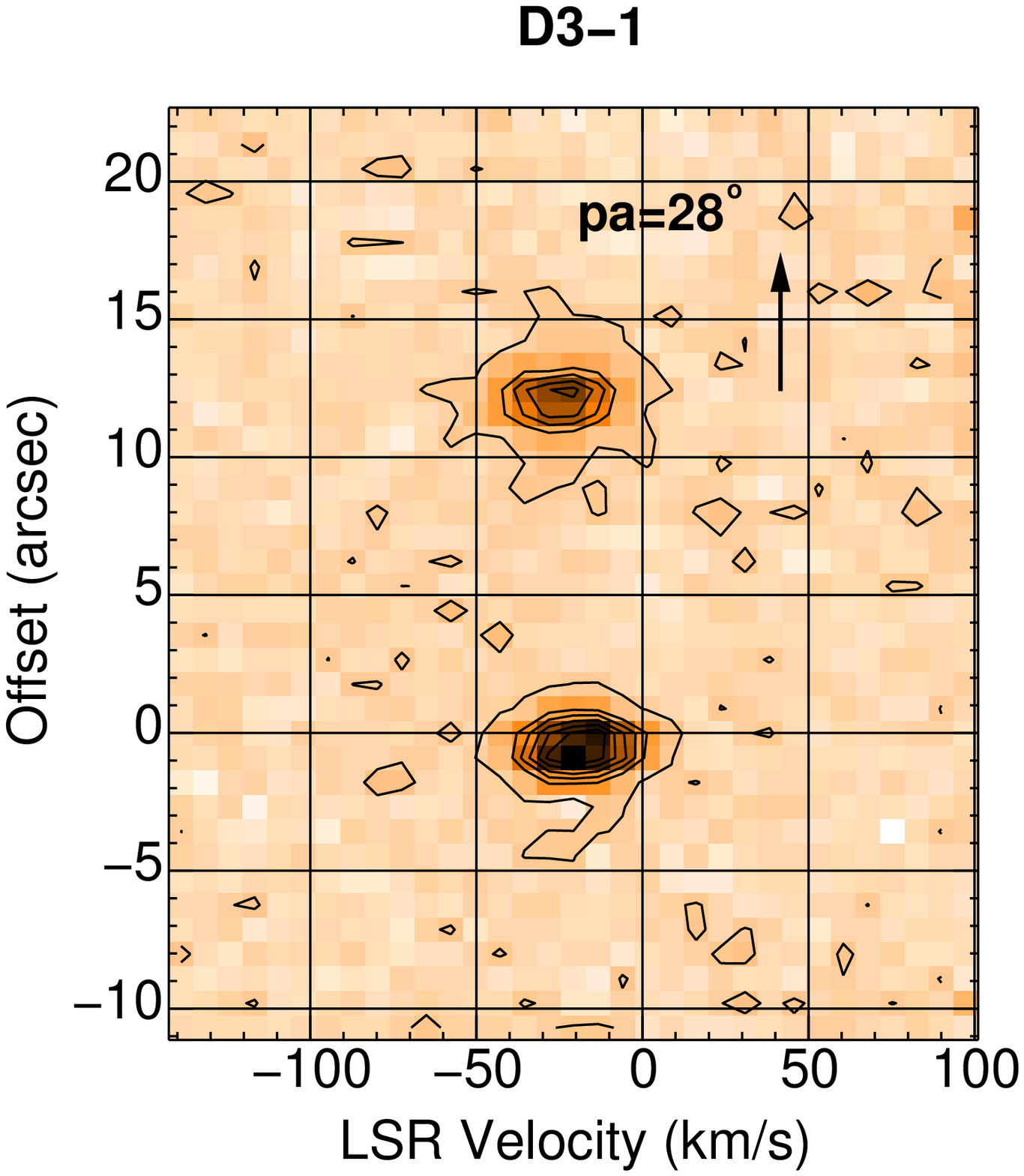}
    \epsfysize=5.3cm         \epsfbox{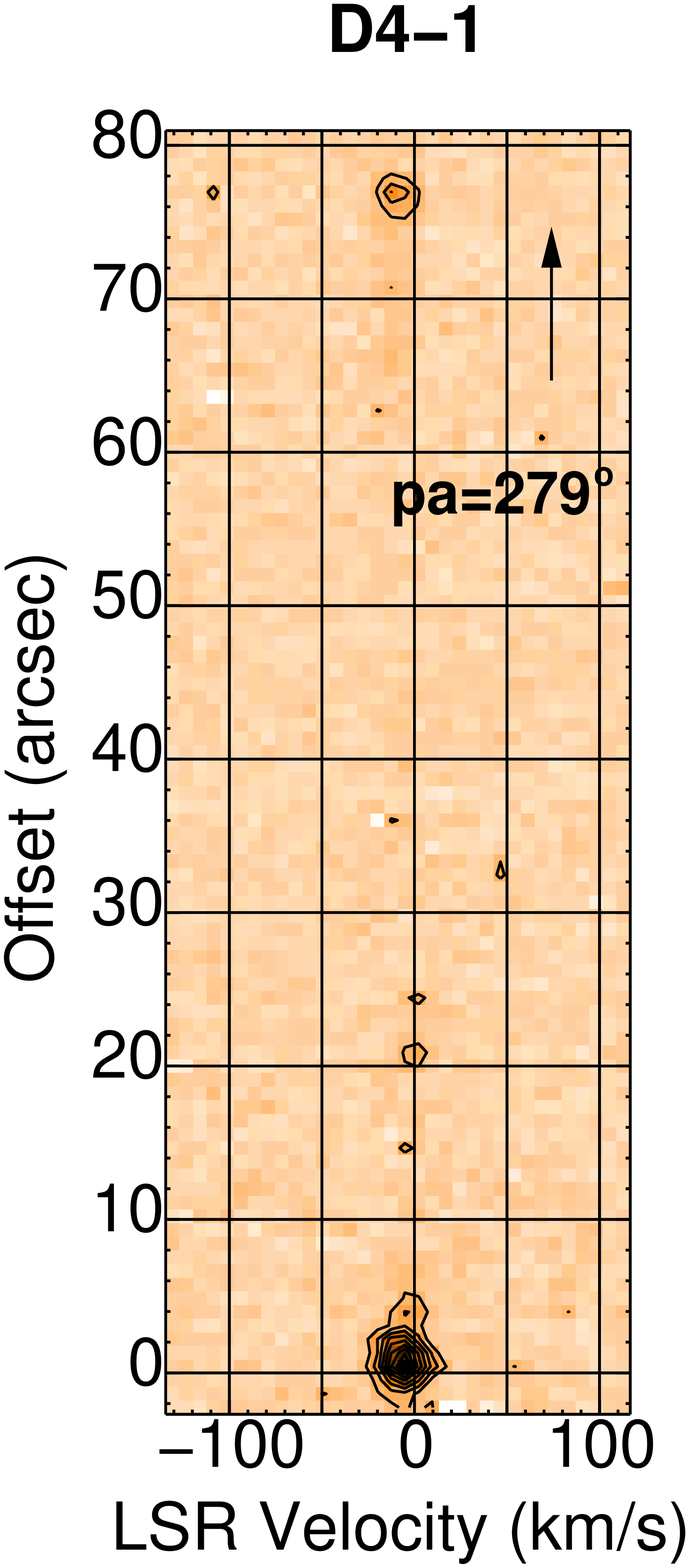}
     \epsfysize=5.3cm        \epsfbox{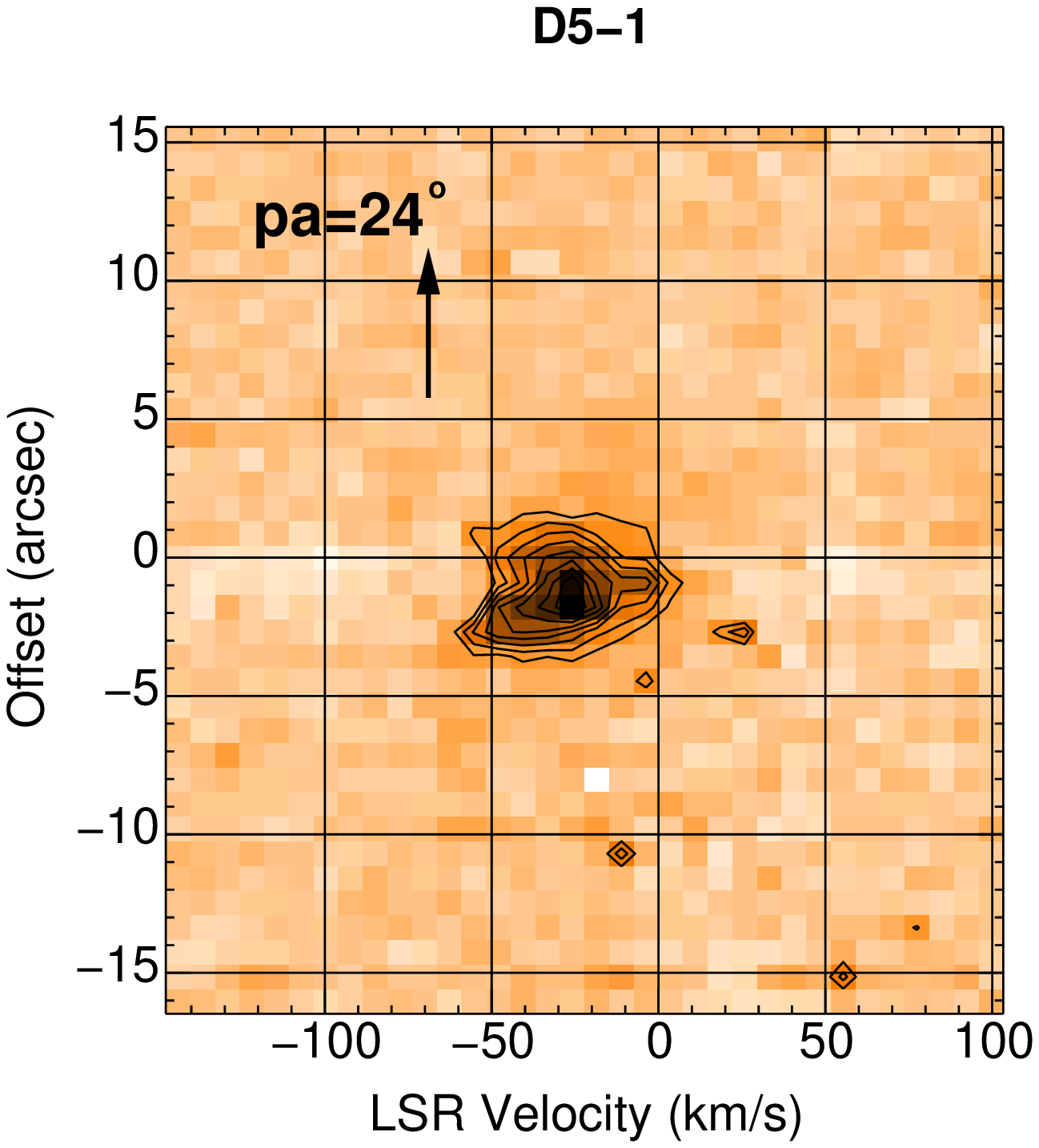}
\caption[]
 {Position-Velocity diagrams for slits in Region D. Contour levels are linearly spaced from an arbitrary lower signal level which is two to three times the noise level.}
    \label{regiond}
  \end{center}
\end{figure*}

\begin{figure*}
  \begin{center}
    \epsfysize=5.3cm          \epsfbox{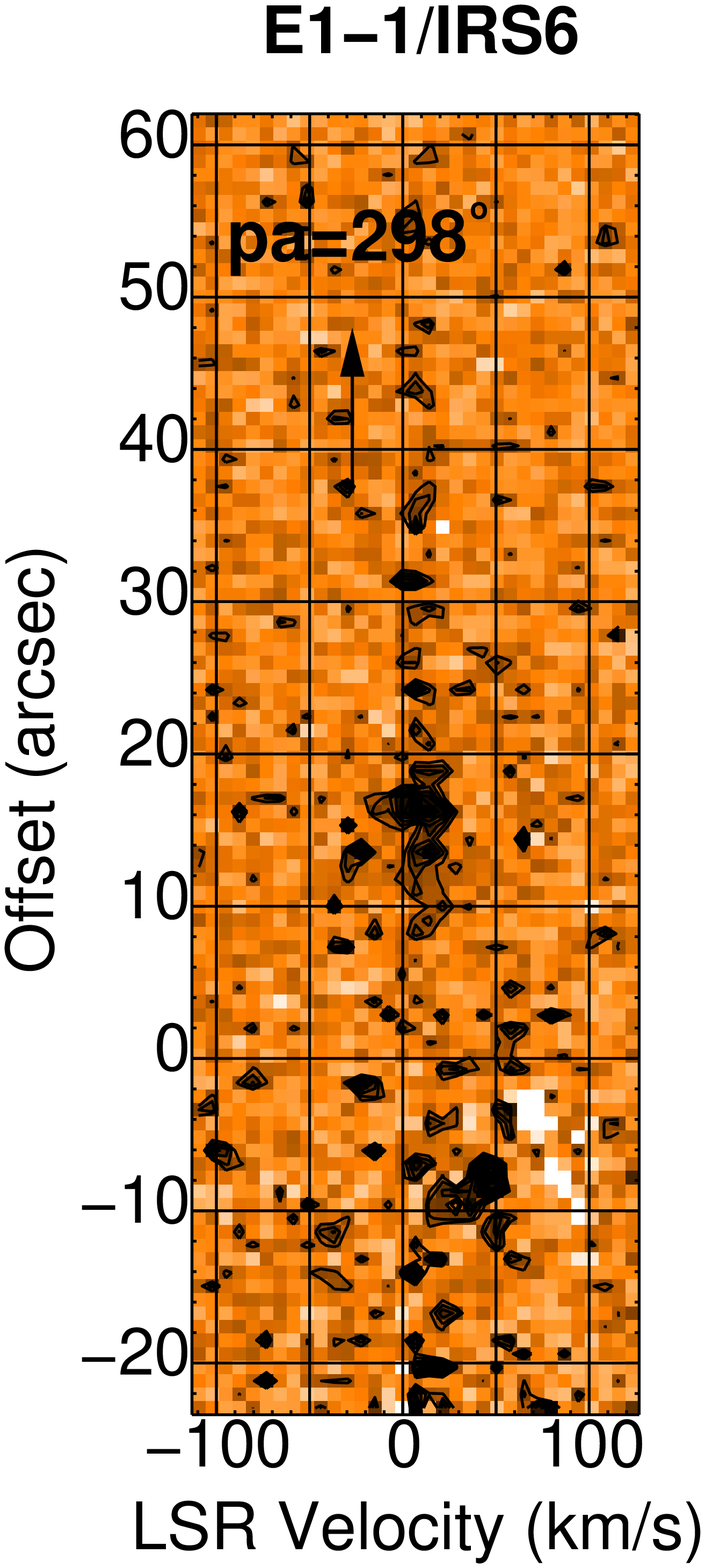}
     \epsfysize=5.3cm         \epsfbox{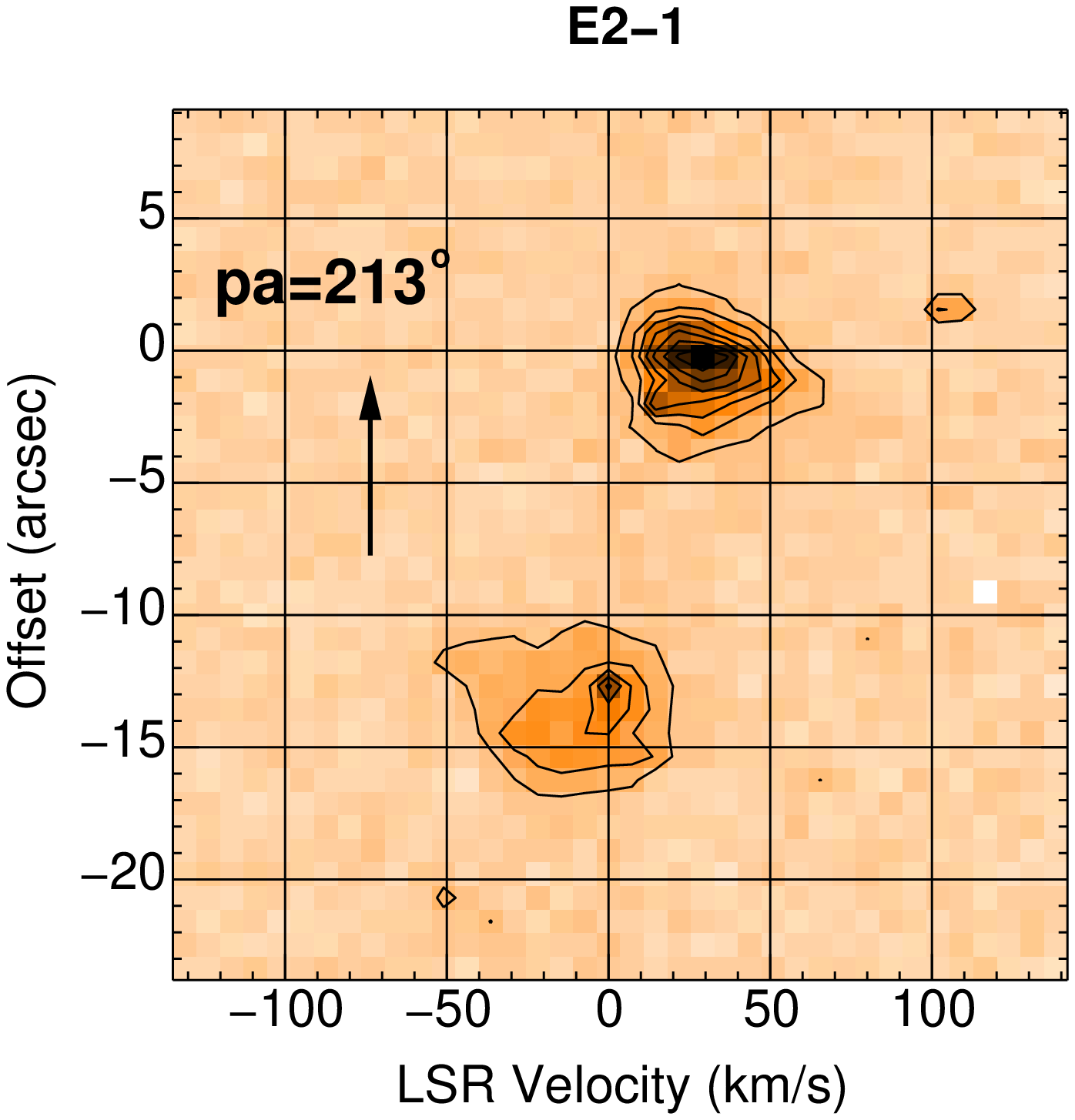}
      \epsfysize=5.3cm        \epsfbox{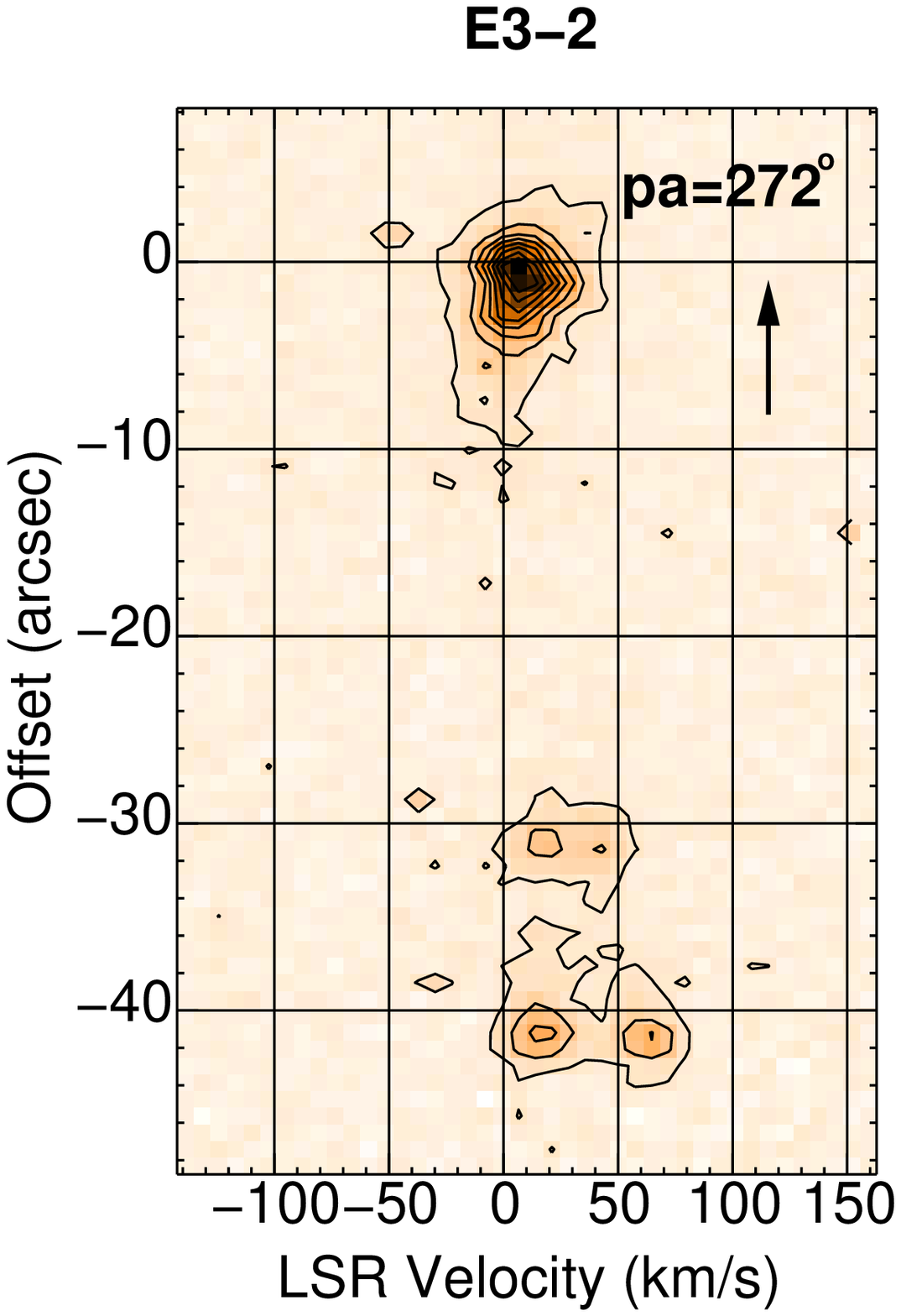}
     \epsfysize=5.3cm         \epsfbox{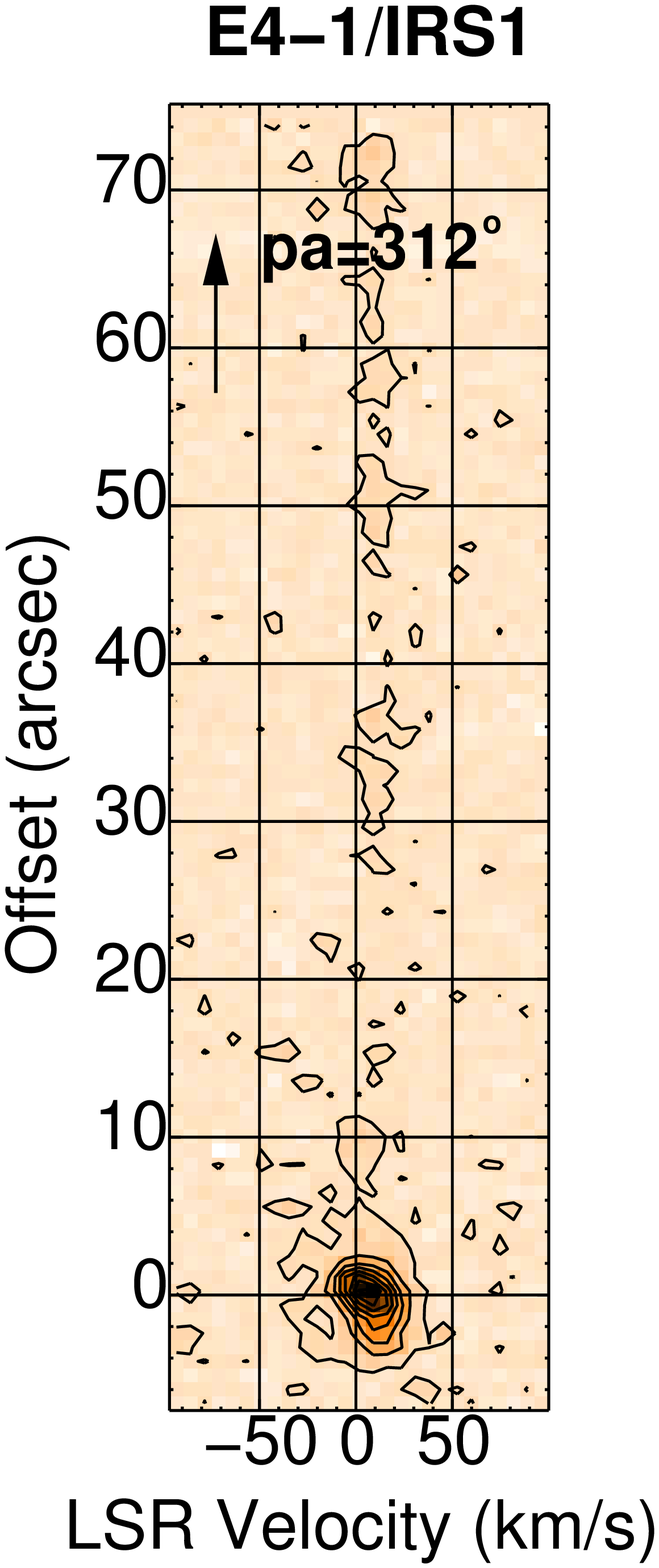}
\caption[]
 {Position-Velocity diagrams for slits in Region E. Contour levels are linearly spaced from an arbitrary lower signal level which is two to three times the noise level.}  
 \label{regione}
  \end{center}
\end{figure*}

The position-velocity diagrams are presented in Figs~\ref{regiona} -- \ref{regione} with the given slit position angles measured East of North.
The radial velocities and flux contributions of individual components within the PV diagrams  are catalogued in Tables~\ref{tab_jets} and \ref{tab_jets2}.
The detailed results for individual objects, including their nature and the possible driving sources in light of the PV information, are discussed in Appendix~B.

\begin{table*}
\caption{\label{tab_sources} List of objects.The Name references refer to the labels assigned by \citet{2007MNRAS.374...29D}  to individual or small groups of knots. The MHO designations typically refer to whole flow lobes. These were assigned as part of the larger Molecular Hydrogen emission-line Object (MHO) catalogue by \citet{2010A&A...511A..24D} }
\begin{center}
\begin{tabular}{rlrlllrrr}
\hline
ID & Name & \multicolumn{1}{c}{MHO} & \multicolumn{1}{c}{Date Observed} &
\multicolumn{1}{c}{RA [2000]} &
\multicolumn{1}{c}{Dec [2000]} & \multicolumn{1}{c}{PA$^a$} &
\multicolumn{1}{r}{Type of Object} & \multicolumn{1}{r}{Possible Source}\\
 & & \multicolumn{1}{c}{number}  & & h~m~s & h~m~s & $\circ $ &  & \\ \hline
1 & A~2-1   & MHO 801 & 08.08.2006 & 20:39:16.2 & 42:16:07 & 47                    & Knots & Protostar {\em a}  \\
2 & A~5-1   & MHO 804 & 22.08.2006 & 20:38:56.49 & 42:18:15 & 11                  & Jets     & Protostar {\em d}/DR21-IRS2\\
3 & A~6-1   & MHO 805 & 22.08.2006 & 20:38:44: & 42:19:08.8 & -105                & Bow-shock and knot & Unknown  in south-east \\
4 & A~9-1   & MHO 808 & 22.08.2006 & 20:38:50.27 & 42:19:09 & 49                   & Collimated flow & Main DR21 outflow\\
5 & A~10-1 & MHO 809 & 08.08.2006 & 20:39:03.7 & 42:20:16 & 47                   & Jet & DR21D\\
6 & B~1-1   & MHO 810 & 29.08.2006 & 20:38:53.1 & 42:20:08 & -69                    & Jets & DR21D/DR21-IRS~4 \\
7 & B~4-1   & MHO 813 & 21.08.2006 & 20:38:58.23 & 42:21:09.4 & -5                 & Bow & DR21-IRS~4\\
8 & B~6-1   & MHO 815 & 08.08.2006 & 20:38:56.87 & 42:23:08.5 & -76               & Jets & Protostars {\em f} \& {\em g }\\
9 & B~7-1   & MHO 816 & 21.08.2006 & 20:38:57.75 & 42:22:48.6 & 32                 & Jets    & Protostar {\em e}/DR21-IRS6\\
10 & B~9-1 & MHO 818 & 08.08.2006 & 20:39:04.1 & 42:26:07 & -90                   & Knots & Protostar {\em h} \\
11 & B~11-1 & MHO 820 & 21.08.2006 & 20:38:44.88 & 42:24:55.4 & -31           & Collimated flow &  DR21-IRS~10\\
12 & B~12-1 & MHO 821 & 29.08.2006 & 20:38:45.9 & 42:24:16 & -116               & Collimated flow? & Not known\\
13 & C~1-3   & MHO 854 & 29.08.2006 & 20:38:26.2 & 42:36:44 & -21                    & Bow-shock & VLA~1 \\
14 & C~1-5   & MHO 854 & 29.08.2006 & 20:38:30.6 & 42:36:51 & -20                    & Bow-shock &W75N-IRS2? \\
15 & C~2-1   & MHO 855 & 29.08.2006 & 20:38:42.5 & 42:37:46.8 & 26                  & Collimated flow + knot & Not known \\
16 & C~6-1   & MHO 857 & 29.08.2006 & 20:38:51.16 & 42:38:36.5 & -10               & Bow-shock and knot &   VLA~1\\
17 & C~8-1   & MHO 829 & 29.08.2006 & 20:38:32.6 & 42:39:40 & -123                  & Jet &   Protostars {\em j} \& {\em k}\\
18 & D~1-1   & MHO 832 & 30.08.2006 & 20:37:44.8 & 42:37:11 & -7                       & Bow & W75N-IRS10/11 \\
19 & D~1-5   & MHO 832 & 29.08.2006 & 20:37:46.5 & 42:38:32 & 40                      & Bow-shock?  & W75N-IRS10/11\\
20 & D~1-8   & MHO 832 & 30.0832.2006 & 20:37:49.3 & 42:40:01 & -109                   & Bow?  &  W75N-IRS10/11? \\
21 & D~3-1   & MHO 834 & 9.08.2006 & 20:38:0.04 & 42:41:32 & 28                       & Knots?  & W75N-IRS7/8 ({\em m}) or W75N-IRS9 \\
22 & D~4-1   & MHO 835 & 29.08.2006 & 20:37:45 & 42:43:47 & -81                         & Collimated jet?  &  Not known\\
23 & D~5-1   & MHO 836 & 30.08.2006 & 20:38:09.8 & 42:38:08 & 24                       & Collimated jet?  & W75N-IRS7/8 (protostar {\em m})? \\
24 & E~1-1   & MHO 844 & 23.08.2006 & 20:36:57.67 & 42:11:28.2 & -62                 & Knot?  & L906E-IRS6\\
25 & E~2-1   & MHO 845 & 23.08.2006 & 20:37:02.19 & 42:11:48.8 & -47                 & Knotty jet?  &  Not known\\
26 & E~3-2   & MHO 846 & 23.08.2006 & 20:36:57.3 & 42:13:18 & -88                       & Knotty jet?  & Not known \\
27 & E~4-1   & MHO 847 & 23.08.2006 & 20:37:19.4 & 42:16:16 & -48                       & Collimated jet  &   L906E-IRS1 (protostar {\em n})  \\
\hline
\end{tabular}
\end{center}
$^a$PA is the Position Angle
\end{table*}


\begin{table*}
\caption{\label{tab_jets} The radial velocities and fluxes of the H$_2$ jets and knots in Regions A and B with low-velocity (LVC) and high-velocity (HVC) components listed.} 
\begin{center}
\begin{tabular}{rllrrrrr}
\hline
  ID & Image Name & Object Name &  \multicolumn{4}{c}{Radial Velocities,  km s$^{-1}$}  &   Fluxes [Low, High] \\ 
       &                         &                           & Blue HVC & Blue LVC & Red LVC  & Red HVC       & \% of total \\ \hline
\hline
1 	& A~2-1	& A~2-1	         & -83.1  &   -3.4    &    -       &    -    &     {52, 31} \\
	&		& A~2-2		& -          &    -         &  28.8   &   -    &     {17, -}    \\
2 	& A~5-1 	& A~5-1		& -43.7  &   -23.3   &  -      	&   -     &    {36.5, 28} \\
	&		& A~5-2		& -85.9  &   -7.3     &   - 	&   -     &   {12, 23.5} \\	
3  	& A~6-1  	& A~7-1		&   -         &  -1.5      &  -	&     -    &    {47, -} \\
	&		& A~6-1		&   -         &  -4.4      &  -	&   -      &    {25, -} \\
	&		& A~6-1		&   -         & -10.3     &  -	&   -      &   {28, -} \\
4 	& A~9-1 	& A~9-1		&   -         &   -4.9     &  -	&   -      &    {53, -} \\
	&		& A~9-2		&   -         &    -         &  0.9    &  -      &     {23, -} \\
	& 		& A~9-3		&  -          &    -         &  5.2    &   -      &  {25, -} \\
5	& A~10-1	& A~10-1	         &  -           &    -0.3   &  -	&   -       & {22.5, -} \\
	& 		& A~10-1	         &  -           &  -18.6   & -	&   -      &  {40.5, -} \\
	& 		& A~10-1     	& -            &  -16.8   & -        &    -     &  {37, -} \\	
6 	& B~1-1 	& B~1-1		&  -           &  -14.0   &  -       &    -     &  {25, -} \\
	& 		& B~1-1		&  -           &  -16.9   &  -       &    -     &  {37, -} \\
	& 		& B~1-3		&  -           &  -22.7   &   -      &    -     &  {23, -} \\
	& 		& B~1-3		&  -           &  -11.0   &  -	&    -      & {15, -} \\
7	& B~4-1 	& B~4-1		&  -           &   -           &  7.9  &  -       & {44, -} \\
	& 		& B~4-1		&  -           &   -8.1     &    -     &  -       & {56, -} \\
8	& B~6-1 	& B~6-1		&   -           &  -12.5   &   - 	 &  -       & {73,  - } \\
	&		& B~6-2		&   -           &   -4.5     &   -     &   -       & {14.5,  - } \\
	&		& B~6-3		&  -            &    -         &   4.3  &  - 	      &   {12.5,  - } \\
9 	& B~7-1/IRS6 & B~7-1	&  -            &   -8.5    &   -   	&      -     &     {64.5,  - } \\
	& 		& B~7-2		&   -            &  -7.1    &  -  	&     -      &    {35.5,  - } \\
10	& B~9-1 	& B~9-1		& -85.3      & -19.2   &  - 	&      -     &     {14, 7} \\
	& 		& B~9-1		& -68.7    &  -24.7    &  - 	&      -     & {14.5, 13.5} \\
	& 		& B~9-1		& -105.4  & -39.4     &   - 	&      -     & {17, 15} \\
	&		& B~9-1		& -96.3     & -28.4    &   -  	&      -     &  {9, 11} \\
11 	& B~11-1/IRS10 & B~11   &  -            &  -           &   -      &  30.0  & { -,50 } \\
	& 		& B~11-2    	&    -           &   -         & 24.2  &    -      &  {50,  - } \\
12	& B~12-1 	& B~12-1a 	&    -           &  -23.8  &  -      &     -      &    {2,  - } \\
	& 		& B~12-1b 	&  -             &  -18.0  &  - 	&      -     &   {7.5,  - } \\
	& 		& B~12-2a 	& -90.6     & , -12.2 &  - 	&      -     & {13, 3} \\
	& 		& B~12-2b 	&  -              &    -0.6  &   -    &      -     & {74.5,  - } \\
	\hline	
\end{tabular}
\end{center}
\end{table*}

\begin{table*}
\caption{\label{tab_jets2} The radial velocities and fluxes of the H$_2$ jets and knots in Regions C, D and E with low-velocity (LVC) and high-velocity (HVC) components listed.} 
\begin{center}
\begin{tabular}{rllrrrrr}
\hline
  ID & Image Name & Object Name &  \multicolumn{4}{c}{Radial Velocities,  km s$^{-1}$}  &   Fluxes [Low, High] \\ 
       &                         &                           & Blue HVC & Blue LVC & Red LVC  & Red HVC       & \% of total \\ \hline
\hline
13	& C~1-3 	& C~1-2		&  -            &    -         & 6.6   &  -        & {25,  - } \\
	&		& C~1-3		&  -            &    -         & 0.7   &  -        & {30,  - } \\
	& 		& C~1-3		&  -            &    -         & 3.7,   &  -       & {45,  - } \\
14 	& C~1-5 	& C~1-1		&  -            &  -          &  16.0  &  -      &     {70.5,  - } \\
	& 		& C~1-5		&  -            &  -         &      2.6  &  -      & {26.5,  - } \\
	&		& C~1-4		&  -            &  -         &    10.0  &  -      & {3,  - } \\
15	& C~2-1 	& C~2-1		&  -            &  -2.8    &       -     &  -      &  {70.5,  - } \\
	& 		& C~2-1		&  -            &  -         &   13.4   &   -      & {29.5,  - } \\
16 	& C~6-1 	& C~6-1		&  -            &  -         &     0.1    & 32.6 & {44, 32.5} \\
	& 		& C~7-1		&  -            &   -5.8   &      -       &  -       &     {24,  - } \\
17	& C~8-1 	& C~8-1		&  -            &    -6.1   & - 	    &  -       &  {100,  - } \\
18 	& D~1-1 	& D~1-1		&   -            &  -15.0  & - 	    &  -       & {16.5, -} \\
	& 		& D~1-2		&   -            &   -26.8 & - 	    &  -       & {37, -} \\
	& 		& D~1-2		&   -            &  -19.4   & - 	    &  -       & {16.5, -} \\
	& 		& D~1-2		&  -34.2     &      -        &  - 	    &  -       & {30, -} \\	
19 	& D~1-5 	& D~1-5		& -115.6    &  -34.4    &   -     &   -     & {36, 13} \\
	& 		& D~1-5		& -102.3    &  -21.2     &  -     &   -     & {38, 13} \\
20 	& D~1-8 	& D~1-8		& -              &  -4.2        &    -   &  -     & {16,  - } \\
	& 		& D~1-8		&  -             &   -3.1       &   -    &  -	    & {25,  - } \\
	& 		& D~1-8		&   -           &    -            &  3.2 &   -    & {29.5,  - } \\
	& 		& D~1-8		&  -            &    -            &  1.1  &  -     & {29.5,  - } \\
21 	& D~3-1 	& D~3-1		&   -            &  -20.0	&     - 	    &  -       &    {55,  - } \\
	& 		& D~3-2		&   -            &  -20.0	&  - 	    &  -       &{45,  - } \\
22 	& D~4-1 	& D~4-1		&   -            &  -4.5	&  - 	    &  -       &{80,  - } \\
	&		& D~4-5		&   -            &  -8.9	&  - 	    &  -       &{20,  - } \\
23 	& D~5-1 	& D~5-1		&   -            &  -25.9	& - 	    &  -       &{100,  - } \\
24	& E~1-1/IRS6 	& E~1-1	&   -            & -              & 3.2	&  -       &{100,  - } \\
25 	& E~2-1 	& E~2-1		&  -            &  -               & 0.7	&  -       &{43.5,  - } \\
	&		& E~2-2		&  -            &   -             & 29.8	&  -       &{56.5,  - } \\
26 	& E~3-2 	& E~3-2		&   -            &  -             &16.0        &  65.4	      & {12.5, 10} \\
	& 		& E~3-2		&  -            &   -              & 17.4     &  40.7    &{6.5, 5} \\
	& 		& E~3-3		&   -            &  -              & 7.3	&  -       &{66,  - } \\
27 	& E~4-1/IRS1 	& E~4-1	&  -            &   -              & 8.2	& -       &{100,  - } \\	
\hline
\end{tabular}
\end{center}
\end{table*}

\section{ Analysis}
\label{regions}

\subsection{Object Analysis}
\label{specific}

We will make the assumption that particle collisions within shock waves are responsible for the vibrational excitation.
This is consistent with the extended nature of most structures where there is no potential source of ultraviolet in the vicinity \citep{2007MNRAS.374...29D}.
Extreme ultraviolet radiation could, however, be generated locally  from high-speed sections of curved shocks. Although the molecules 
would be destroyed in these shock sections, H$_2$ in adjacent weaker sections could be excited.   
Nevertheless, fluorescent emission following pumping, reformation or Lyman resonance, is unlikely to dominate over collisional excitation to the lower vibrational state as supported 
by observations of bow shocks in other locations \citep{2000A&A...359.1147E}.

To interpret the PV diagrams, we need to interface the radiative emission process with the physics of the shock wave. Most importantly,
molecular hydrogen emission from a shock wave is expected to peak at a velocity corresponding to that of the pre-shock medium. This is because the H$_2$ is heated to typical 
emitting temperatures of 2,000K well before being fully accelerated by the driving medium. This applies to both C-shocks \citep{1990MNRAS.245..108S} and J-shocks 
\citep{2004A&A...419..975O}, and thus differs from optical line emission from atoms  which can  generate emission lines as wide as the bow speed with peaks at 
widely separated radial velocities \citep{1987ApJ...316..323H}. For this reason, the FWHM widths of H$_2$ emission lines even from bow shocks is very limited, typically to 10--20~km~s$^{-1}$ while FWZI are $\sim$ 45~km~s$^{-1}$ for both C and J-type bow shocks \citep{2000AJ....120.1974Y,2004A&A...419..975O}.
Therefore, we can take  the H$_2$ radial velocities to correspond within these limits to the motion of the pre-shock gas. 

Certain PV  morphologies  recur quite often. We thus attempt here to group these, as follows.

Type A. A number of PV diagrams display dominant single components (e.g. B\,12-1, C\,6-1, C\,8-1, D\,5-1) or well spatially-separated compact components (A\,6-1, C\,6-1, D\,3-1).  Generally extended in velocity space by up to  $\sim$ 45~km~s$^{-1}$, the peaks are
adjacent to but generally not coincident with the LSR velocity although some emission usually extends to the rest speed. Such components thus could correspond to bow shocks driven forwards into an ambient medium by jets which themselves do not emit significantly (such as atomic jets or low density molecular jets). Similar structures were found in the OMC2/3 study
(YBD4, YBD5, 5, YBD12; \citet{2000AJ....120.1974Y}).

Type B. A second type is distinguished by two H$_2$ emission components which are spatially coincident but 
widely separated  in radial velocity. Objects A\,2-1 and  D\,1-5 display the two components. 
Similar structures were found in a spectroscopic study of jets and outflows in the OMC 2/3 region.
 (YBD-17N and YBD-39) \citep{2000AJ....120.1974Y}. We recognise that these objects are consistent with classic twin-shock configurations. 
The highly-shifted component is identified with a Mach shock disc, corresponding to a reverse shock in a high speed molecular jet being brought to a halt.
This drives a forward bow shock  into the slow-moving ambient medium. 

Type C. Other velocity-pairs are found to be spatially coincident but appear to be part of more complex aligned  structures extended along the slit (A\,5-1, B\,9-1; see also YBD-17S of 
\citet{2000AJ....120.1974Y}).
This may represent a class of object which consists of an internal shock in a molecular jet which drives oblique shocks into a surrounding molecular medium. 
The jet is thus deflected or constricted by the ambient obstacle rather than being terminated. 

Type D. Several PV diagrams display emission at low speed along an extended  section of the slit. (A\,9-1, B\,7-1, C\,2-1, D\,1-8, E\,1-1, E\,4-1; see also YBD67/68). Some of these are clearly a chain of 
connected knots while others are harder to classify due to the low level of emission. While the bright knots may represent oblique shocks propagating into the ambient molecular cloud,
the emission inbetween the knots is enigmatic. It could correspond to the limb-brightened edge emission from a shock driven by an expanding cavity inflated by a wide wind (or wide-angle precessing jet)  or  could be from a turbulent interface layer between a jet and the ambient cloud (see \citet{1999A&A...348..584B})

\subsection{Jet deceleration}
\label{deceleration}

Here we evaluate evidence from PV diagrams for outflowing material impacting into the ambient medium. This can be gauged by looking for deceleration of the molecular hydrogen, away from the suspected source (so the material closest to the source has the highest velocity).

Object B~9-1 is a good illustration of systematic deceleration. The material appears to be decelerating as one looks across the four components further away from the likely source which is a Spitzer identified protostar 
(object `h' of citet{2007MNRAS.374...29D}) which is located close to the western knot (i.e. towards the top of the slit as displayed in Fig.~\ref{regionb}). In the future (a few thousand years),  the four knots may accumulate to form a single massive knot.
The PV diagram for A~5-1 also displays strong deceleration from object A~5-2 to A~5-1. The likely source (protostar `d') is not far from A~5-2.

The PV diagram for B~12-1 appears to show some deceleration going from B~12-1 to B~12-2. However, the source driving this object is not known. The source would have to be in the east for this to be the case. The PV diagram for C~1-3 shows some deceleration away from the likely source VLA~1.

The PV diagram for E~2-1 shows deceleration going from object E~2-2 to E~2-1. This points to a possible source in the north-east. However no source is known as the \textit{Spitzer} IRAC images used by~\citet{2007MNRAS.374...29D} did not cover all of the region. A couple of pre-main sequence objects are in the correct vicinity.
Similarly, Object E~3-2 shows material possibly decelerating as it impacts with the ambient medium. However the source driving this object is not known from current data (not covered by the \textit{Spitzer} IRAC images). 

\subsection{Distribution}
\label{distribution}

Excluding the distinct high-speed components, the average radial velocity of the low-speed components for each region can be calculated relative to the LSR. We find these to be as follows.

\begin{description}
                  \item[Region A:] ~~~-6.5~km~s$^{-1}$                  
                  \item[Region B:]~~-10.8~km~s$^{-1}$
                    \item[Region C:] ~~~~~2.8~km~s$^{-1}$
                   \item[Region D:] ~~-16.0~km~s$^{-1}$
                   \item[Region E:] ~~~~11.8~km~s$^{-1}$
 \end{description}

These can be compared to the individual LSRs of approximately -2.5~km~s$^{-1}$ (A \& B),  10.3~km~s$^{-1}$ (C \& D),  -1.4~km~s$^{-1}$ (D) and  16.9 ~km~s$^{-1}$ (E) \citep{2006A&A...458..855S,2013A&A...551A..28T}. 

The low velocity H$_2$ in  Regions A and B is clearly blueshifted.  This could be due to dust extinction from the DR21 cloud which is obscuring corresponding redshifted
objects behind the cloud. However, the outflows are not large and, unless the cloud was relatively thin, it is not clear that the small flows would thus appear asymmetric. On the other hand if the extinction was local to individual cores in which the outflows     are driven, then the asymmetry would be expected. Indeed, the cloud is described as a ridge with filament and sub-filaments
in tracers of the cold dense gas \citep{2010A&A...520A..49S,2012A&A...543L...3H}. The typical filament column of 10$^{22}$~cm$^{-2}$ \citep{2012A&A...543L...3H}.is sufficient to provide the required near-infrared obscuration. 

For Region C, we note that slits C 1-3 and 1-5 pass through the bow that envelopes the western red-shifted lobe of the W75N outflow, while C 2-1 and C 6-1 dissect the eastern blue lobe \citep{1998MNRAS.299..825D}.  Interestingly, the H$_2$ velocities are all low and close to the systemic LSR velocity of $\sim$ 10 km s$^{-1}$, even though these sweeping bow shocks are known to be associated with well-defined blue and red-shifted outflow lobes.  The only real exception is the high-velocity feature seen at offset $\sim$ 33\arcsec in the C 6-1 PV plot, which is associated with a separate outflow to the north of W75N. The low velocities in Fig.~\ref{regionc}support the idea that H2 radiates before it is accelerated to high speeds.  Note that the high-speed molecular gas, traced in CO J=3-2 emission by \citet{1998MNRAS.299..825D}, sits inside the large bow shock in the blue lobe (see their Figure 2), as expected.

Region D displays higher systematic blueshifted H$_2$ motions. The faster moving objects tend to be those roughly aligned such as D~1-1 (i.e. D\,1-1, D\,1-5, D\,3-1 and D\,5-1). This suggests a common, driving source. Suggestions for this source are W75N-IRS10/11 and W75N-IRS7/8~\citep{2007MNRAS.374...29D}.

In Region E, the low speed H$_2$  is generally blueshifted relative to the LSR of the DR017/L906 cloud. Moreover,   all objects in this region are  roughly aligned (see Fig.~\ref{fiveregions}). Possible sources are suggested to be L906E-IRS\,1 or L906E-IRS\,6 \citep{2007MNRAS.374...29D}. 

\begin{figure}   \begin{center}
  \epsfysize=6.0cm   \epsfbox{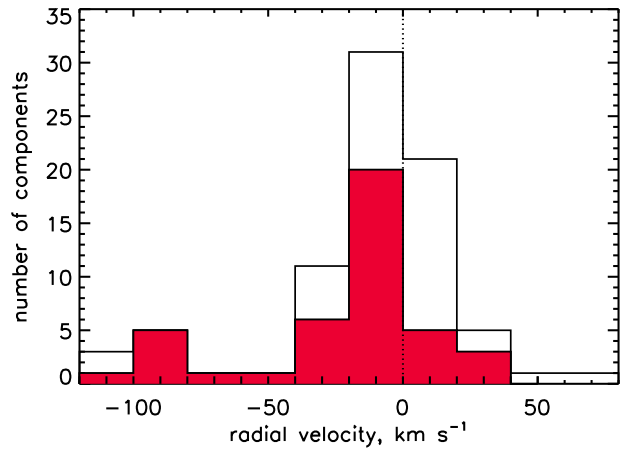}
\caption[Histogram of the radial speed of all discrete velocity components.]{Histogram of the number of components with peak radial speed binned into 20~km~s$^{-1}$ of all discrete velocity components  in all Regions (unshaded) and in Regions A and B (shaded, red).}
\label{histo2}
  \end{center}
\end{figure}

The distribution of peak radial velocities for all the objects, put into bins of 20~km~s$^{-1}$, is shown in 
Fig.~\ref{histo2}. This displays a peak in the -20 to 0~km~s$^{-1}$ bin,  with the average peak for all objects in the five regions blueshifted by  10~km~s$^{-1}$. This low-velocity peak suggests that the pre-shocked H$_2$ in the ambient medium is outflowing before the exciting shock arrives. This would be consistent with the existence of CO outflows (of order 10,000 years) with much longer lifetimes than the H$_2$ outflows/jets (dynamical times of a few thousand years).

In support of this jet-outflow scenario, is the existence of high-speed H$_2$. This is shown in Fig.~\ref{histogram} with the peak at -80-100~km~s$^{-1}$. Given that this is the radial component, there is 
clear evidence for molecular  jets with  speeds exceeding 100~km~s$^{-1}$. All these purported high-velocity  jets are blueshifted. This could at least partly be explained through an orientation effect
since the high speeds would be associated with an angle well out of the plane of the sky. Hence the receding jet is likely to be obscured by the core associated with the driving protostar. 

The alternatives to the interpretation as  molecular jets, are models that invoke acceleration of the ambient  cloud. The fact that we are observing vibrationally-excited molecular hydrogen at these speeds places severe constraints on these models. Single C-shocks have been invoked as more effective that J-shocks in both channelling energy into the molecular lines and in accelerating the molecules relatively gently. However, high Alfv\' en speeds (i.e. strong magnetic fields) would be required to provide acceleration to the observed speeds  exceeding   80~km~s$^{-1}$ \citep{1991MNRAS.248..730S}.

Acceleration of molecular cloud material through entrainment and multiple shock acceleration, followed by vibrational excitation, could produce high speed H$_2$.
However,  such complex turbulent acceleration mechanisms  require further study to determine how well they can operate. For example, one could expect a wide range of speeds,
from zero up to the highest velocity  such as found in the main OMC-1 and DR21 outflows. This is in contrast to the observed discrete component at high speed in the present sample. 
We thus favour the high-speed jet model as the interpretation here.

The two prominent asymmetries in the radial velocity distribution suggest that extinction is very significant. However, it is not obvious how cloud extinction would remove all high-speed components
with absolute radial speeds exceeding 80~km~s$^{-1}$  except  the eight blueshifted objects.   There appears to be a selection effect in operation which we explain as follows.
Dense molecular jets are associated with high accretion rates corresponding to the Class 0 stage of protostars. The jets are believed to propagate through a narrow conical cavity which widens with time as the accretion rate falls and the jet becomes lighter, entering the Class 1 stage for the protostar. Therefore, 
if these high-velocity components  correspond to shocks within or terminating dense molecular jets with typical speeds of $\sim$ 100-120~km~s$^{-1}$, then one can also expect that these components are within the conical Class 0 cavities. Therefore, we would only detect the high speed emission knots from cavities in which the cavity is open to our line of sight. 

In this interpretation, Class 1 outflows are driven by lighter jets which may be dominated by neutral atomic gas. Therefore, H$_2$ emission is mainly generated at the locations where the ambient medium is swept up and shocked by the lighter outflow. 
The above scenario is not easy to test. The prediction is that there are many more high speed Class 0 flows hidden from view due to an unpropitious cavity opening axis.    

\section{Models}
\label{models}
\subsection{Bow Shock Modelling}
\label{shocks}

Models for bow shock structures have provided successful interpretations for many isolated components of H$_2$  protostellar flows. \citep{1998MNRAS.299..825D,1999MNRAS.308..539D,2004A&A...418..163K,2010A&A...513A...5G}. 
The objects A~6-1, B~4-1, C~1-3, C~1-5, C~6-1, D~1-1, D~1-5 and D~1-8 were indeed  believed to be bow shocks, based just on the  previous imaging~\citep{2007MNRAS.374...29D}.
Position-velocity diagrams for bow shocks permit a more detailed comparison and so provides a means of confirmation \citep{2004A&A...419..975O}. We  have therefore simulated PV diagrams for some of the prominent bow shapes.

We first employ the CBOW and JBOW models as developed by \citet{1991MNRAS.252..378S,2007A&A...466..565S}.   These models assume a three-dimensional shock surface 
such as a paraboloid and then dissect the surface into planar elements each treated as a separate one-dimensional steady shock wave. Cooling and chemistry are included and 
the shock thickness taken into account with the approximation that the cooling time is much shorter than the flow time across the entire bow. 
 
\begin{figure*}
  \begin{center}
      \epsfysize=3.5cm        \epsfbox{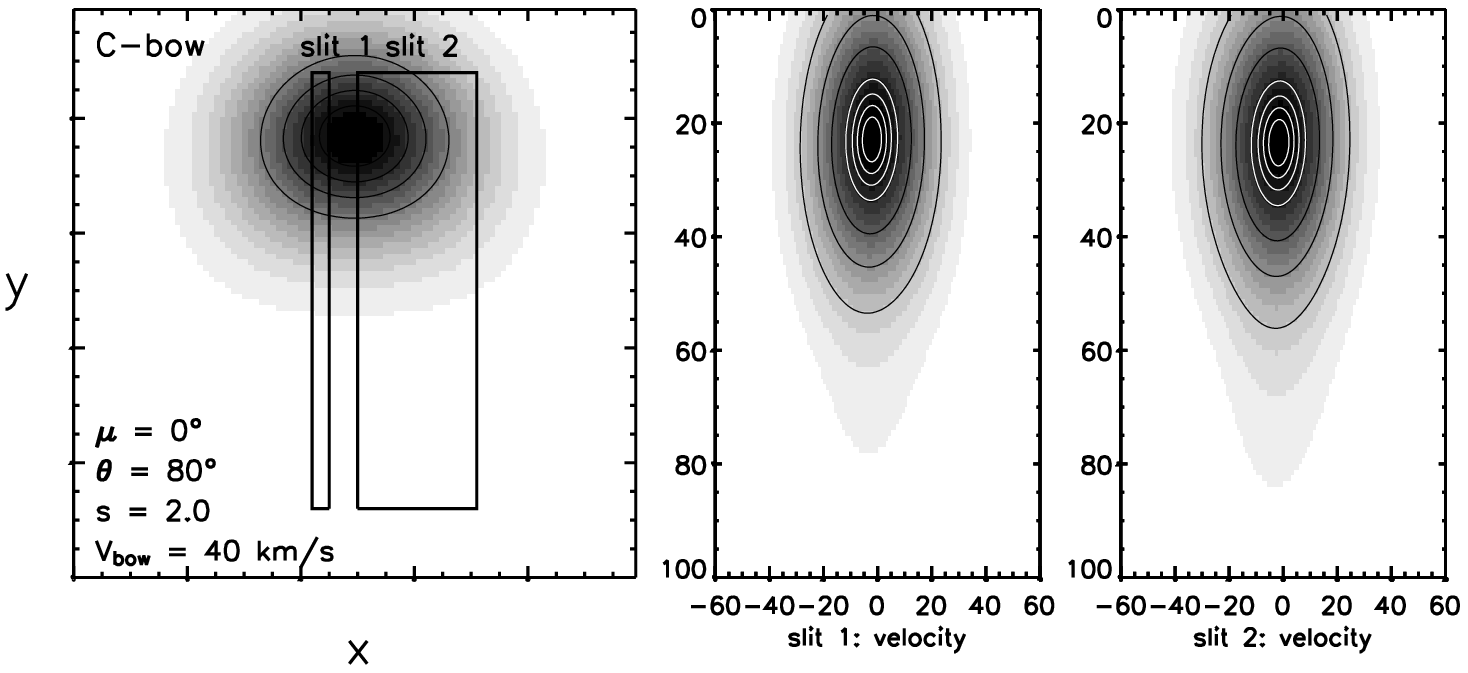}
      \epsfysize=4.0cm          \epsfbox{4B4_1.eps}
     \epsfysize=4.0cm         \epsfbox{4C1_3.eps}  
      \caption{A  C-type bow shock generated with a CBOW model, showing the image (left) and PV diagrams for or Slit~1 (second panel) and Slit~2 (third panel). Chosen bow speed is
V$_{bow}$ = 40~km~s$^{-1}$,  the shape is a paraboloid ($s = 2$), angle to the line of sight is $\theta = 80^\circ$, and a uniform magnetic field along the bow axis ($\mu = 0^\circ$)   Comparison is to bows such as found in Objects B~4-1 and C~1-3. }
\label{compare2}
  \end{center}
\end{figure*}

\begin{figure*}
  \begin{center}
     \epsfysize=3.5cm        \epsfbox{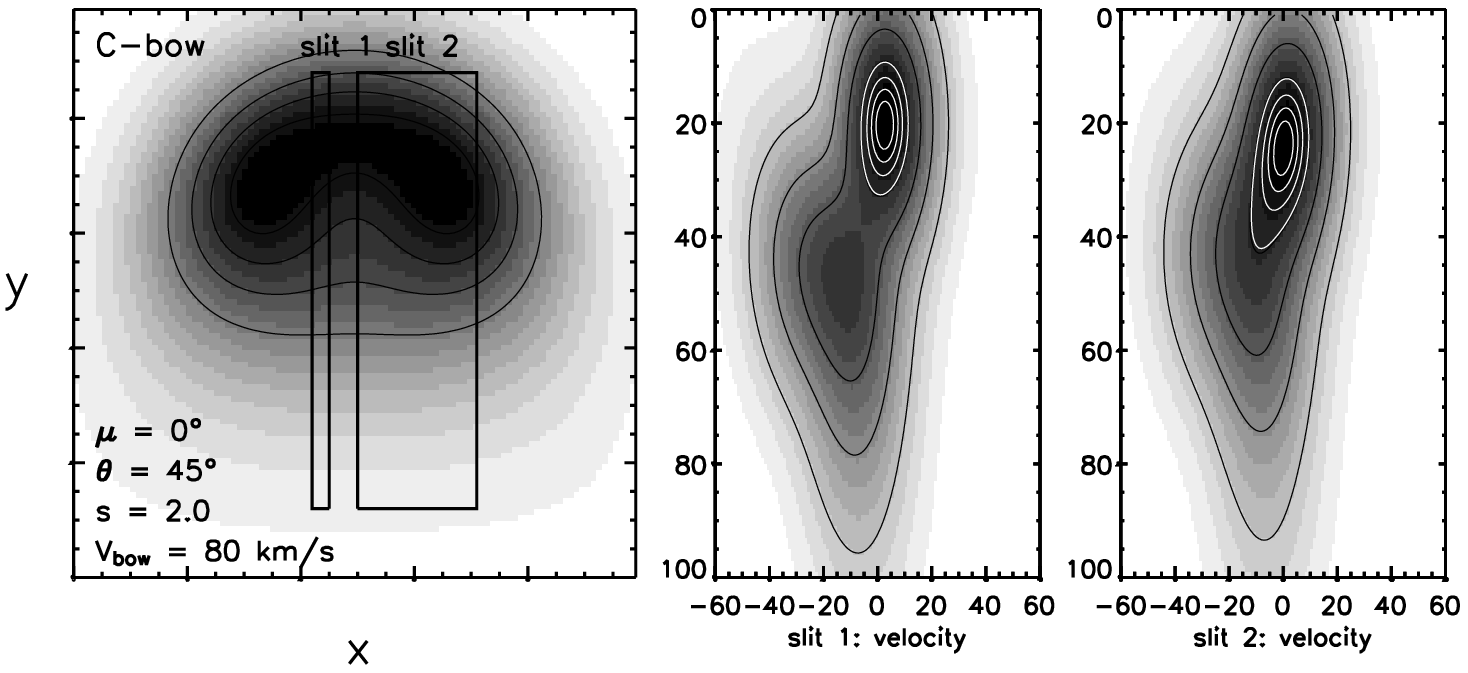}
      \epsfysize=4.0cm          \epsfbox{4B1_1.eps}
     \epsfysize=4.0cm         \epsfbox{4D1_8.eps}         
\caption{A  C-type bow shock generated with a CBOW model, showing the image (left) and PV diagrams for or Slit~1 (second panel) and Slit~2 (third panel). Chosen bow speed is
V$_{bow}$ = 80~km~s$^{-1}$,  the shape is a paraboloid ($s = 2$), angle to the line of sight is $\theta = 45^\circ$, and a uniform magnetic field along the bow axis ($\mu = 0^\circ$)   Comparison is to bows such as found in Objects B~1-1 and D~1-8.}
\label{compare4}
  \end{center}
\end{figure*}

A large number of such simulations were undertaken and compared to the Type~A components described above. The model is particularly useful for an interpretation in terms of magnetohydrodynamic C-shocks.  Figure~\ref{compare2} displays a synthetic image and two synthetic PV diagrams for a paraboloidal C-type bow of  low speed (40~km~s$^{-1}$) moving almost transverse to the line of sight ($\theta = 80^\circ$).
This generates a compact bow shock which would not be resolved in the present observations, consistent with the compact knots along several flows as shown in Figure~\ref{compare2}. In contrast, Figure~\ref{compare4} displays a higher speed bow at 80~km~s$^{-1}$ moving at an angle of $\theta = 45^\circ$ towards the line of sight. This generates PV diagrams with a small redshift at the leading edge and a large blueshift 
immediately behind, followed by an unshifted tail. Again, comparable features are observed, as shown in the right two panels, although more observed detail is needed before being convincing.

The two synthetic PV diagrams correspond to both a narrow slit and a wide slit relative to the bow size with the slit locations shown on the image. The velocity has been convolved with a Gaussian with a standard deviation of 8~km~s$^{-1}$. Interestingly,  with this resolution, there is little structural difference between the slits. .Even though the slit covers a wide lateral area that extends into the wings of the bow shock, and the bow has a high speed and is tilted out of the plane of the sky at an angle of 45$^\circ$, the PV diagram is hardly affected.


\subsection{Jet Simulations}
\label{simulations}

PV diagrams of Type~B display well-resolved components. It is clear from the above  that single bow shock models will be inadequate.  The question now is: can we explain the observed components
with models involving multiple shocks associated with both the jet and the ambient medium? To answer this, we have run a set of numerical simulations
with a new version of the adapted ZEUS code \citep{2010ApJS..187..119C}, which has already been extensively employed to model the evolution of protostellar outflows \citep{2005MNRAS.357..579S,2007MNRAS.378..691S}.

We confine the present study  to that of hydrodynamic and axisymmetric simulations, explored previously in two dimensions by \citet{2006MNRAS.371.1448M} and \citet{2008MNRAS.386.2091M}.
The physics includes the cooling and chemistry as described by \citet{2003MNRAS.339..133S}.
We set up a simulation regime of size 10$^{16}$\,cm~$\times$~3\.10$^{16}$\,cm covered by   1500~$\times$~500 zones. To ensure that the emission is recorded when rotating this solid cylinder,
the (optically thin) emission is projected on to a grid 2082 zones long (with the jet origin fixed at coordinates (0,0) in the figures displayed).

We impose outflow boundary conditions except where we  inject  a uniform collimated jet of radius 10$^{15}$\,cm from the inner boundary of the long axis. 
We take a uniform ambient medium which is fully molecular of hydrogen nucleon density 10$^4$~cm$^{-3}$ and temperature 10\,K.
For the molecular  jet, we take a fixed temperature of 100\,K. 

\begin{figure}
  \begin{center}
     \epsfysize=7.5cm        \epsfbox{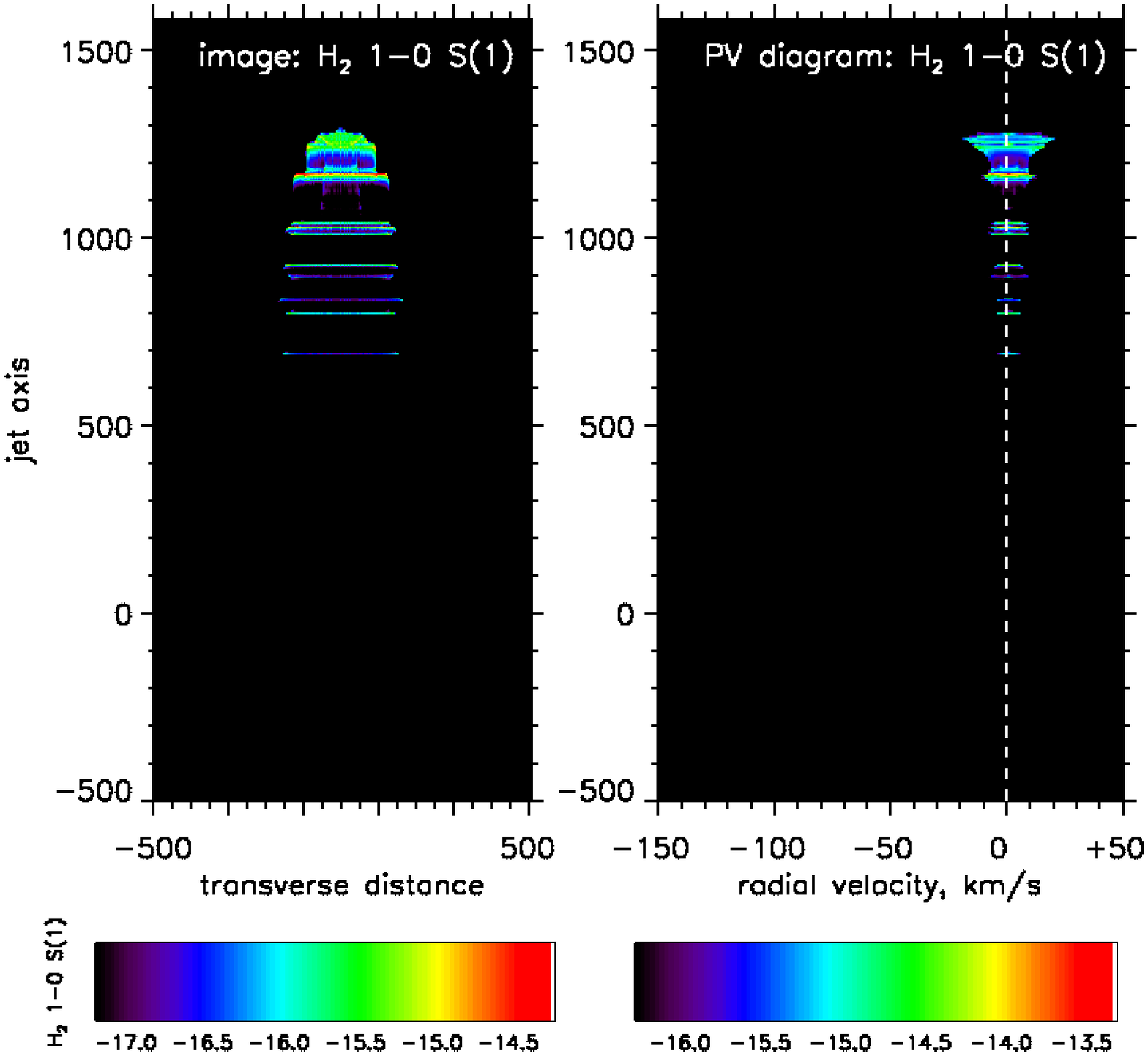}
      \epsfysize=7.5cm       \epsfbox{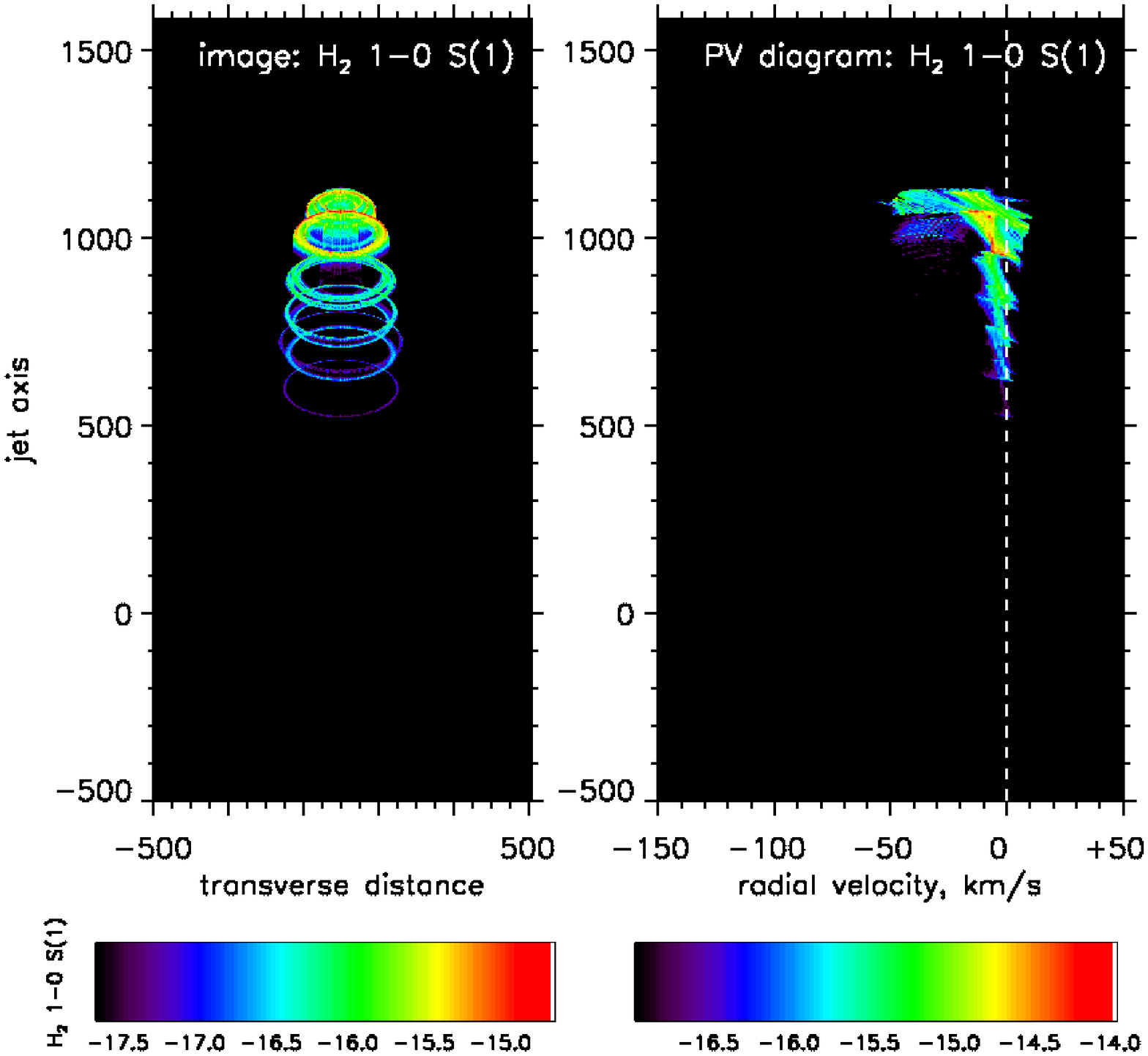}
     \epsfysize=7.5cm         \epsfbox{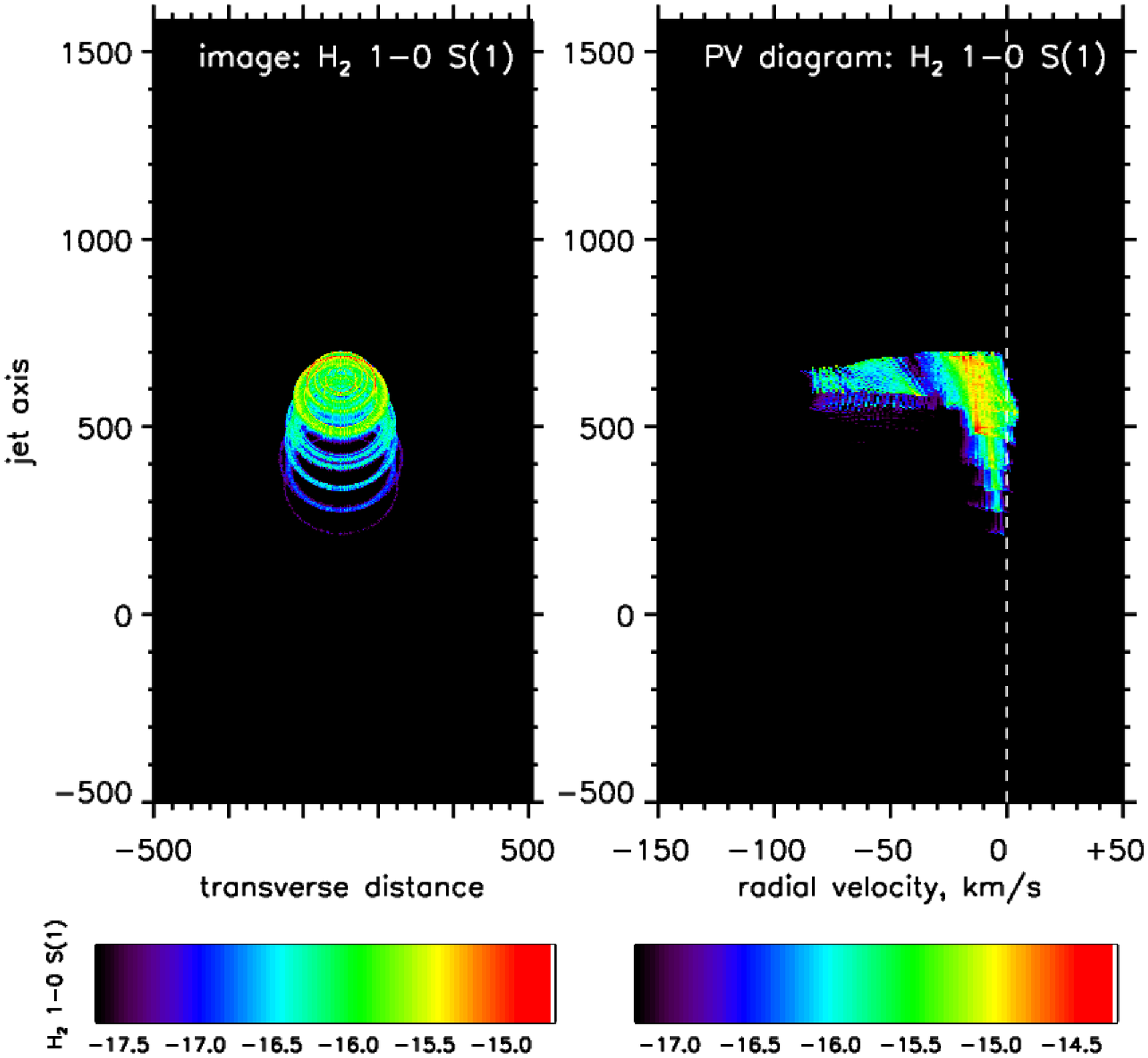}         
\caption{ Images (left) and PV diagrams (right) for the H$_2$ 2.12$\mu$m emission derived from a numerical simulations of a jet of 100~km~s$^{-1}$ directed at an angles of 90$^]circ$ (top),  60$^]circ$ (middle)  and 30$^]circ$ (bottom) to the line of sight. Flux units are in erg~cm$^{-2}$~s$^{-1}$. See text for details.}
\label{zeusequald}
  \end{center}
\end{figure}

We have run simulations with a range of jet  densities and velocities. The basic simulation is that of a 100~km~s$^{-1}$ jet in which the initial density is equal to that of the ambient medium.
The resulting images and PV diagrams are displayed in Fig.~\ref{zeusequald} for the three angles of 90$^]circ$ (top),  60$^]circ$ (middle)  and 30$^]circ$ (bottom) to the line of sight.  

It is first important to note that the bow shock which propagates through the ambient medium fragments into rings here in two dimensions. However, these rings are the quinine of  mini-bow shocks in three dimensions as found by \citet{1997A&A...318..595S}. Here,  we are limited to two dimensional simulations for this work since we are able to better resolve the interface between the forward and reverse shock sandwich. This does not greatly alter the flux distribution across the PV diagrams as can be seen by comparison to \citet{2004A&A...413..593R} and \citet{2007MNRAS.378..691S}. 

Figure~\ref{zeusequald} demonstrate that when moving well out of  the sky plane, the shocked jet and ambient bow do indeed produce two distinct structures on H$_2$ PV diagrams. 
The two components are of comparable flux. The peak of the high-speed component is just below the expected value corresponding to the radial component of the jet velocity while 
the low-speed component corresponds to the  radial component of the post-shock speed.  Notably, the low-speed component is spatially extended while the high-speed component 
is compact and positioned just behind the bow's leading edge. Note also that the low-speed component is blueshifted and moderately wide when well out of the sky plane, consistent with 
the observed distribution. However, the trailing wings  of the bow shock generate a trail of emission at radial speeds close to the local rest frame speed.

\begin{figure}
  \begin{center}
     \epsfysize=7.5cm        \epsfbox{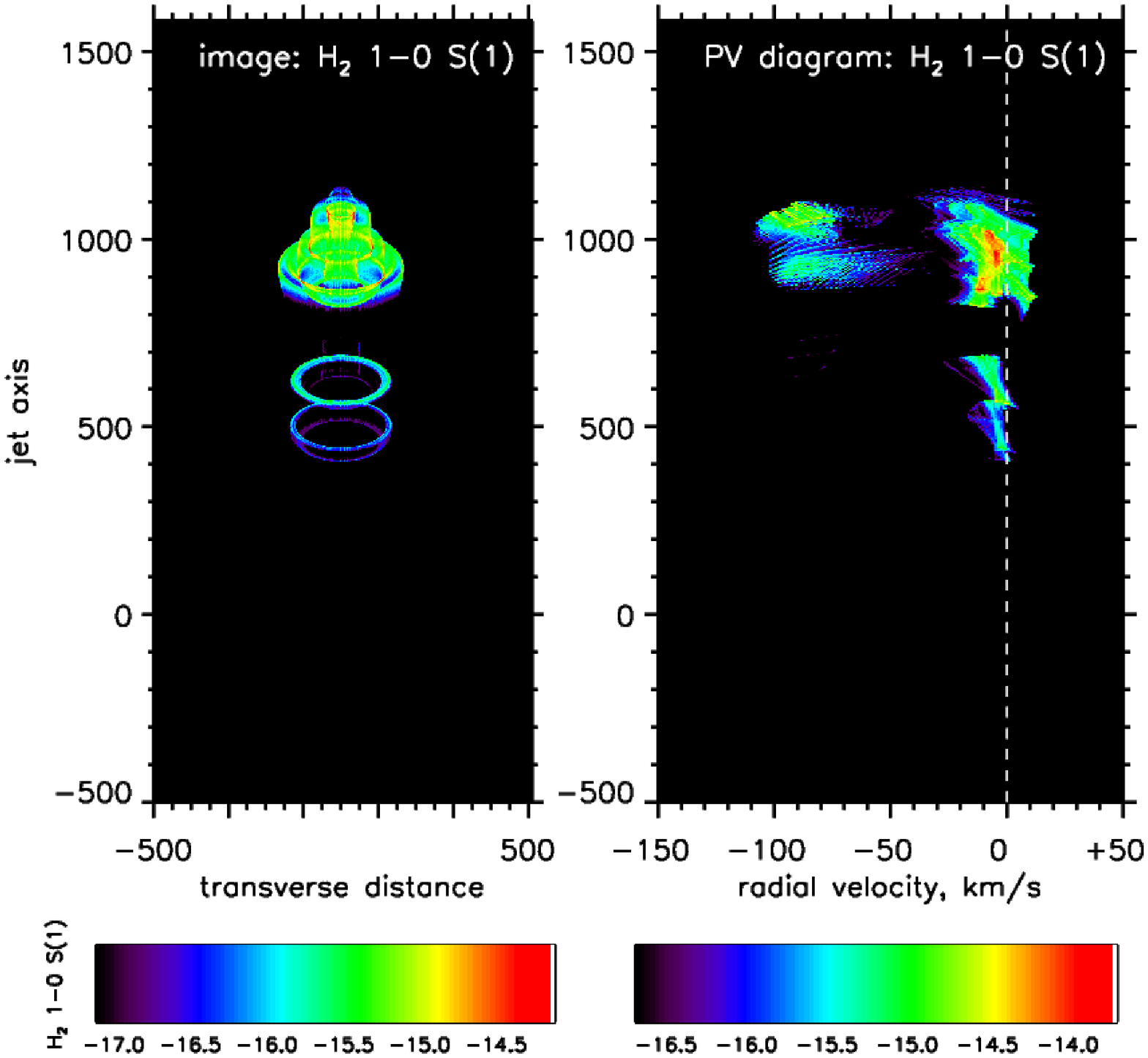}
      \epsfysize=7.5cm       \epsfbox{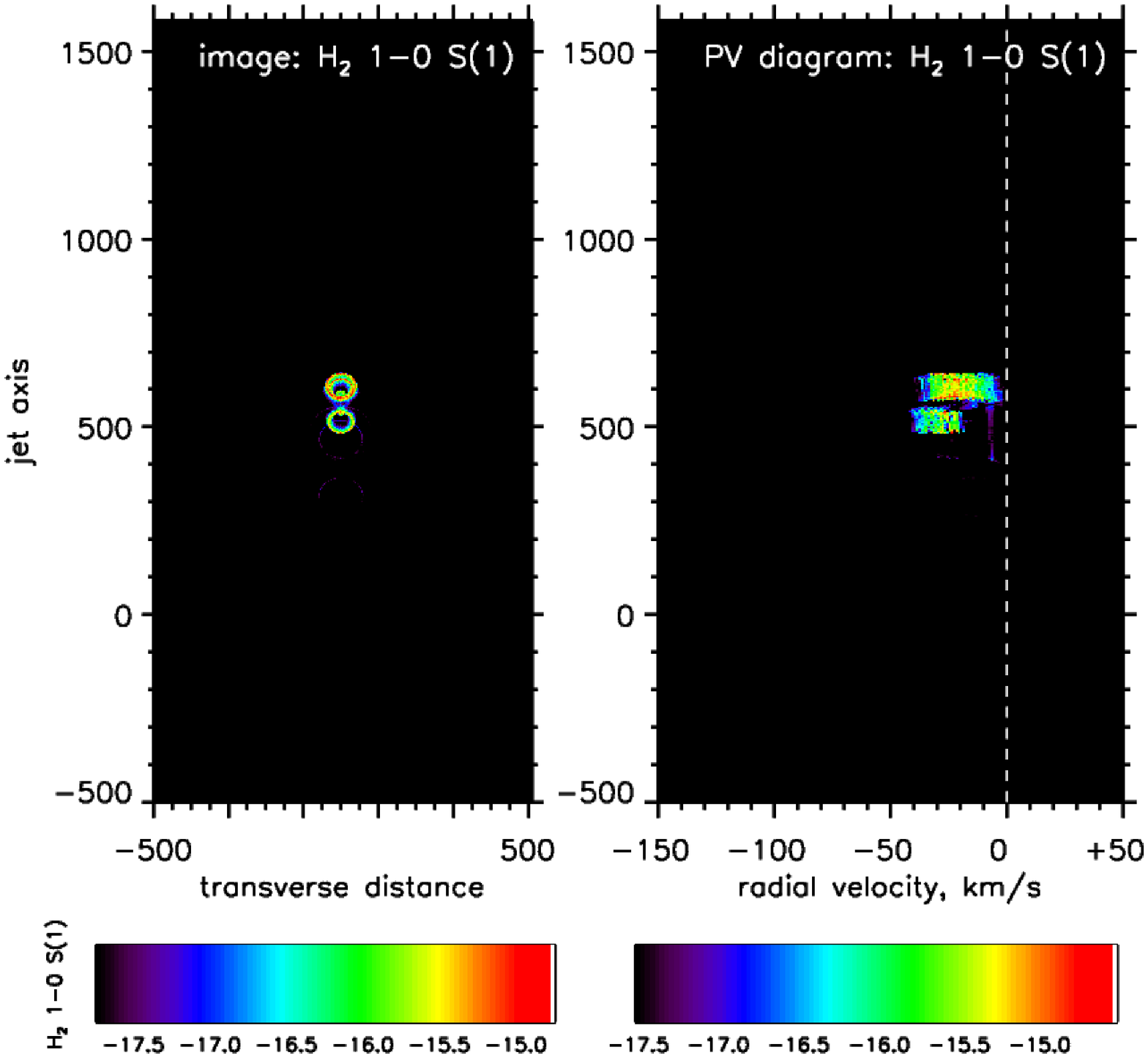}       
\caption{ Images (left) and PV diagrams (right) for the H$_2$ 2.12$\mu$m emission derived from a numerical simulations of a jet of 200~km~s$^{-1}$ directed at an angles of  30$^]circ$
(top),  and 50~km~s$^{-1}$ directed at an angles of  60$^]circ$  (bottom) to the line of sight. The snapshots are taken at simulation times of 80 years (top) and 280 years (bottom). See text for details}
\label{zeusjetspeed}
  \end{center}
\end{figure}

To determine if these encouraging results are sensitive to the chosen jet speed, we present a high speed and low speed jet in Figure~\ref{zeusjetspeed}. The viewing angle has been altered to determine which angles can correspond to the radial velocities detected in the presented observations. The high-speed jet now
generates well separated radial velocity components at higher angles to the line of sight. Furthermore, the low and high speed components are now much more distinct: the shock deceleration and molecule excitation occur in two very different reference frames. The high speed structure can also be bisected spatially, a structure observed on some PV diagrams. This is a result of a particular phenomenon yielded in previous simulations: the jet is not brought to a halt with one terminating shock but can undergo an oblique shock in a recollimation region well before the final shock. 
Also interesting is the low-speed jet which yields a jet shock and ambient shock separated in space: the emission from the bow shock occurs from the apex of the slow bow, being pushed ahead by the jet piston. With such an analysis, these characteristics are potentially able to act as diagnostics for outflows.

We conclude that (1) PV diagrams displaying two distinct velocity components are consistent with molecular jets impacting a molecular ambient medium, (2) trails of emission at low speeds can be generated by slits covering unresolved bows and (3) more complex patterns can be expected from global outflow simulations.

\section{Conclusions}
\label{conclusions}

We have investigated protostellar outflows through near-infrared echelle spectroscopy within the DR21/W75N star forming field. We have,  analysed the emission resulting from vibrationally-excited molecular hydrogen
by studying position-velocity diagrams. We find that these data can be characterised by a few distinct properties which suggest a classification into  four types. These four types correspond to: (A) single components with a low radial velocity, (B) two spatially coincident components separated in velocity-space by several tens of kilometres per second, (C)  multiple components which are spatially aligned but with a range of velocities and (D) low speed
trails of emission, narrow in velocity width. 

Many components are
consistent with individual bow shocks although the limited spatial resolution at the distance of DR21 does not permit a detailed interpretation. Their radial velocities are consistent
with the bow shocks being driven into a stationary ambient medium. The high number with a peak blueshift in the range 0-20~km~s$^{-1}$ have been shown to be consistent with expectations
from a steady bow shock model and numerical simulations. The relatively few low-velocity redshifted components suggest that extinction is significant. 

Theosition-velocity diagrams involving spatially resolved structures all at low radial velocities are identified. We suggest that these are produced by gas-dynamic interaction of jets 
with the molecular ambient medium. 

The overall radial velocity distribution shows a clear excess of high speed blueshifted components. Such PV diagrams were also found previously in the OMC survey of \citet{2000AJ....120.1974Y}. We demonstrate here that these can be generated by twin shocks, one reverse shock associated with the deceleration of the jet and one forward shock corresponding to the bow shock. Simulations show that the bow tends to fragment into many mini-bow shocks which can generate emission at low speeds back along the flow as they develop and then slow down. 

The observations of the high-speed components are roughly consistent with jets of speed 100-140 ~km~s$^{-1}$ moving at 
30- 45$^\circ$ to the line of sight. Higher jet speeds would produce some objects with higher radial speeds, while lower jet speeds would generate more intermediate radial velocity components than found.
We thus interpret this as evidence for propagation of molecular jets within conical-shaped cavities; when observed outside this cone dust extinction will obscure the outflow from view. In this manner we account for the surfeit of high-speed blueshifted flows. 

Dense molecular jets are indeed associated with Class~0 protostars while more evolved sources may drive preferentially atomic jets \citep[e.g.][]{2014A&A...563A..28D}.
A conical opening is also suspected, with a dense molecular core still to accrete and disperse \citep{2013ApJ...771..128H} although 
 such cores are likely to possess very complex dynamical structures \citep{2010ApJ...712.1010T}. The origin of the molecules in the jet, however, remains open to question. Possibilities include pre-existence and survival within a disc wind on scales within 0.1\,AU, formation at high density on dust within 10\,AU, 
entrainment from a core through fluid dynamic instabilities followed by gradual acceleration, or acceleration by a series of non-dissociative shock waves. 

The high speed components can be identified directly here, and are probably related to  oblique Mach stems or transverse Mach discs within a complex impact region. Without velocity data,  
a few candidates for Mach discs have also been  identified  in H$_2$, based on imaging and excitation
\citep{2002ApJ...576..313K,2003MNRAS.338...57K,2004AJ....128.2917H}. Clearly, more high-resolution velocity studies will elucidate the jet-ambient interaction physics.


\section{Acknowledgements}
\label{acks}

At the time of the observations the United Kingdom Infrared Telescope was operated by the Joint Astronomy Centre  on behalf of the Science and Technology Facilities Council of the U.K. Some of the data reported here were obtained as part of the UKIRT Service Programme.

\bibliography{jetbib}

\section*{Appendix A}
\label{appa}

\begin{figure*}
  \begin{center}
  \epsfxsize=7.4cm       \epsfbox{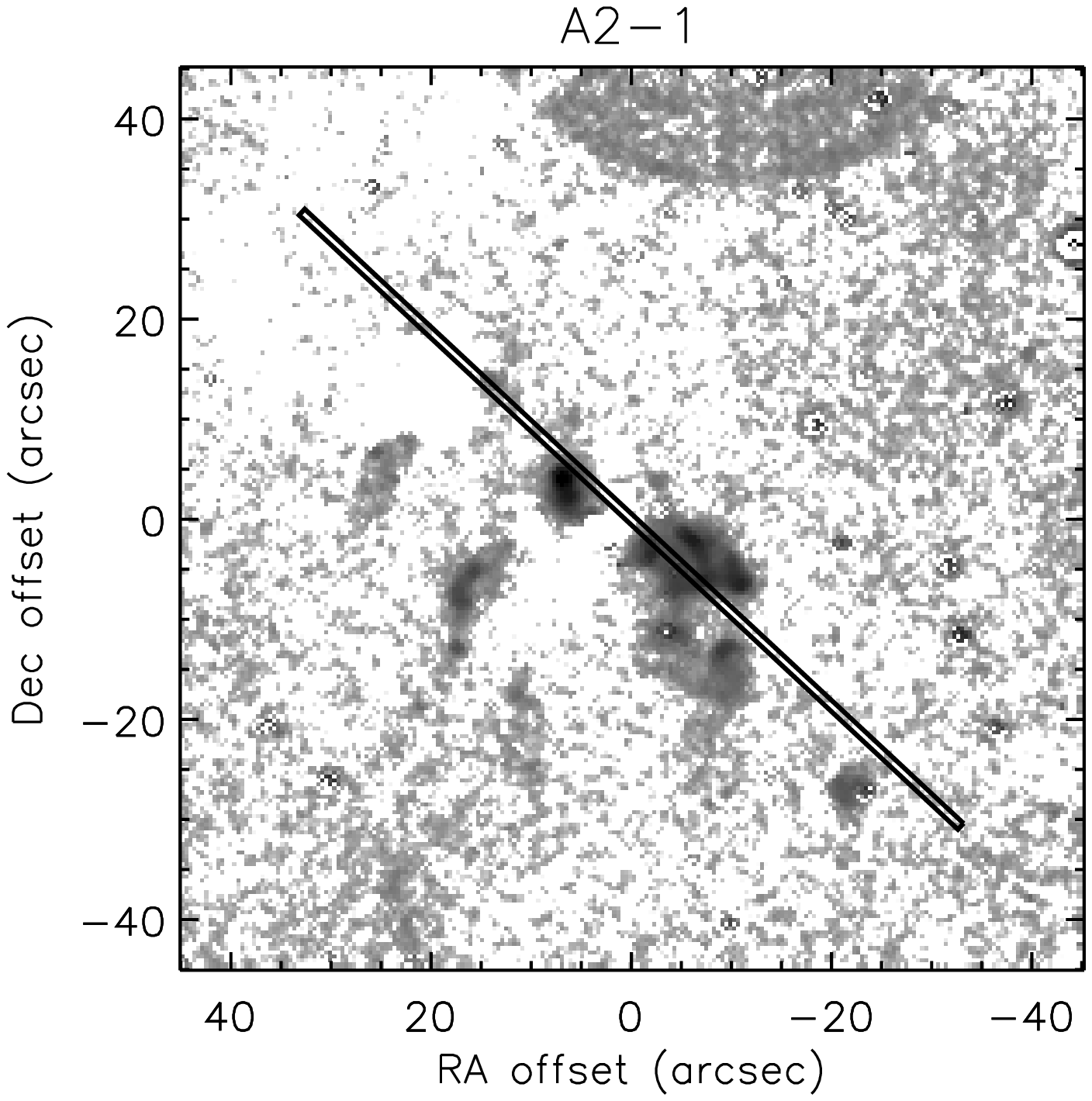}
  \epsfxsize=7.4cm       \epsfbox{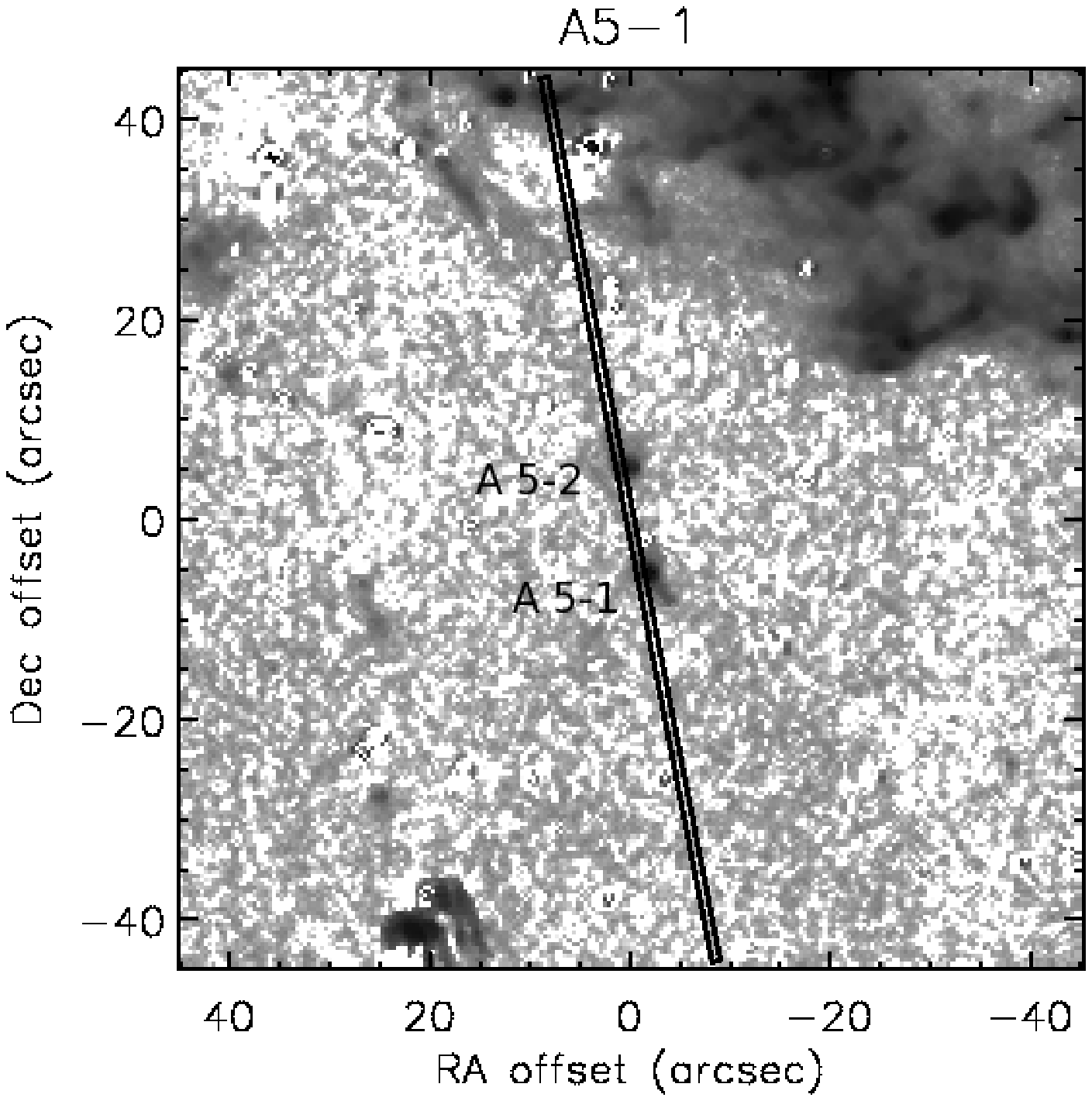}
  \epsfxsize=7.4cm       \epsfbox{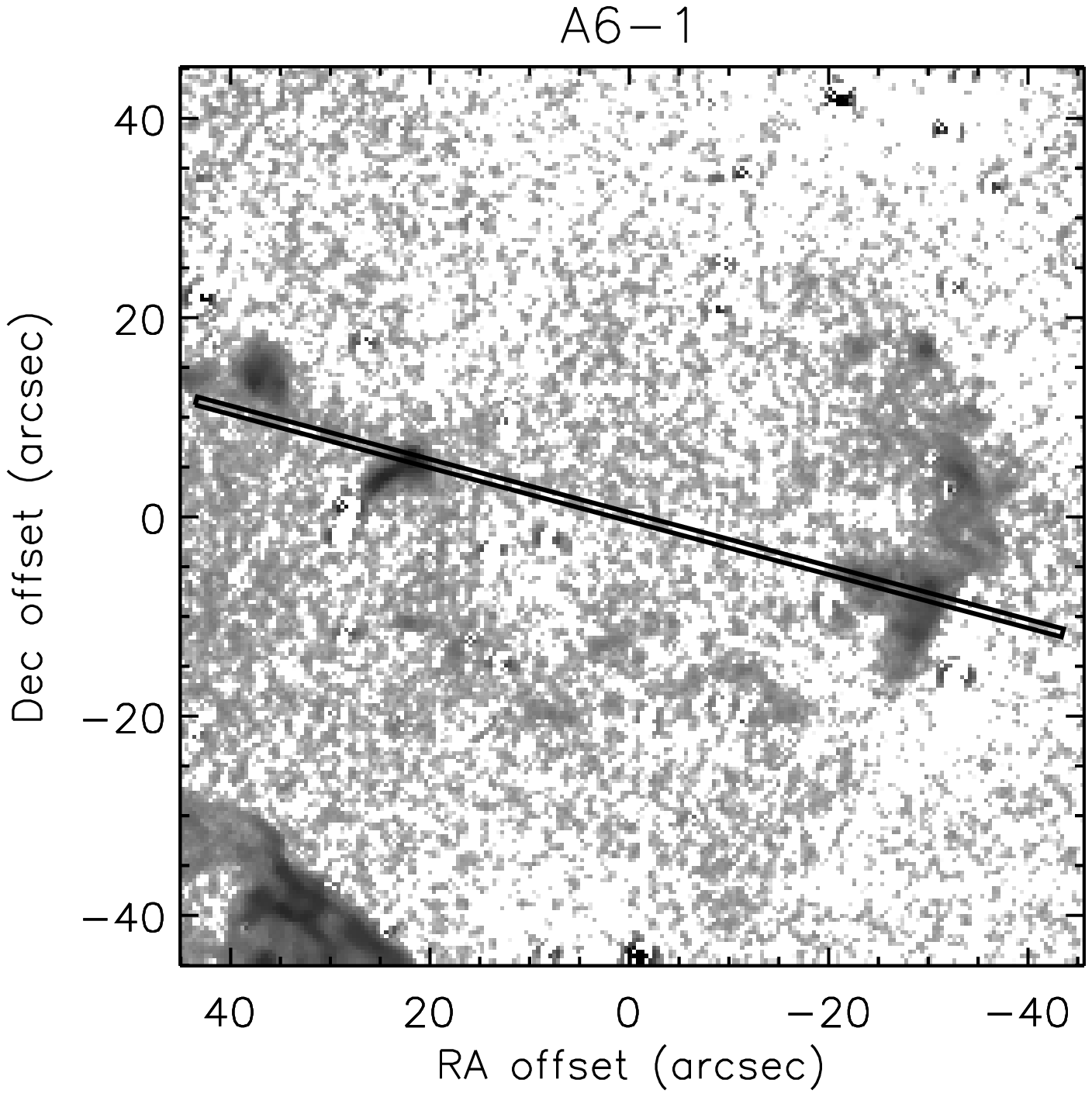}
  \epsfxsize=7.4cm       \epsfbox{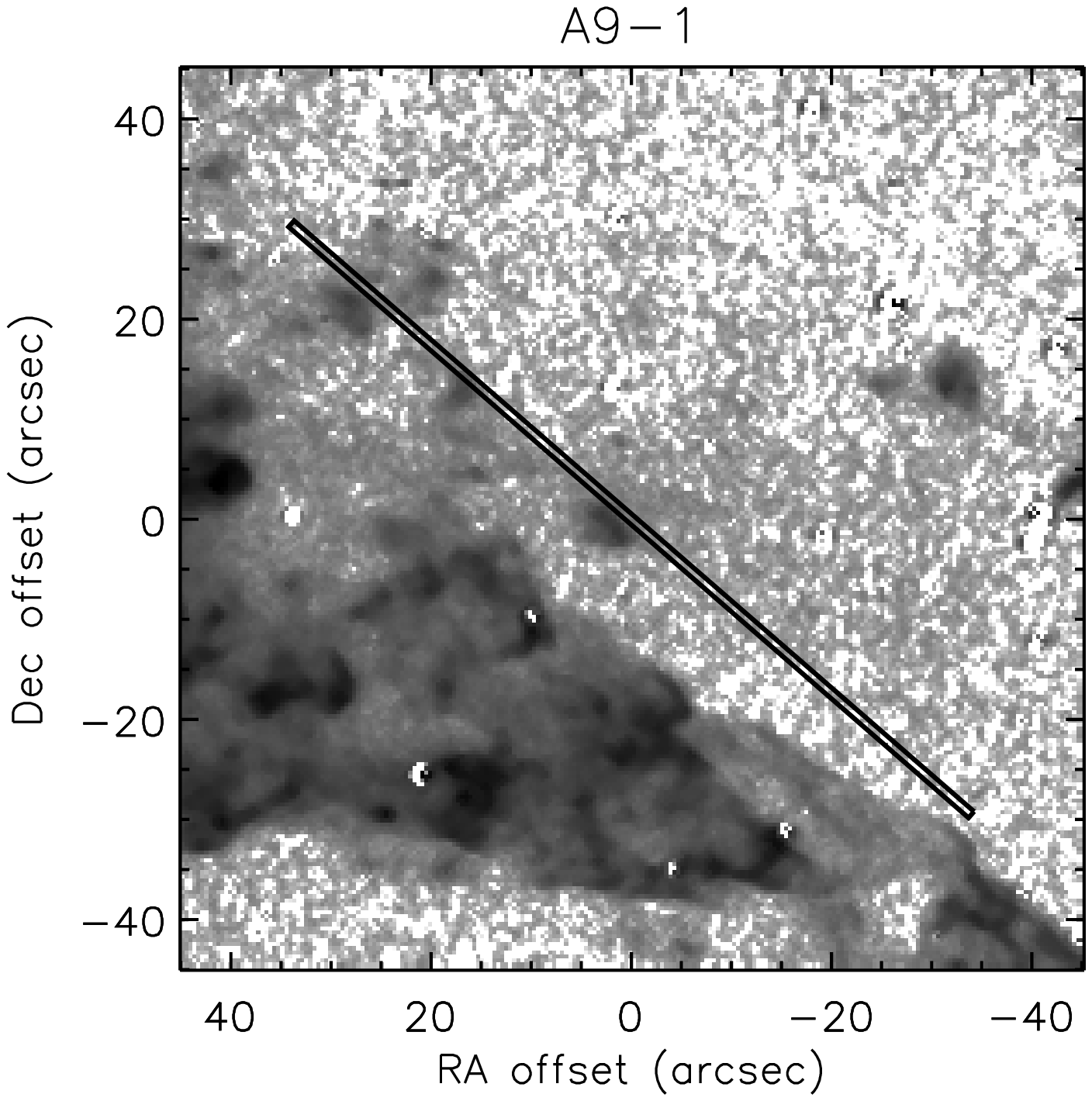}
  \epsfxsize=7.4cm       \epsfbox{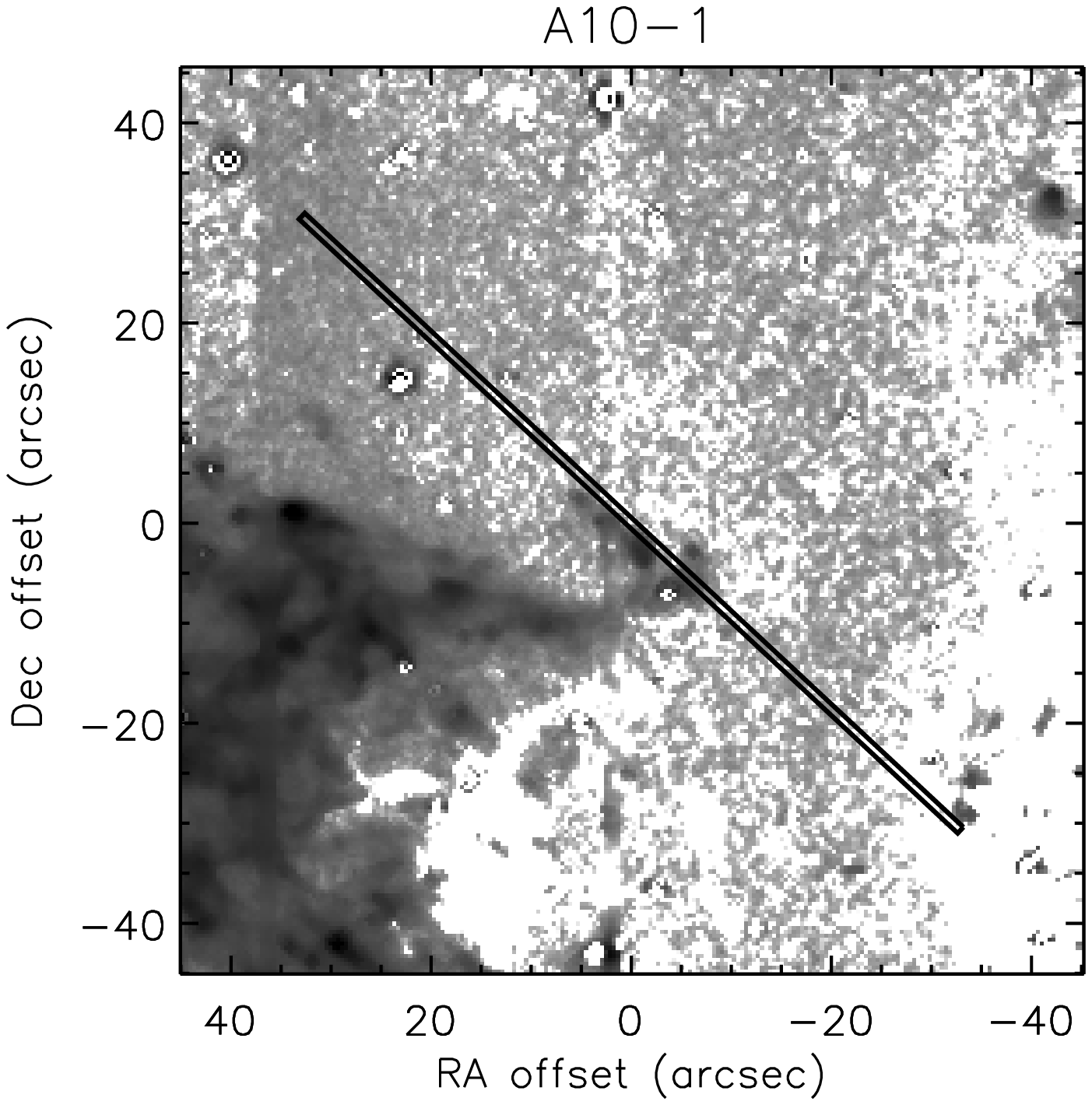}
  \epsfxsize=7.4cm       \epsfbox{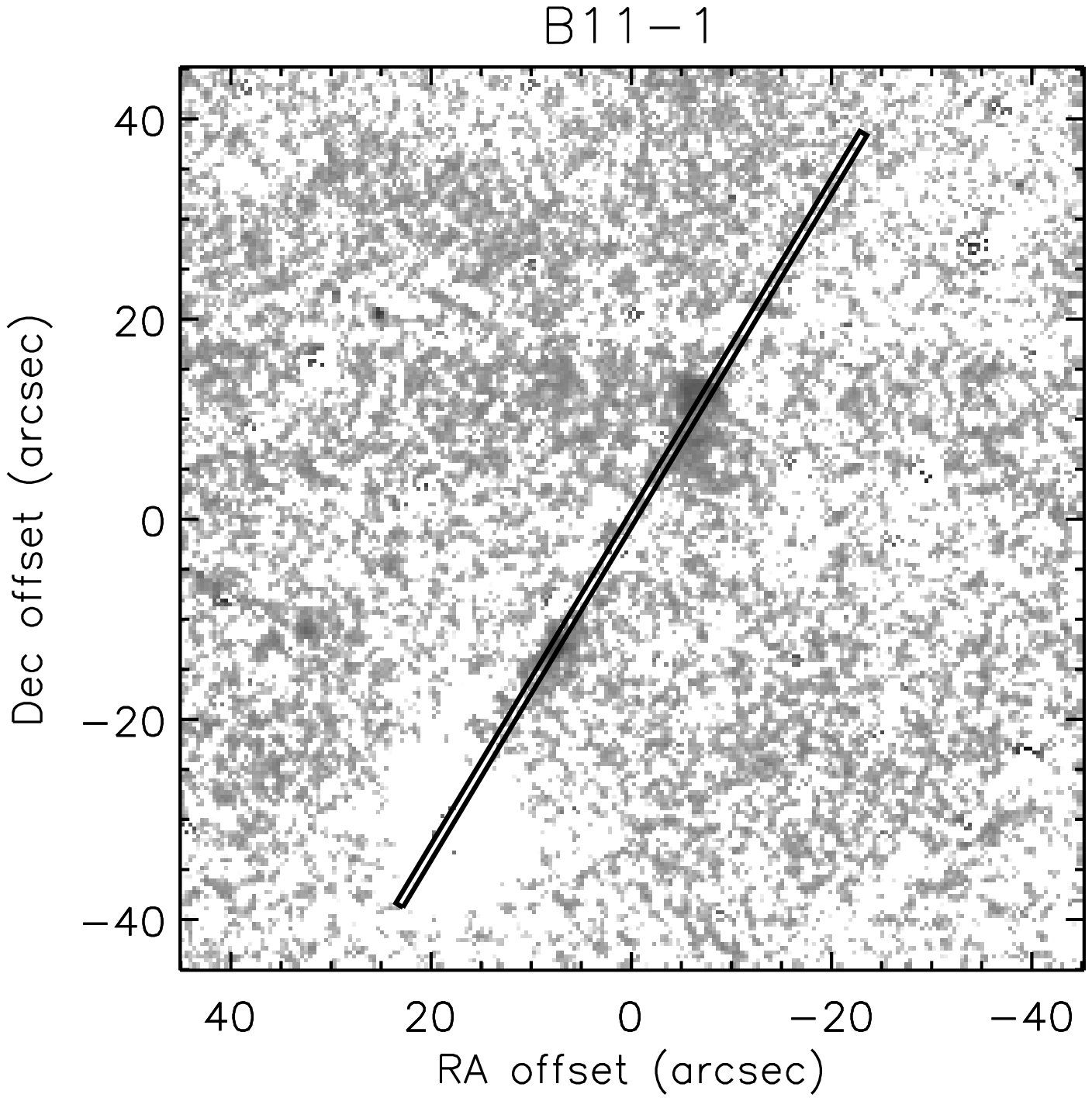}
\caption{\label{zoomplots1} Close-up images displaying the actual areas covered by the slits for objects A\,2-1 to B\, 11-1.}
  \end{center}
\end{figure*}

\begin{figure*}
  \begin{center}
  \epsfxsize=7.4cm       \epsfbox{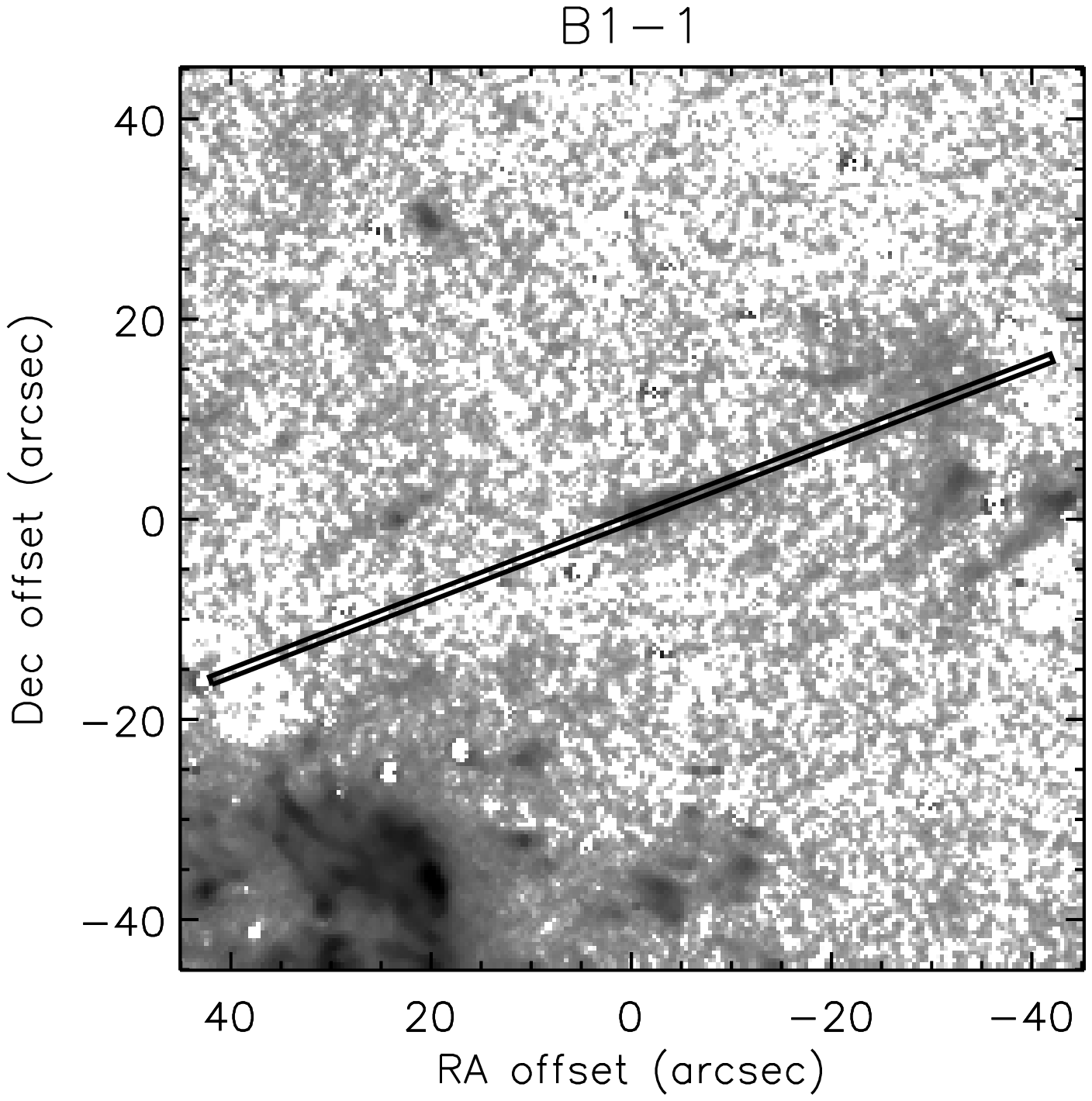}
  \epsfxsize=7.4cm       \epsfbox{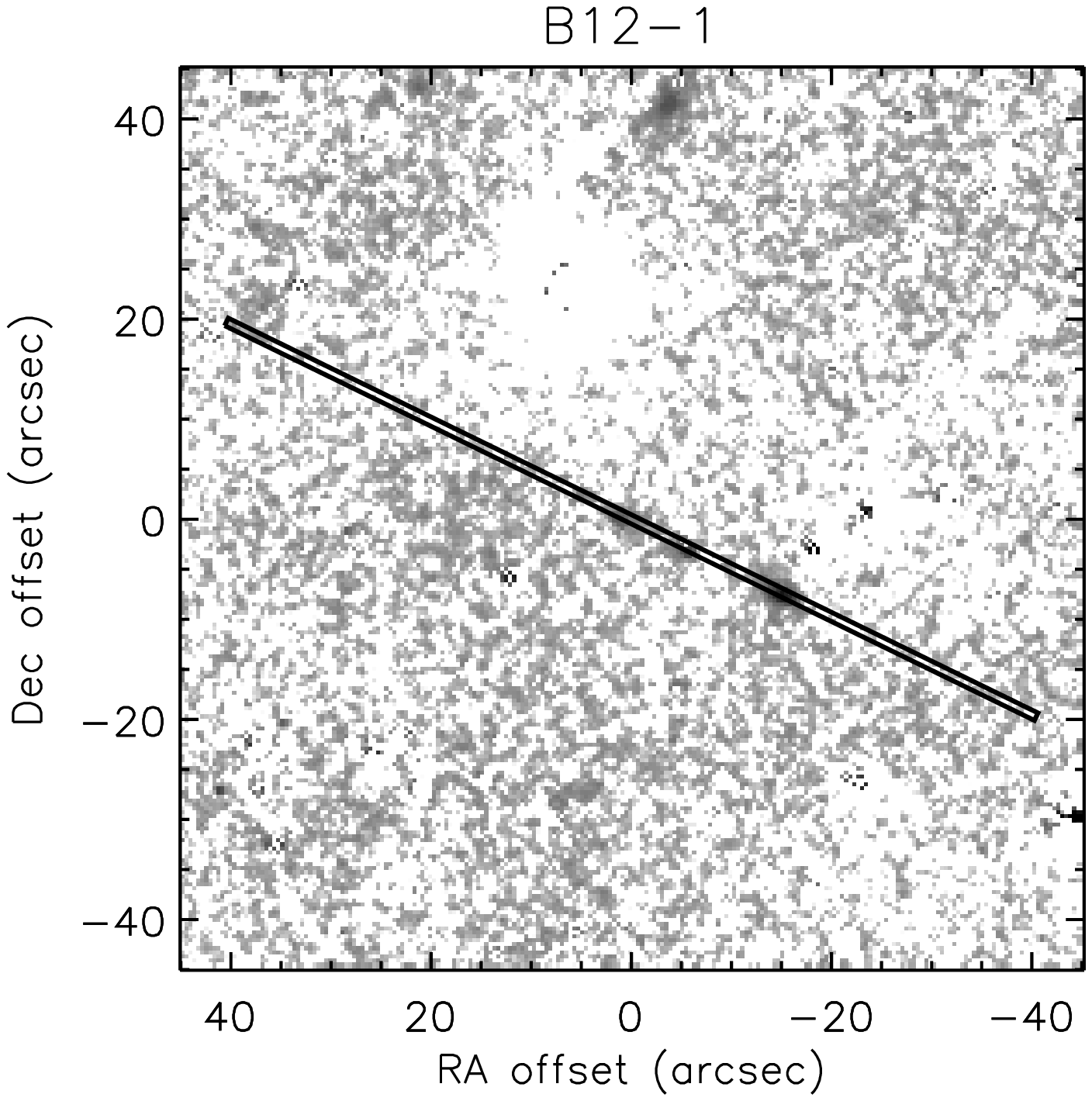}
  \epsfxsize=7.4cm       \epsfbox{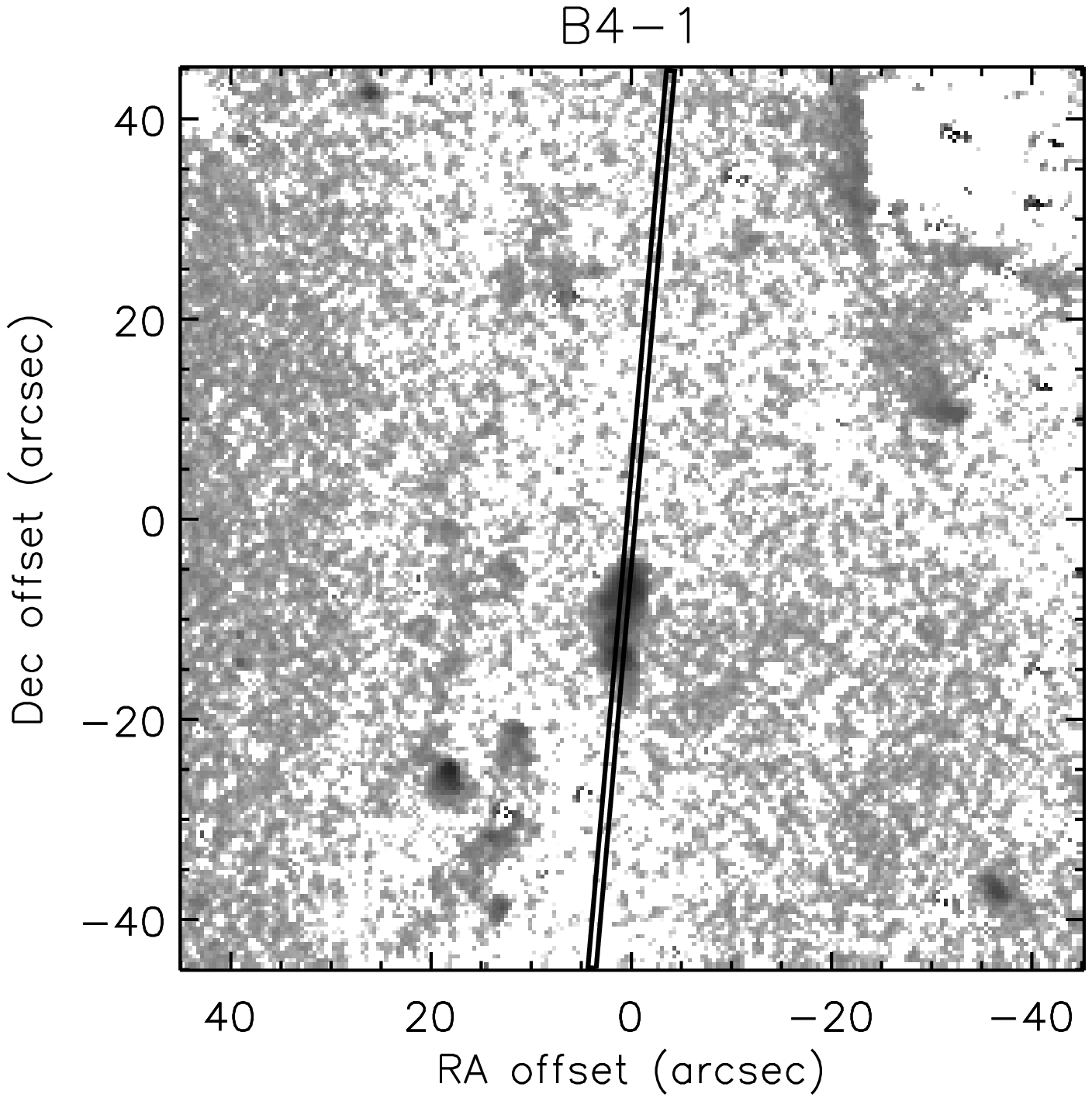}
  \epsfxsize=7.4cm       \epsfbox{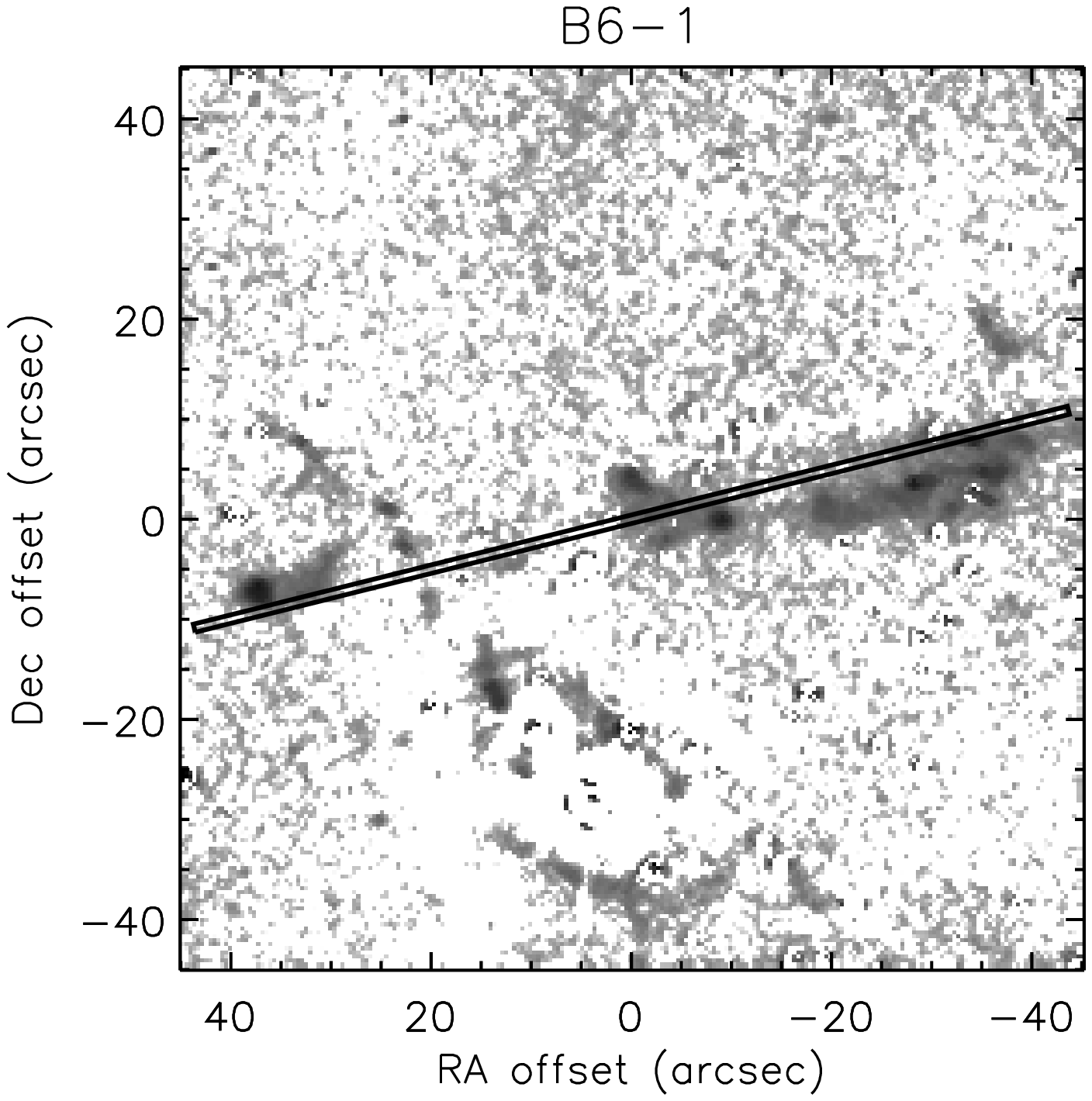}
  \epsfxsize=7.4cm       \epsfbox{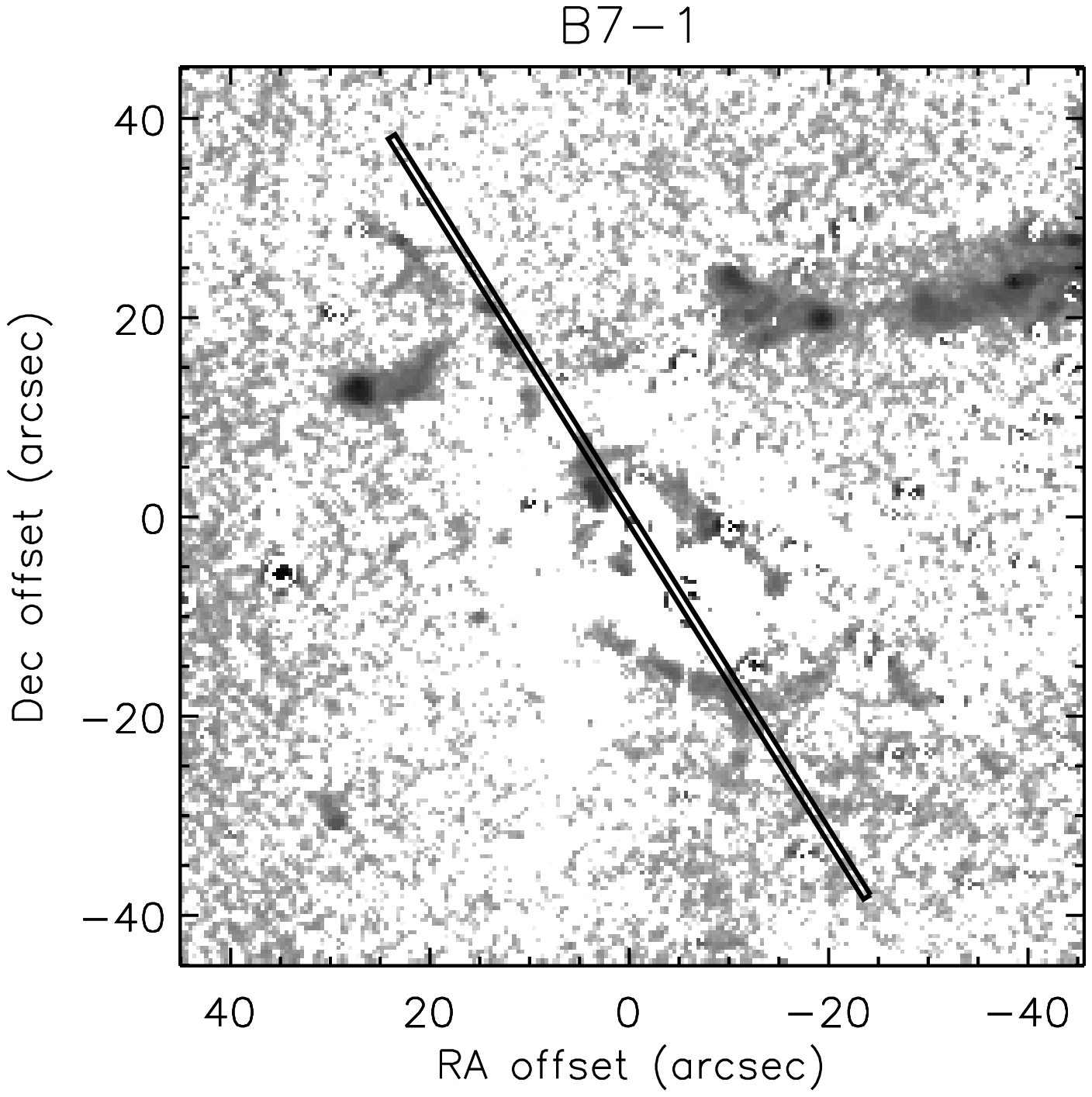}
  \epsfxsize=7.4cm       \epsfbox{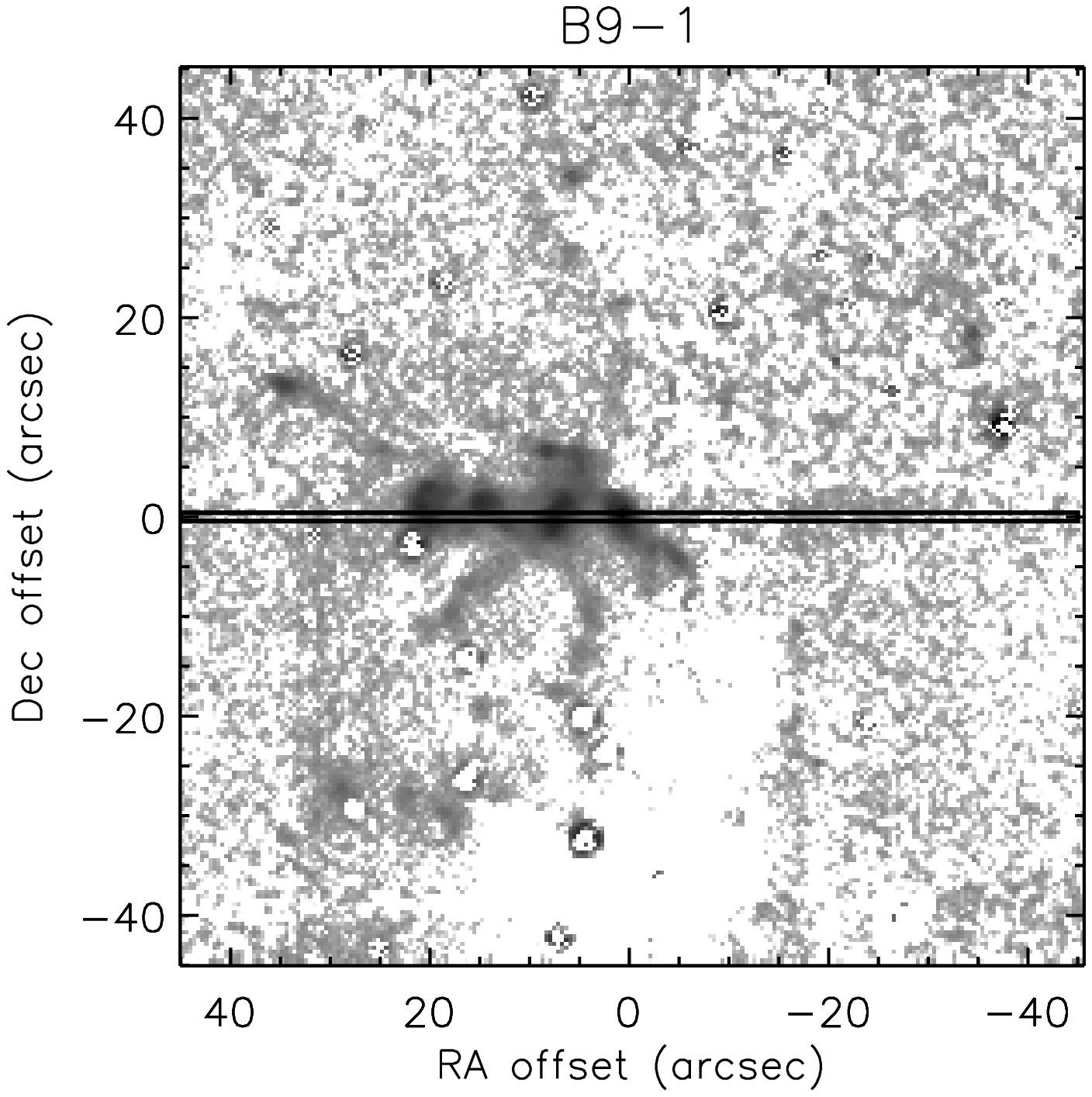}
\caption{\label{zoomplots2} Close-up images displaying the actual areas covered by the slits for objects B\,1-1 to B\,9-1 .}
  \end{center}
\end{figure*}

\begin{figure*}
  \begin{center}
  \epsfxsize=7.4cm       \epsfbox{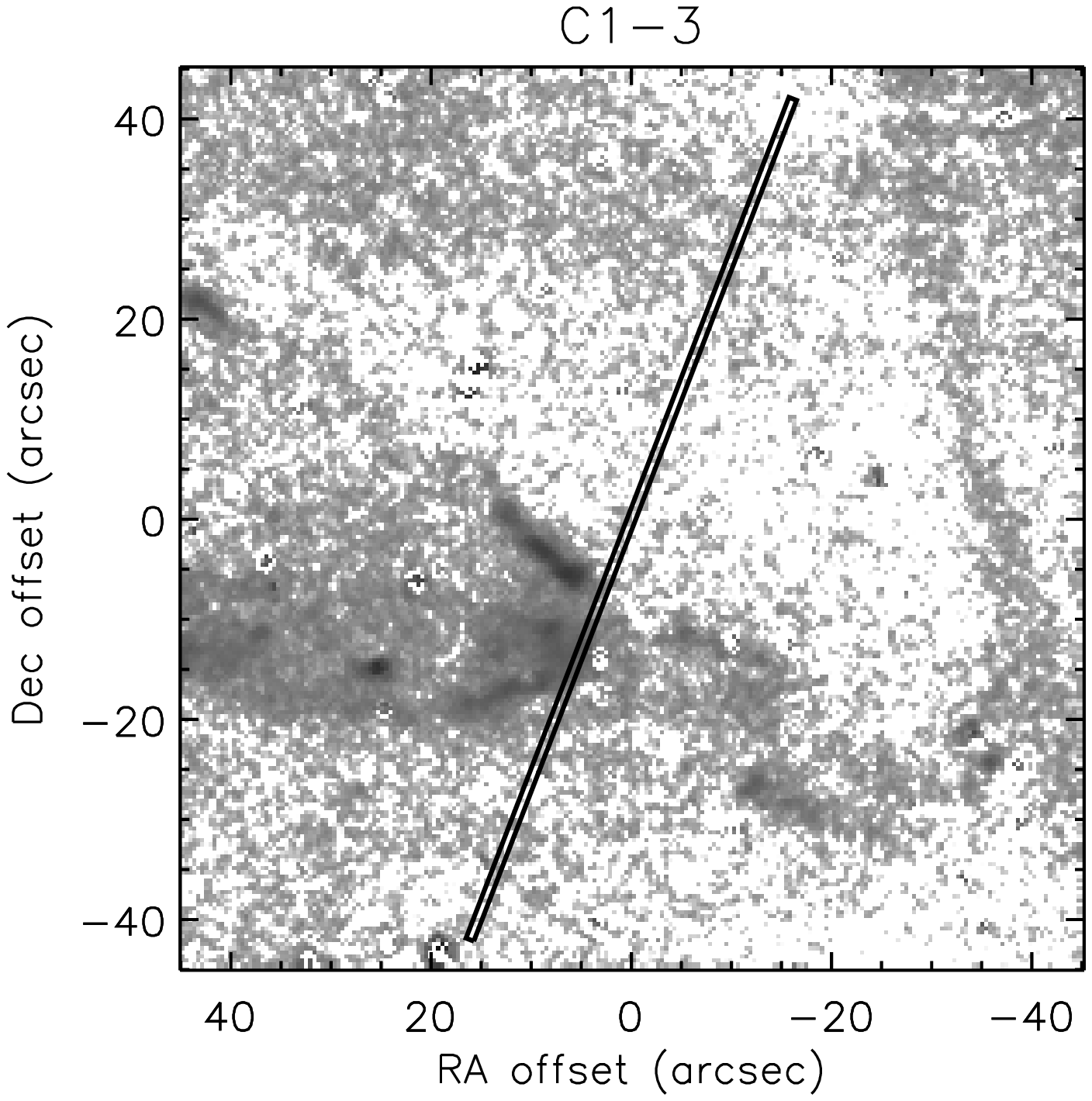}
  \epsfxsize=7.4cm       \epsfbox{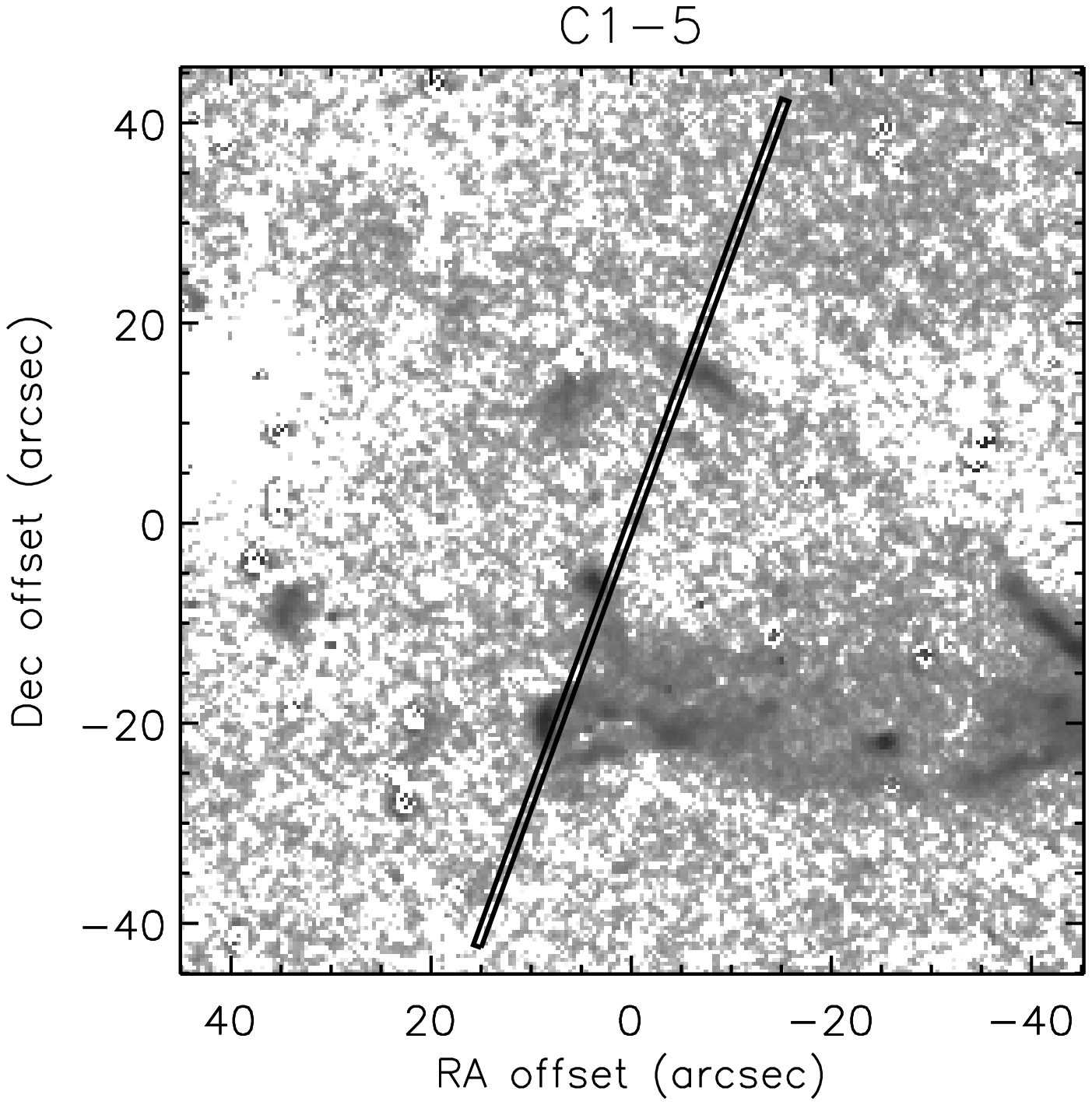}
  \epsfxsize=7.4cm       \epsfbox{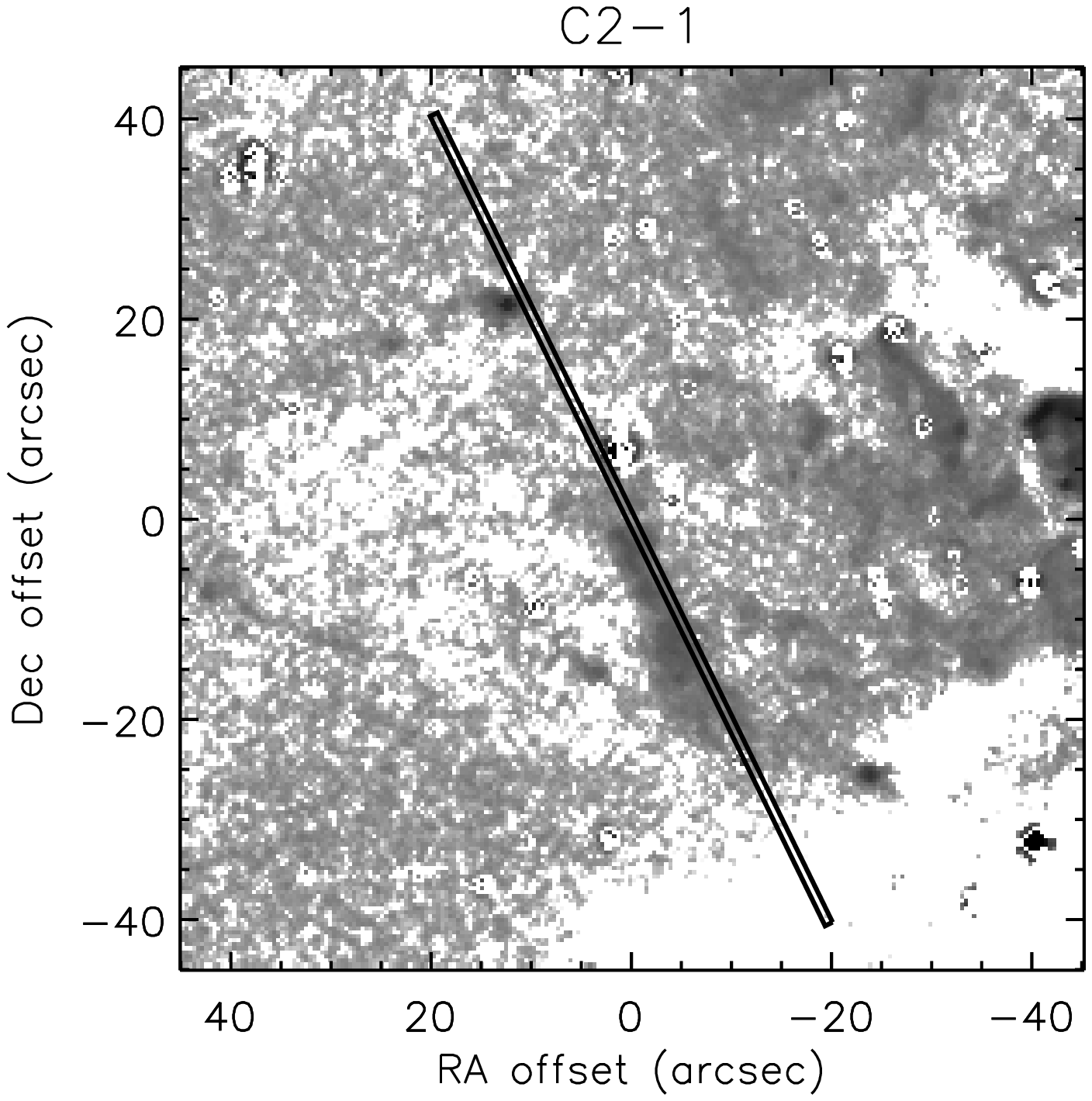}
  \epsfxsize=7.4cm       \epsfbox{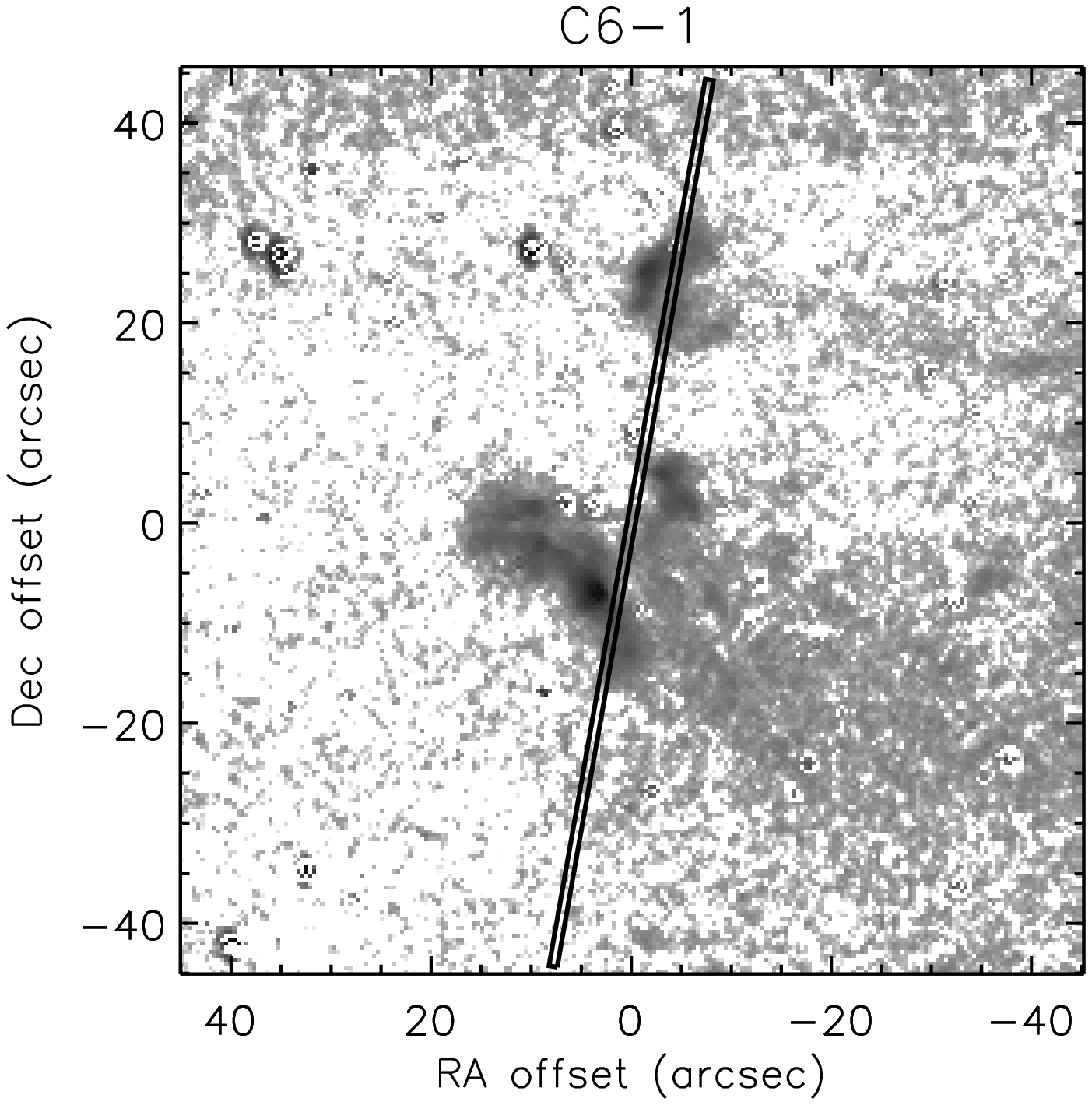}
  \epsfxsize=7.4cm       \epsfbox{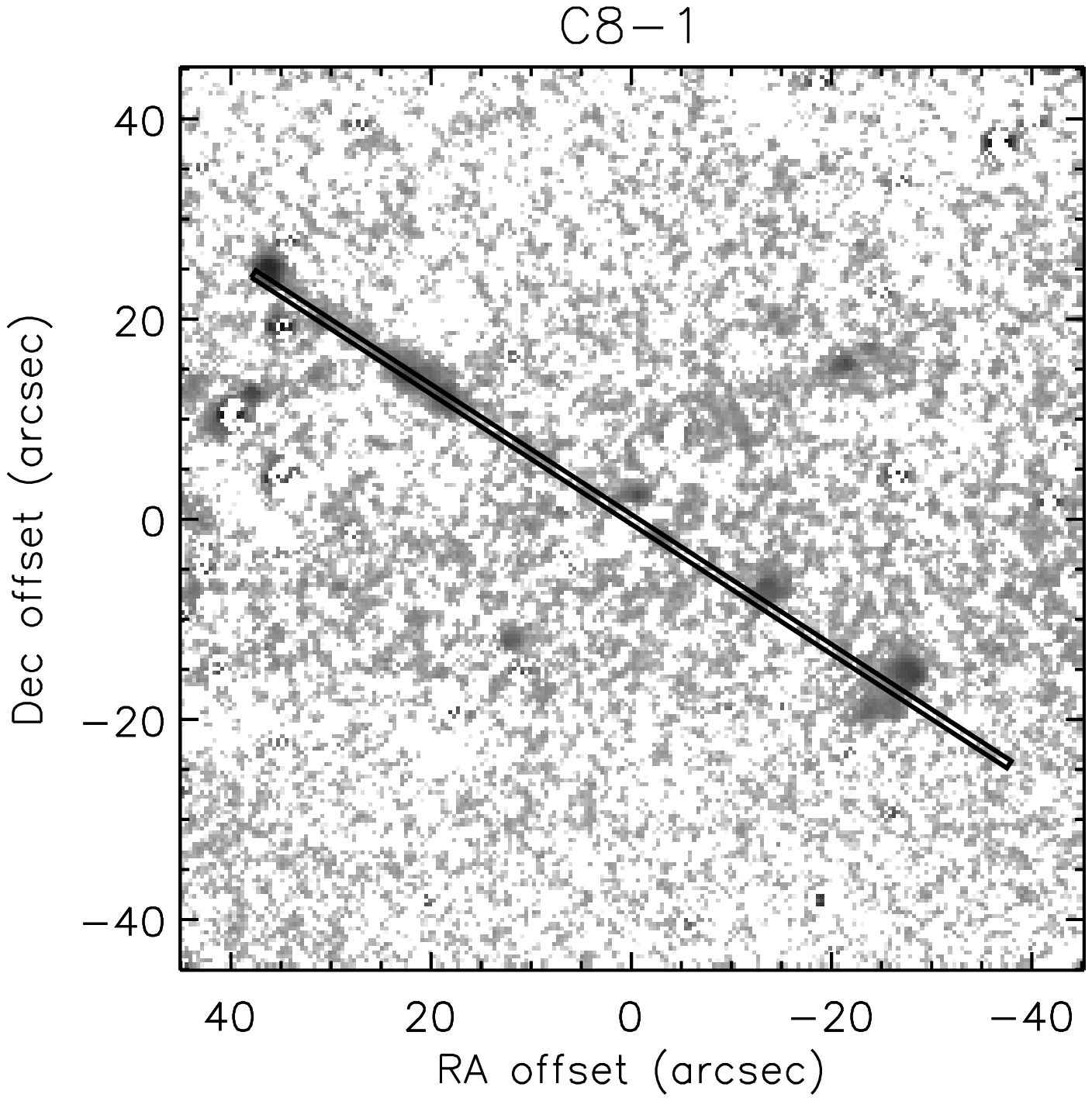}
  \epsfxsize=7.4cm       \epsfbox{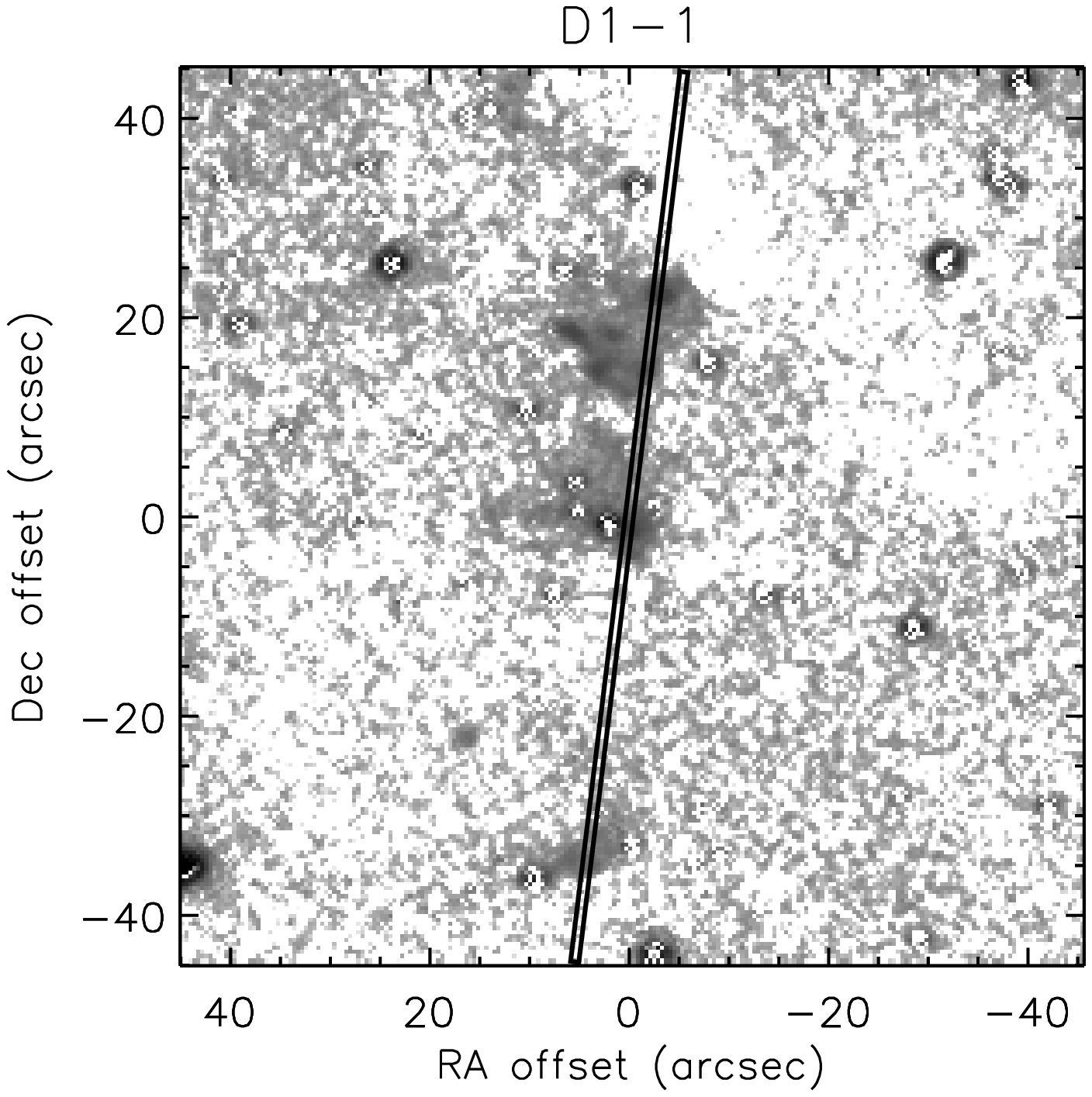}
\caption{\label{zoomplots3} Close-up images displaying  the slits for objects C\,1-3 to D\,1-1 (taken from a larger J-K-H$_2$ composite).}
  \end{center}
\end{figure*}

\begin{figure*}
  \begin{center}
  \epsfxsize=7.4cm       \epsfbox{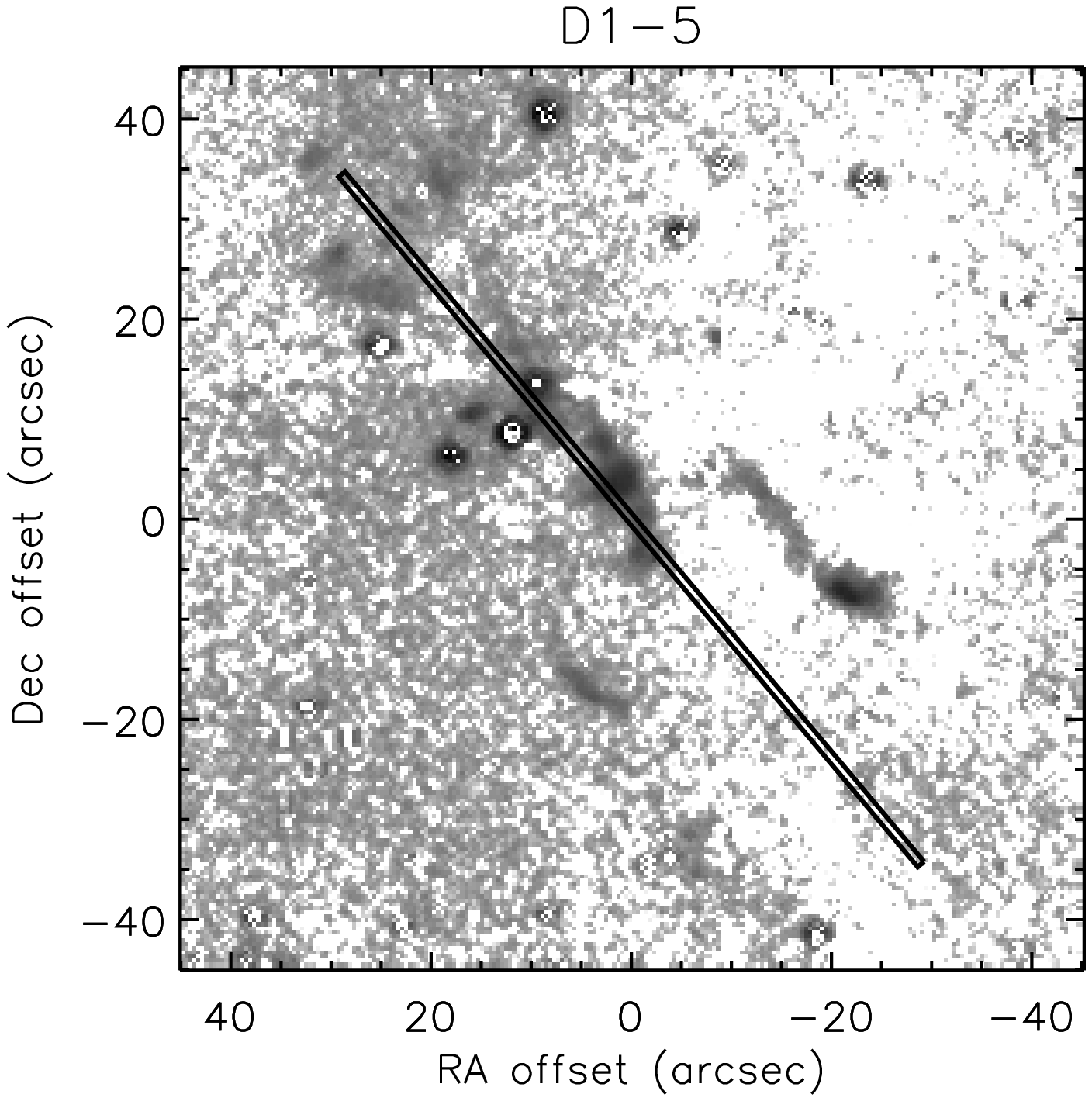}
   \epsfxsize=7.4cm      \epsfbox{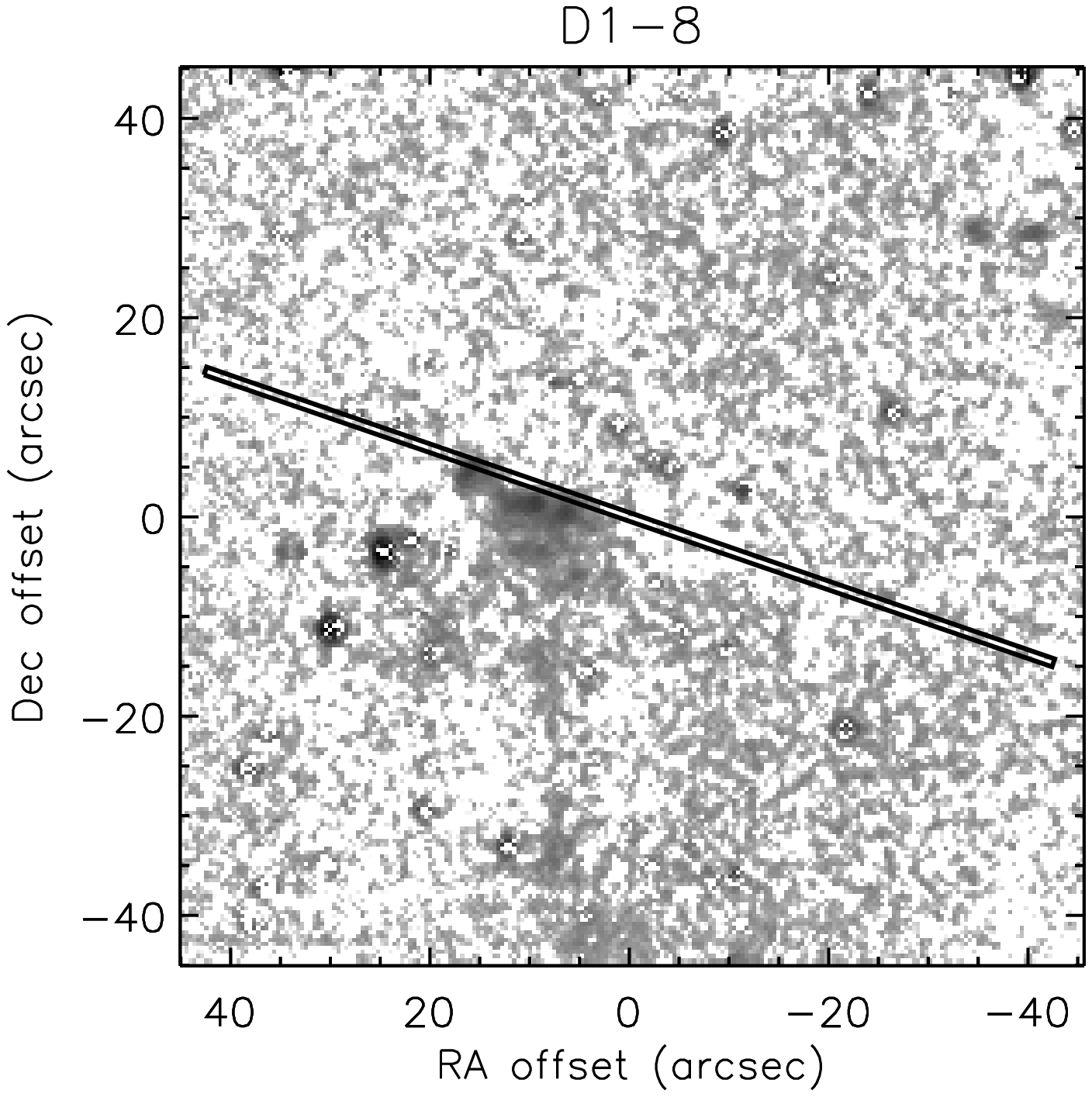}
  \epsfxsize=7.4cm       \epsfbox{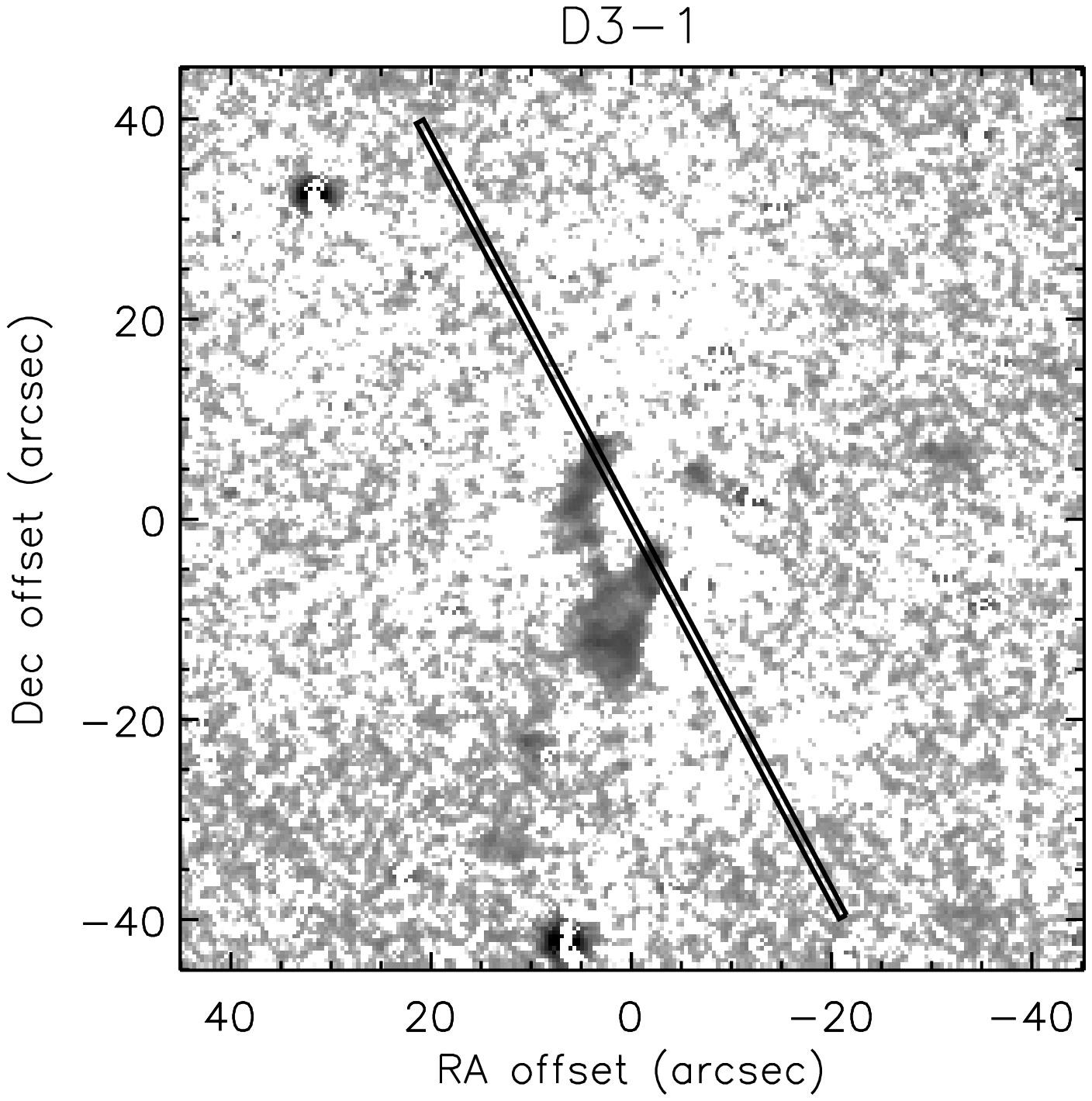}
   \epsfxsize=7.4cm      \epsfbox{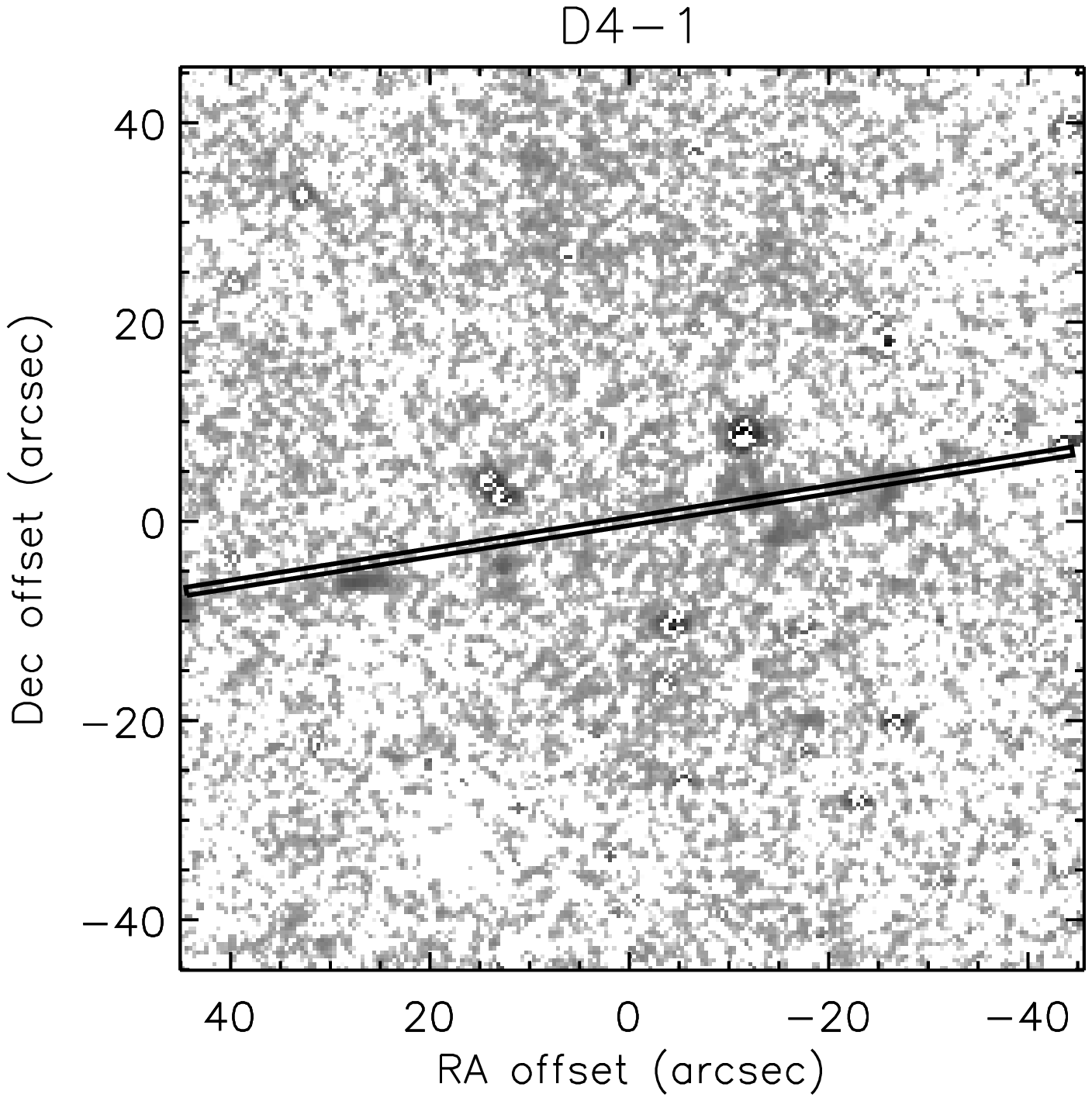}
   \epsfxsize=7.4cm      \epsfbox{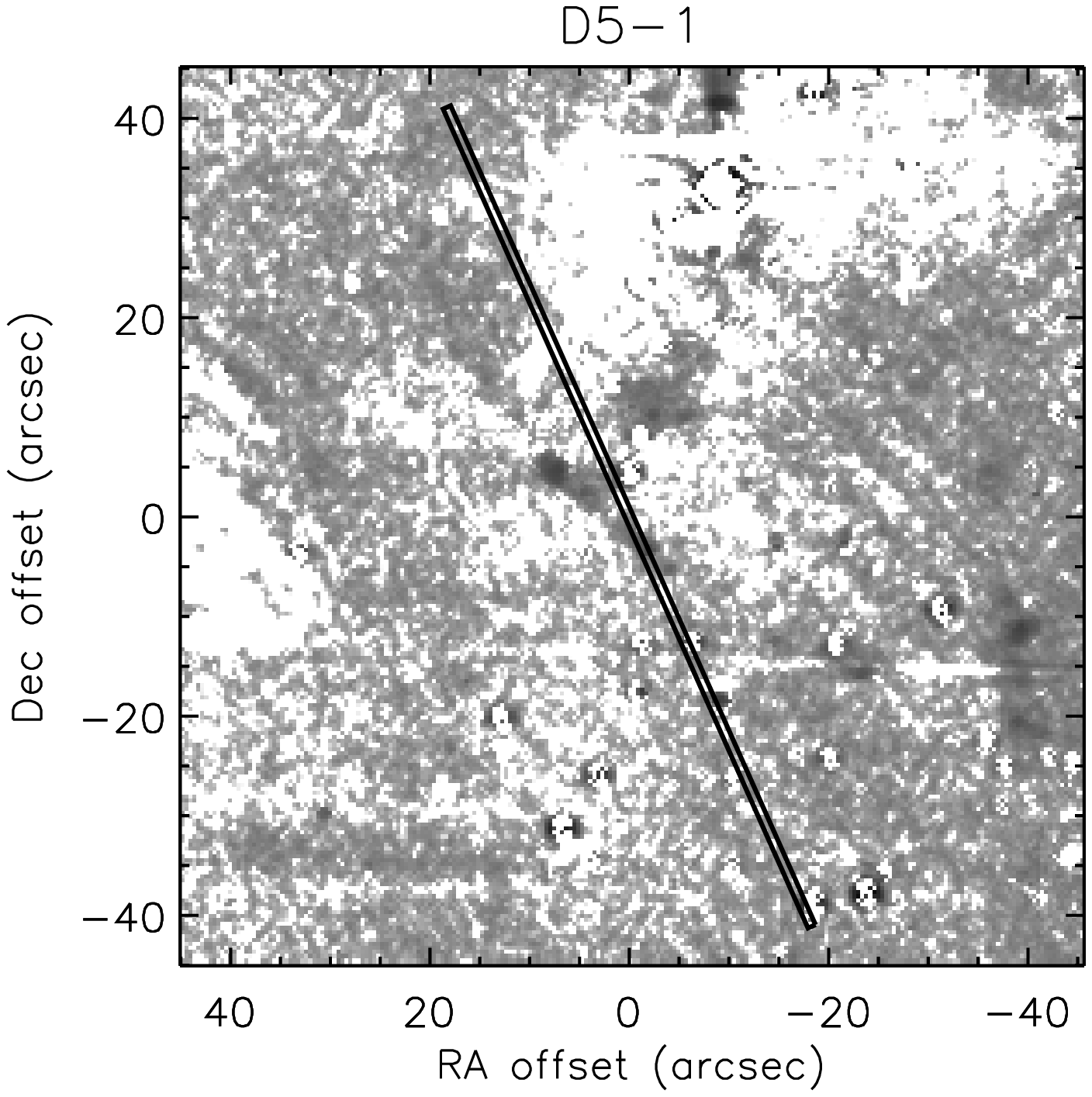}
  \epsfxsize=7.4cm       \epsfbox{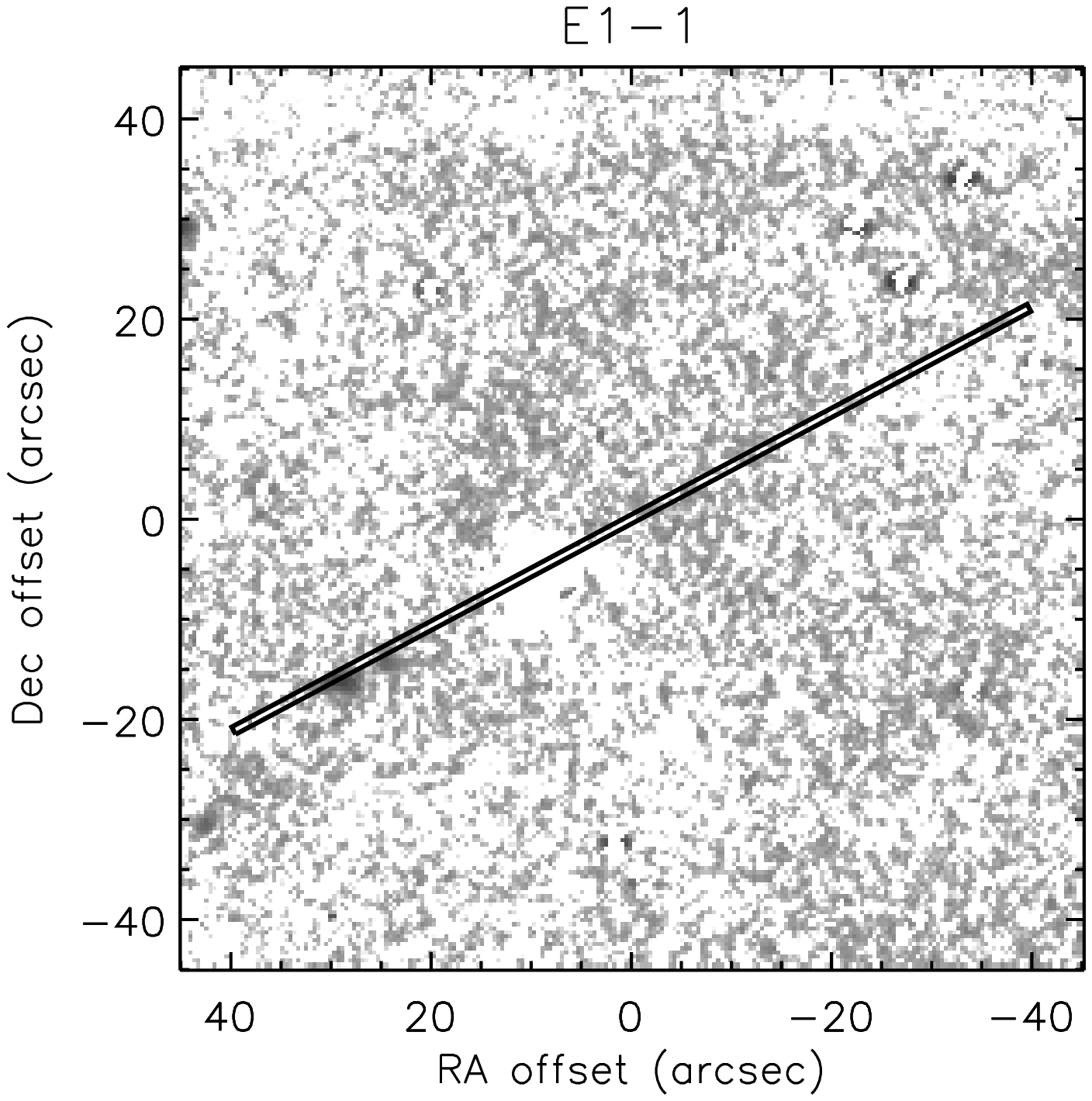}
\caption{\label{zoomplots4} Close-up images displaying  the slits for objects D\,1-5 to E\,1-1.}
  \end{center}
\end{figure*}

\begin{figure*}
  \begin{center}
  \epsfxsize=7.4cm       \epsfbox{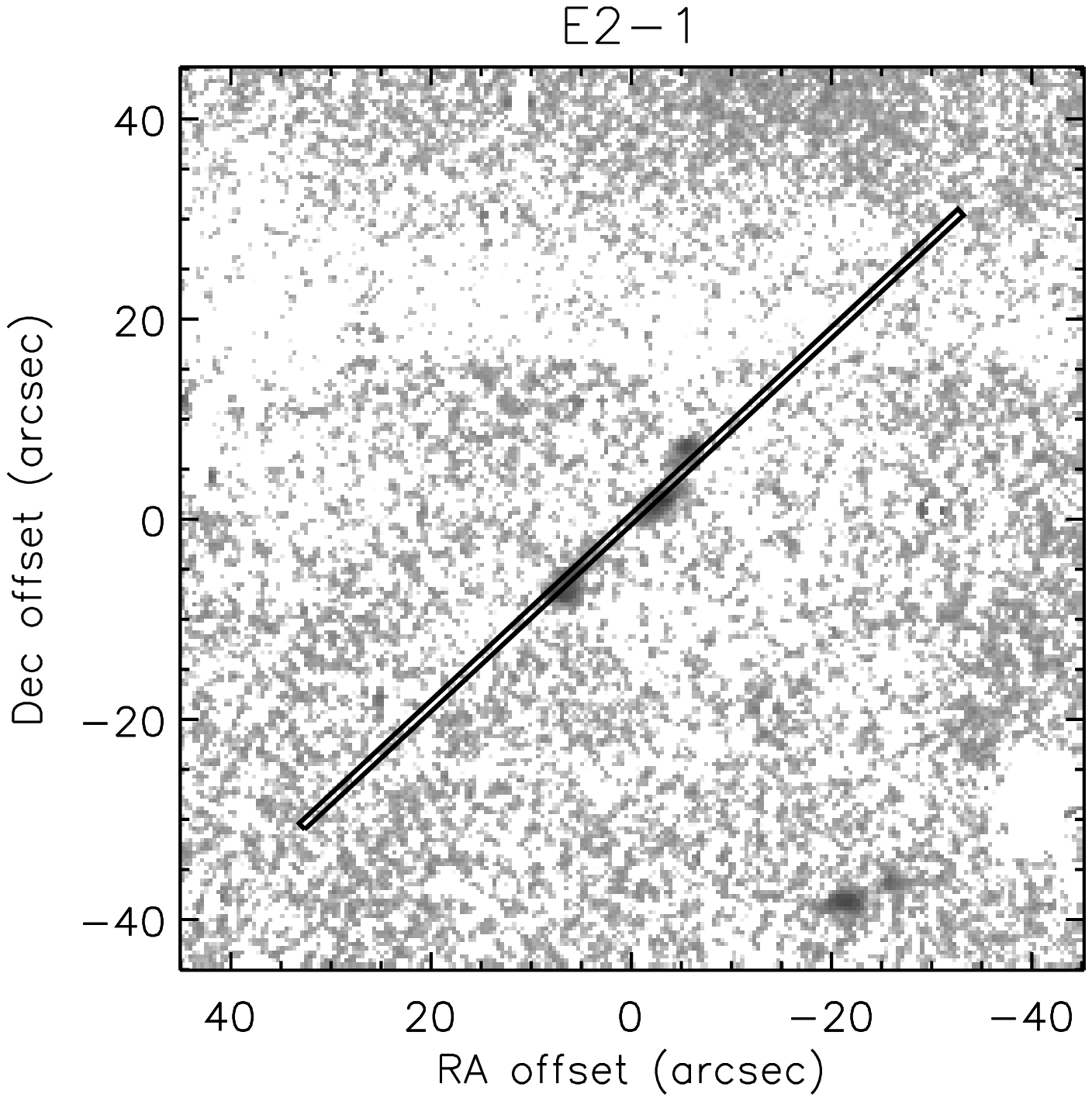}
   \epsfxsize=7.4cm      \epsfbox{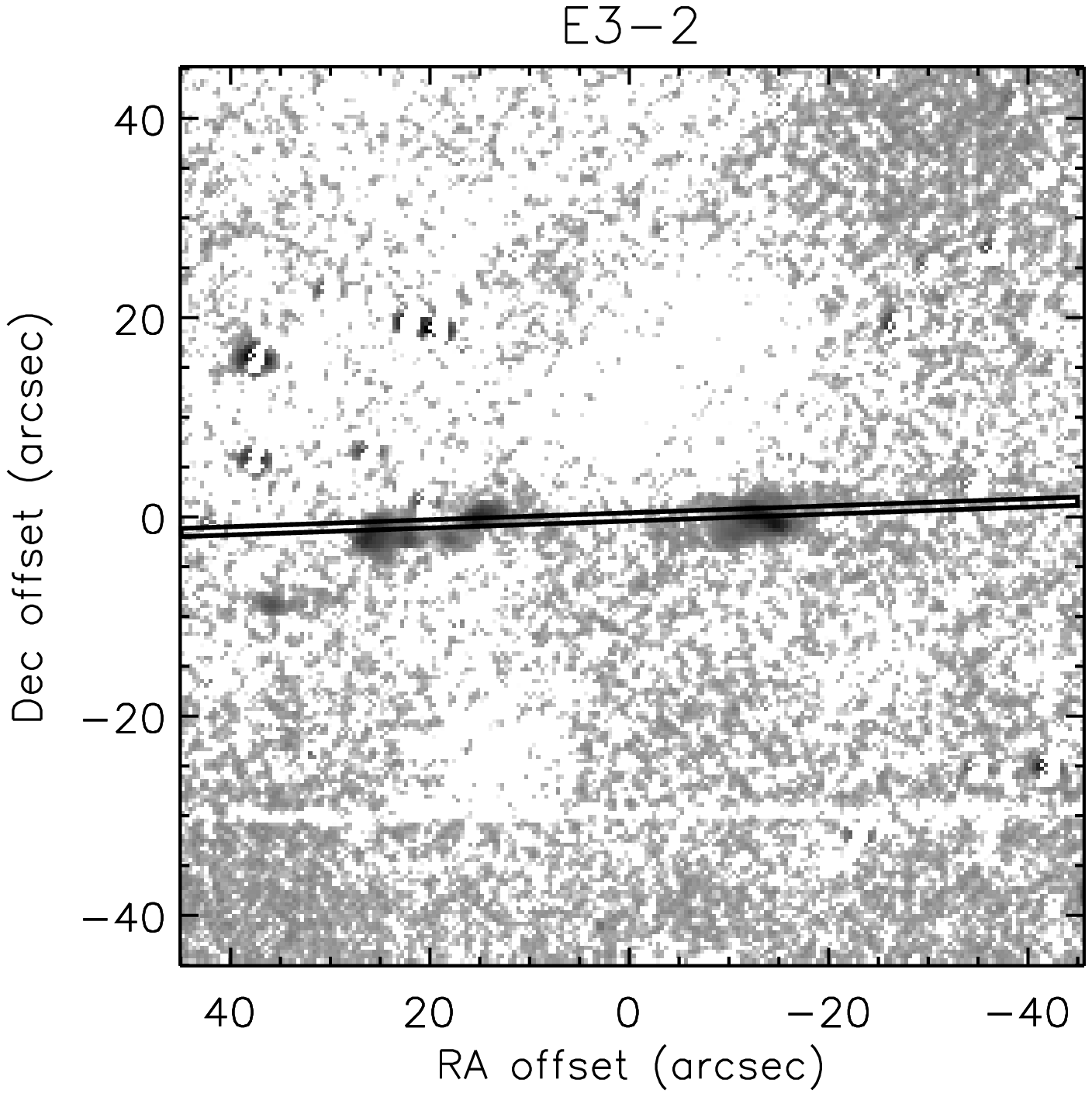}
  \epsfxsize=7.4cm       \epsfbox{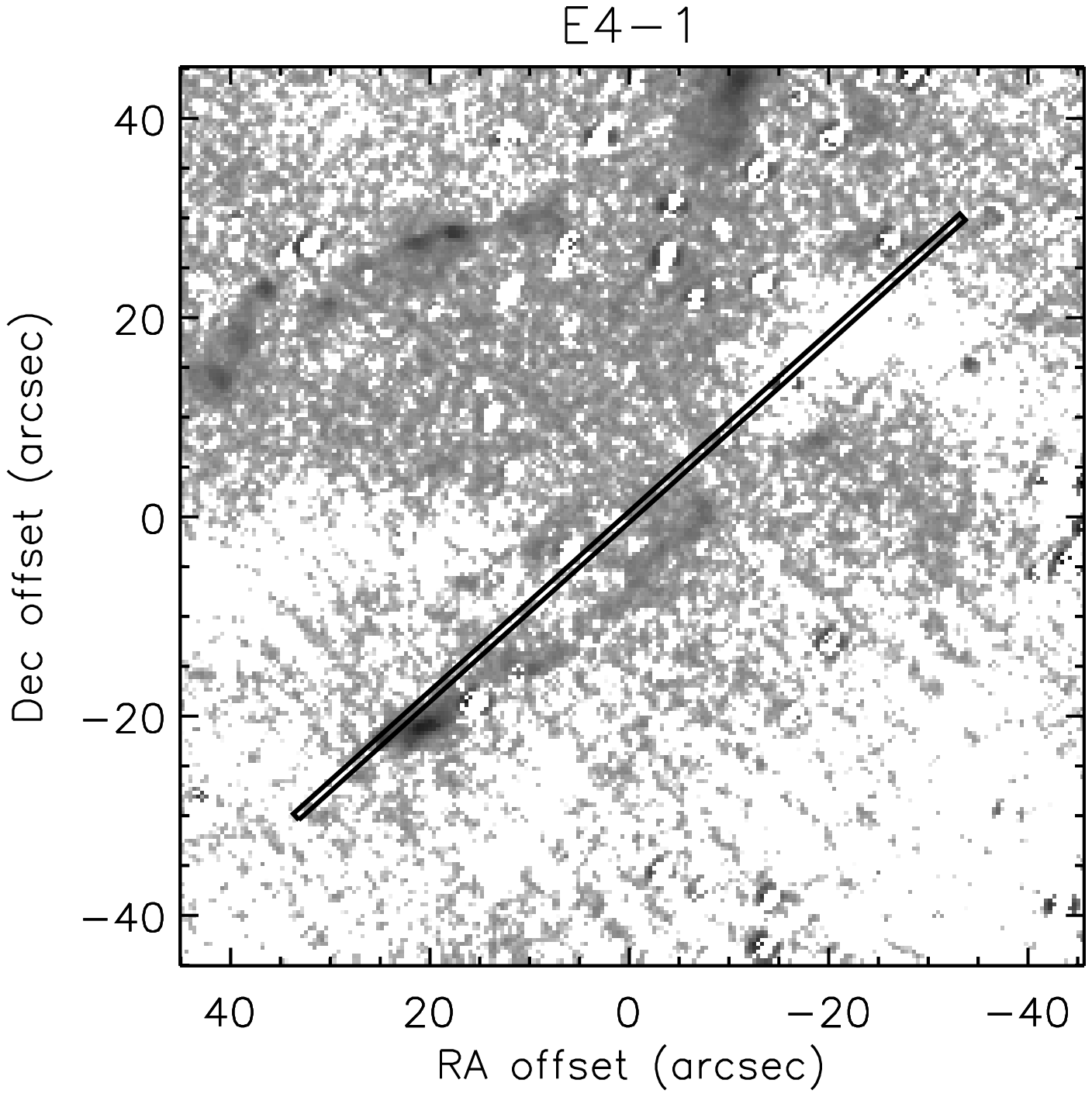}
\caption{\label{zoomplots5} Close-up images displaying  the slits for objects E\,2-1 to E\,4-1 (taken from a larger J-K-H$_2$ composite).}
  \end{center}
\end{figure*}

Figures~\ref{zoomplots1}~--~\ref{zoomplots5} display the precise areas covered by the slits using the correct width and length. This allows us to see exactly which parts were 
traversed by each slit.

\clearpage
\clearpage
\clearpage
\clearpage

\section*{Appendix B}
\label{appb}
Individual objects, their nature and the possible driving sources are discussed here in light of the radial velocity information. The sources 
were identified and demarcated on Figures A1 to A5 of \citet{2007MNRAS.374...29D} and the reader is referred to those maps.

\subsection*{Region A}
\label{intregiona}

Object~A\,2-1 has two distinct radial velocity components with a high blueshift (-83.1~km~s$^{-1}$) and low redshift (3.4~km~s$^{-1}$). On the other hand, 
A\,2-2 to the west has a moderately high redshift (28.8~km~s$^{-1}$). However, there is evidence for weak emission extending to 100~km~s$^{-1}$. This would suggest that the source `a' (ERO1) which lies inbetween the two objects  (see Fig.~A1 of \citet{2007MNRAS.374...29D}) is indeed the driving source. The object A\,2-2 forms part of a conical structure with A\,2-3 and A\,2-4, concave towards ERO1. Object A\,2-1 is tracing a counter flow to this structure. The high radial speed favours an interpretation in which the outflow is at  a skall angle to the line of sight  with the eastern component A\,2-1 moving towards the observer. 

The source driving A\,5-1 and A\,5-2 could be either protostar `d', a source in the DR21-IRS~1 infrared cluster~\citep{2007MNRAS.374...29D},   DR21-IRS~2 (a K-band source), or a combination of both. The objects appear to be aligned with DR21-IRS~2 but are in closer proximity to `d', found midway between knots A\,5-1 and A\,5-3.  Both objects, separated by $\sim $10\arcsec, have H$_2$ emitting material strongly blueshifted (at -43.7~km~s$^{-1}$ and -85.9~km~s$^{-1}$ respectively) as well as at rest which suggests both a jet and ambient shocks are present, and that the material is moving towards the observer. An outflow deceleration would suggest a source to the north.

The PV diagram for object A\,6-1 has three spatial components with A\,6-1 separated by $\sim $50$\texttt{"}$ from A\,7-1, and A\,6-2 5$\texttt{"}$ further.  A\,6-1 is a good candidate for a bow shock. However, there is no source in the \textit{Spitzer} data associated with A\,6-1, but the suggestion is a source to the south-east~\citep{2007MNRAS.374...29D} due to the shape. A\,6-1 has a small radial velocity (-10.3~km~s$^{-1}$) but this is greater than A~7-1 (1.5~km~s$^{-1}$).
This is consistent with an interpretation that these components are remnants of some earlier outflow stemming from DR21, now inactive or re-orientated.

Objects A\,9-1, A\,9-2 and A\,9-3 appear as a collimated flow parallel to the main DR21 outflow~\citep{2007MNRAS.374...29D}. These objects are also close to rest velocity (-4.9~km~s$^{-1}$, 0.9~km~s$^{-1}$ and 5.2~km~s$^{-1}$) with a radial velocity gradient that is consistent with a bipolar flow  quite close to the plane of the sky. A pre-main sequence object which lies adjacent to A~9-2 could be the driving source. 

The source for A\,10-1 is possibly  the luminous, accreting infrared source DR21D, located to the south-east as seen on Figure~A1 of  \citet{2007MNRAS.374...29D}. The object has three components, two of which (furthest from the source) are slightly blueshifted (-18.6~km~s$^{-1}$ and -16.8~km~s$^{-1}$) suggesting motion towards the observer at a small angle to the plane of the sky. However the component nearest to the probable source has the lowest velocity of the three (-0.3~km~s$^{-1}$).

\subsection*{Region B}
\label{intregionb}

Proposed driving sources for jet B\,1-1 are DR21D or DR21-IRS\,4.  \citet{2013ApJ...765L..29Z} note that B\,1-1 to B\,1-3 could be directed away from the origin of an explosion which gave rise to the entire DR21 outflow  10,000 years ago,  associated with DR21D. The blueshift of both components of B\,1-1 (-14.0~km~s$^{-1}$ and -16.9~km~s$^{-1}$) and B\,1-3 (-22.7~km~s$^{-1}$and -11.0~km~s$^{-1}$), with these speeds maintained along the linear spatial extension, is consistent with a jet motion towards the observer moderately out of the plane of the sky. There is no discernible trend in the radial velocity to point to an explosive origin or other  velocity gradient involving acceleration or deceleration.

The two distinct  PV components of B\,4-1 are slightly redshifted and blueshifted (7.9~km~s$^{-1}$and -8.1~km~s$^{-1}$ respectively). However, there is no detected source between these two components (separated by 7\arcsec\  or 10,000\,AU) to support a symmetric twin jet system. The PV diagram does not support a bow shock scenario either, suggesting a more complex scenario possibly involving source DR21-IRS\,4 to the south as seen on Figure~A2 of \citet{2007MNRAS.374...29D}. ). 

B\,6-1 has H$_2$ material blueshifted at over -50~km~s$^{-1} $ (though the peak of this is at -12.5~km~s$^{-1}$), and material is also at rest. Therefore, this 
could be an unresolved  jet Mach shock and bow shock  combination driven from protostar `f', Objects B\,6-2 and B\,6-3, found  at $\sim $42$\texttt{"}$ and $\sim $70$\texttt{"}$ 
along the slit, exhibit only minor blue (-4.5~km~s$^{-1}$) and red (4.3~km~s$^{-1}$) shifts, with the slit position indicating that a shock may be driven into ambient cloud material.

The probable source for B\,7-1 is DR21-IRS~6 (also known as protostar `e',~\citet{2007MNRAS.374...29D}) through which the slit cuts. The extended H$_2$ emission to the north is slightly blueshifted (-8.5~km~s$^{-1}$). There is more material on the south side of the source which is also blueshifted. This does not appear very clearly on the continuum-subtracted H$_2$~1-0~S(1) image as the whole area around the source is very bright.  B\,7-1 also possesses faint emission across all velocities on the PV diagram, indicative of a stellar source.

The PV diagram for B\,9-1 displays remarkable structure. The four components all have blueshifted H$_2$ emission at moderate and at high velocity (from zero offset, high speed components are  -85.3~km~s$^{-1}$, -68.7~km~s$^{-1}$, -105.4~km~s$^{-1}$ and -96.3~km~s$^{-1}$; low velocity components are -19.2~km~s$^{-1}$, -24.7~km~s$^{-1}$, -39.4~km~s$^{-1}$ and -28.4~km~s$^{-1}$).  The proposed west to east  motion is most likely being driven by protostar `h'~\citep{2007MNRAS.374...29D} located very close to the west component (alternates are ERO3, DR21-IRS~7 and DR21-IRS~8).
The radial velocity generally decreases with distance from the source. The source just happens to be in the line-of-sight with the end of the object (probably inclined at a high angle to the observer) and there is a gap between the source and the observed emitting components. This would allow for a model where the H$_2$ is entrained from the medium by the jet and accelerated to moderate velocity before being shock heated. On the other hand, the high-speed H$_2$ could be generated within a dense jet within 100\,AU of the source.

For B\,11-1 and B\,11-2 the H$_2$ emission is redshifted with the source DR21-IRS\,10 which appears as the continuous emission across the bottom of the PV plot. B\,11-1 shows the higher velocity (30.0~km~s$^{-1}$),  consistent with material being accelerated by the source then slowed by the ambient medium.

Object B\,12-1 has three faint and one bright  spatial components. They are separated (lower to upper) by about 8, 8 and 12~km~s$^{-1}$. The lowermost component is blueshifted at $\sim $-24~km~s$^{-1}$. The second component is blueshifted at $\sim $-18~km~s$^{-1}$. The third component has two elements; one is blueshifted at -90~km~s$^{-1}$, the other at -12~km~s$^{-1}$. The uppermost component (B\,12-2) is virtually at local rest and contains 74.5$\%$ of the total flux.

\subsection*{Region C}
\label{intregionc}

Objects C\,1-2 and C\,1-3 are slightly redshifted (3.7--6.6~km~s$^{-1}$), suggesting  a bow motion away from the observer. The source here is thought to be VLA\,1 in the centre of Region~C \citep{1997ApJ...489..744T}, which produces a large-scale bipolar CO outflow first identified by~\cite{1998AJ....115.1118D}. 
On PV plot C\,1-4, objects C\,1-1  ($\sim $15$\texttt{"}$), C\,1-4  ($\sim $15$\texttt{"}$) and C\,1-5  ($\sim $38$\texttt{"}$) are suspected to be part of this large-scale bow shock. They show slight redshifts in our PV diagrams. The middle knot C\,1-5 may be a part of a collimated jet driving the bow shock, driven by W75N-IRS\,2~\citep{2007MNRAS.374...29D}. However, we find here no evidence for radial motion.

Object C\,2-1 has two spatial components, the lower one being resolved in space by about 30$\texttt{"}$.  It is at rest  (blueshifted at -2.8~km~s$^{-1}$), though the compact arc $\sim $20$\texttt{"}$ ahead of it is redshifted (13.4~km~s$^{-1}$). No source is obvious for this collimated structure. 
If the source lies toward the more active region to the south-west, then C\,2-1 would represent shocked ambient material while the arc could be the working surface. 

The radial velocity of  C\,6-1 varies from rest to 50~km~s$^{-1}$, over a short distance of about 1.5\arcsec. This could be the result of the slit cutting through two shocks with a distance between the two fronts of 2~$\times$~10$^{16}$~cm. Thus,  C\,6-1 could be a bow shock and jet knot as proposed by \citet{2007MNRAS.374...29D}. The source is thought to be the same as for C\,1-3 - VLA\,1. In which case, both lobes of the W75N bow display motion away from the observer. 

Object C\,8-1 is being driven either by protostars `j' or `k' and shows H$_2$ emission from material at rest to -50~km~s$^{-1}$, blueshifted. This looks like a well collimated jet but unfortunately the slit just missed most of the material e.g. C\,8-5 through to C\,8-2.

\subsection*{Region D}
\label{intregiond}

The objects in the PV diagrams for D\,1-1, D\,1-5 and D~1-8 may belong to the same large-scale flow. The PV plot  for object D\,1-1 has four spatial components. The three upper components correspond to D\,1-2. Objects D\,1-1 and D\,1-2 are both blueshifted (-15.0~km~s$^{-1}$ and -26.8~km~s$^{-1}$) with a further small arc of blueshifted emission 35\arcsec\  to the south.
 D\,1-1 has material varying in radial velocity from 0~km~s$^{-1}$ to -50~km~s$^{-1}$, while D\,1-2 is moving at velocities of -20~km~s$^{-1}$ to -60~km~s$^{-1}$. The higher velocities of D\,1-2 can be explained by it being located nearer to the probable sources, with D\,1-1 being further ahead and decelerated by the ambient medium. The possible sources are W75N-IRS\,10 and~11~\citep{2007MNRAS.374...29D}. 
 
Object D\,1-5 has material at all velocities from 0~km~s$^{-1}$ to -120~km~s$^{-1}$. These high velocities could be related to the proximity of the suggested sources W75N-IRS\,10 and~11 (much closer than D\,1-2). 

Molecular hydrogen object D\,1-8 varies in velocity across the slit  from a blueshifted velocity of -20~km~s$^{-1}$ to a redshifted velocity of 30~km~s$^{-1}$. The object is predominantly 
redshifted. This could suggest that the entire W75N-IRS\,10/11 flow is orientated such that the D\,1-1 lobe points towards, and the D\,1-8 lobe points away, from the observer. D\,1-8 is also a promising candidate to compare with bow shock models.

The diagram for object D\,3-1 has two spatial components, the upper of which is D\,3-2.  D\,3-1 and D\,3-2 are both blueshifted (both -20.0~km~s$^{-1}$). They appear as elongated knots. Possible sources are W75N-IRS\,7 and~8 (including protostar `m')~\cite{2007MNRAS.374...29D} or W75N-IRS\,9 which is also in close proximity. Although the peak velocity for D\,3-2 is the same as D\,3-1,
D\,3-2 does not possess  H$_2$ at rest. Hence it is not clear why this appears as the most advanced component.

D\,4-1 is a collimated jet with an unknown source~\citep{2007MNRAS.374...29D}. The parts that can be seen in the PV diagram, D\,4-1 (bottom) and D\,4-5 (top of slit), are both slightly blueshifted (-4.5~km~s$^{-1}$ and -8.9~km~s$^{-1}$) which indicates that they are unlikely to represent the opposite bows of a bipolar outflow with the source located centrally.

D\,5-1 could be part of a collimated flow with D\,5-2 and D\,5-3 (not covered by these observations). If this is the case then no source is readily apparent. However it could be associated with sources W75N-IRS\,7 and~8 (including protostar `m') due to the apparent proximity. It has material with H$_2$ radial velocity ranging from 0~km~s$^{-1}$ to -50~km~s$^{-1}$.

\subsection*{Region E}
\label{intregione}

The features in Region E lie on the eastern edge of the Lynds dark cloud L906.
The H$_2$ emission in object E\,1-1  is slightly redshifted (3.2~km~s$^{-1}$). The source is thought to be associated with a nebulous K-band source L609E-IRS\,6~\citep{2007MNRAS.374...29D}, which lies close by. The material detected along the slit is very faint in the PV diagram. The PV plot  for E\,2-1 has two components, separated by $\sim $12.5$\texttt{"}$.  E\,2-1 has H$_2$ emission material varying in velocity from 0~km~s$^{-1}$ up to -50~km~s$^{-1}$, while E\,2-2 is redshifted from 5~km~s$^{-1}$ to 60~km~s$^{-1}$ with the main flux concentration at $\sim $30~km~s$^{-1}$. Object E\,2-3 does not appear on the echelon observation.

The PV diagram for E\,3-2 yields some interesting objects. E\,3-2 itself has two components, each with H$_2$ emission material at rest and at $\sim $50~km~s$^{-1}$. The bright E\,3-3 has material slightly redshifted with a mean at $\sim $10~km~s$^{-1}$.

Object E\,4-1 has H$_2$ emission which is redshifted at $\sim $20~km~s$^{-1}$, driven by the source L609E-IRS\,1 (protostar `n',~\cite{2007MNRAS.374...29D}). There is a faint line of material appearing down the slit, all moving at a slightly redshifted velocity. This could be evidence for it to be part of a collimated jet in the plane of the sky.

\label{lastpage}
\
\end{document}